\def\isarxiv{1}

\newif\ifarxiv
\newif\ifnotarxiv

\ifdefined\isarxiv
\arxivtrue
\fi

\ifarxiv\notarxivfalse
\else\notarxivtrue
\fi

\ifnotarxiv
\documentclass[letterpaper,twocolumn,10pt]{article}
\PassOptionsToPackage{hyphens}{url}
\usepackage{usenix}
\usepackage[available]{usenixbadges}

\else
\documentclass[sigconf]{acmart}
\fi
\ifnotarxiv
\usepackage{amsmath,amssymb,amsfonts}
\usepackage{cite}
\else
\usepackage{amsmath,amsfonts}
\fi

\newcommand{\revise}[1]{{#1}}
\usepackage{pifont}
\usepackage{algorithmic}
\usepackage{textcomp}
\usepackage[dvipsnames]{xcolor}
\usepackage{paralist}
\usepackage{enumitem}
\usepackage{hyperref}
\usepackage{cleveref}
\usepackage{float}
\usepackage{subcaption}
\usepackage{fvextra}
\usepackage[textsize=footnotesize]{todonotes}
\usepackage{glossaries}
\glsdisablehyper
\usepackage{caption}
\usepackage{subcaption}
\usepackage{listings}
\usepackage{xspace}
\usepackage{comment}
\usepackage{fancyvrb}
\usepackage{graphicx,txfonts}
\usepackage{multirow}
\usepackage{tikz}
\usepackage{multicol}
\usepackage{verbatim}
\usepackage{framed}
\usepackage{setspace}
\usepackage{fontenc}
\usepackage{bytefield}
\usepackage{booktabs}
\usepackage[dvipsnames]{xcolor}
\usepackage{mwe}
\usepackage{wrapfig}
\usepackage{color, colortbl}
\usepackage{threeparttable}
\usepackage{array}
\usepackage{adjustbox}
\usepackage{wasysym}
\usepackage{xparse}
\usepackage[htt]{hyphenat}
\usepackage[hashEnumerators,smartEllipses]{markdown}
\usepackage{todonotes}
\usepackage{censor}
\newif\ifabridged
\newif\ifnotabridged
\newif\ifanonymous
\newif\ifnotanonymous
\newif\ifreviewer
\newif\ifnotreviewer
\newif\iftags
\newif\ifnottags

\ifdefined\isabridged
\abridgedtrue
\fi

\ifabridged\notabridgedfalse
\else\notabridgedtrue
\fi

\ifdefined\isanonymous
\anonymoustrue
\fi

\ifanonymous\notanonymousfalse
\else\notanonymoustrue
\fi

\ifdefined\isreviewer
\reviewertrue
\fi

\ifreviewer\notreviewerfalse
\else\notreviewertrue
\fi

\ifdefined\hashtags
\tagstrue
\fi

\iftags\nottagsfalse
\else\nottagstrue
\fi

\newcommand{\colorbitbox}[3]{
  \sbox0{\bitbox{#2}{#3}}
  \makebox[0pt][l]{\textcolor{#1}{\rule[-\dp0]{\wd0}{\ht0}}}
  \bitbox{#2}{#3}
}

\makeatletter
\def\mathcolor#1#{\@mathcolor{#1}}
\def\@mathcolor#1#2#3{%
  \protect\leavevmode
  \begingroup
    \color#1{#2}#3%
  \endgroup
}
\makeatother

\glsaddkey
    {hyphenated}
    {\relax}
    {\glsentryhyph}
    {\Glsentryhyph}
    {\glshyph}
    {\Glshyph}
    {\GLShyph}

\DeclareFontFamily{OT1}{mathc}{}
\DeclareFontShape{OT1}{mathc}{m}{it}{<-> mathc10}{}
\DeclareMathAlphabet{\mathabxcal}{OT1}{mathc}{m}{it}
\newacronym{abi}{ABI}{application binary interface}
\newacronym{adv}{$\mathabxcal{Adv}$}{adversary}
\newacronym{alloc}{$\mathabxcal{Alloc}$}{allocator}
\newacronym{alu}{ALU}{arithmetic logic unit}
\newacronym{api}{API}{application programming interface}
\newacronym{dma}{DMA}{direct memory access}
\newacronym[hyphenated={colored-capability},longplural={colored capabilities}]{cc}{CC}{colored capability}
\newacronym{cisa}{CISA}{Cybersecurity & Infrastructure Security Agency}
\newacronym{cheri}{CHERI}{Capability Hardware Enhanced RISC Instructions}
\newacronym{clc}{\texttt{clc}}{load capability via capability}
\newacronym{cpu}{CPU}{central processing unit}
\newacronym{crg}{\texttt{CRG}}{capability read generation }
\newacronym{csc}{\texttt{csc}}{store capability via capability}
\newacronym{csp}{\texttt{csp}}{stack pointer capability}
\newacronym{csr}{CSR}{control and status register}
\newacronym{cve}{CVE}{common vulnerability enumeration}
\newacronym{cwe}{CWE}{common weakness enumeration}
\newacronym{cw}{\texttt{CW}}{capability write}
\newacronym{df}{DF}{double-free}
\newacronym{eda}{EDA}{electronic design automation}
\newacronym{fpga}{FPGA}{field-programmable gate array}
\newacronym[longplural={instruction set architectures}]{isa}{ISA}{instruction-set architecture}
\newacronym{ir}{IR}{intermediate representation}
\newacronym{ip}{IP}{intellectual property}
\newacronym{jit}{JIT}{just-in-time}
\newacronym{lsu}{LSU}{load-store unit}
\newacronym{lsq}{LSQ}{Load/Store Queue}
\newacronym{gep}{\textsf{GEP}}{\textsf{GetElementPtr}}
\newacronym[hyphenated={provenance-identifier},longplural={provenance identifiers}]{pid}{provenance ID}{provenance identifier}
\newacronym{ncsc}{NCSC}{National Cyber Security Centre}
\newacronym{nist}{NIST}{National Institute of Standards and Technology}
\newacronym{mte}{MTE}{Memory Tagging Extension}
\newacronym{mrs}{MRS}{\texttt{malloc} revocation shim}
\newacronym[hyphenated={operating-system}]{os}{OS}{operating system}
\newacronym{otype}{\texttt{otype}}{object type}
\newacronym{otth}{OTYPETH}{otype threshold}
\newacronym{u}{\texttt{U}}{user mode access allowed}
\newacronym{ucrg}{\texttt{UCRG}}{user capability read generation}
\newacronym{pa}{PA}{pointer authentication}
\newacronym{pc}{PC}{program counter}
\newacronym{pcc}{\texttt{pcc}}{program counter capability}
\newacronym{picasso}{\texttt{PICASSO}}{Provenance Indirection-enabled CHERI Architecture for Scarcely Swept Objects}
\newacronym{poc}{PoC}{proof-of-concept}
\newacronym{pte}{PTE}{page table entry}
\newacronym{ptlb}{PTLB}{provenance-translation lookaside buffer}
\newacronym{pvb}{PVB}{provenance-validity bit}
\newacronym{pvt}{PVT}{provenance-validity table}
\newacronym{pvtr}{PVTR}{provenance-validity table register}
\newacronym{ross}{\texttt{ROSS}}{Revocation-Orchestrating System Service}
\newacronym{rtl}{RTL}{register-transfer level}
\newacronym{rtos}{RTOS}{real-time operating system}
\newacronym{rss}{RSS}{resident set size}
\newacronym{rv64y}{RV64Y}{64-bit RISC-V}
\newacronym{rof}{RoF}{revoke-on-free}

\newacronym{sard}{SARD}{Software Assurance Reference Dataset}
\newacronym{sp}{SP}{stack pointer}
\newacronym{soc}{SoC}{system-on-chip}
\newacronym{tlb}{TLB}{translation lookaside buffer}
\newacronym{toctou}{TOCTOU}{time-of-check to time-of-use}
\newacronym{tps}{TPS}{transactions per second}
\newacronym{uaf}{UAF}{use-after-free}
\newacronym{uar}{UAR}{use-after-reallocation}
\newacronym{unr}{\texttt{unr}}{unit number}
\newacronym{wns}{WNS}{worst negative slack}
\newacronym{qps}{QPS}{Queries Per Second}

\iftags
\newcommand{\hashtagl}[1]{\marginpar{$\!\!\!\!\!\!\!\!\!\!$\color{blue}\em\##1\vspace{-7mm}}}
\newcommand{\hashtagr}[1]{\marginpar{$\!\!$\color{blue}\em\##1\vspace{-7mm}}}
\fi
\ifnottags
\newcommand{\hashtagl}[1]{}
\newcommand{\hashtagr}[1]{}
\fi

\newcommand*\circledblk[1]{\tikz[baseline=(char.base)]{
            \node[shape=circle,fill,inner sep=0pt] (char) {\textcolor{white}{#1}};}}

\newcommand{\dOne}{\ding{182}\xspace}
\newcommand{\dTwo}{\ding{183}\xspace}
\newcommand{\dThree}{\ding{184}\xspace}
\newcommand{\dFour}{\ding{185}\xspace}
\newcommand{\dFive}{\ding{186}\xspace}
\newcommand{\dSix}{\ding{187}\xspace}
\newcommand{\dSeven}{\ding{188}\xspace}
\newcommand{\dEight}{\ding{189}\xspace}
\newcommand{\dNine}{\ding{190}\xspace}
\newcommand{\dTen}{\ding{191}\xspace}
\newcommand{\dEleven}{\circledblk{\scriptsize11}\xspace}
\newcommand{\dCOne}{\ding{192}\xspace}
\newcommand{\dCTwo}{\ding{193}\xspace}
\newcommand{\dCThree}{\ding{194}\xspace}
\newcommand{\dCFour}{\ding{195}\xspace}
\newcommand{\dCFive}{\ding{196}\xspace}
\newcommand{\dCSix}{\ding{197}\xspace}
\newcommand{\dCSeven}{\ding{198}\xspace}
\newcommand{\dCEight}{\ding{199}\xspace}
\newcommand{\dCNine}{\ding{200}\xspace}

\newlength{\dingwidth}
\definecolor{mGreen}{rgb}{0,0.6,0}
\definecolor{mGray}{rgb}{0.5,0.5,0.5}
\definecolor{lGray}{rgb}{0.9,0.9,0.9}
\definecolor{mPurple}{rgb}{0.58,0,0.82}
\definecolor{backgroundColour}{rgb}{0.95,0.95,0.92}

\newcommand\LSTSize{\fontsize{7}{7.2}\selectfont}
\newcommand*\LSTfont{\LSTSize\ttfamily\SetTracking{encoding=*}{-0}\lsstyle}

\newcommand\realnumberstyle[1]{#1}

\makeatletter
\newcommand{\zebra}[2]{%
    {\realnumberstyle{#2}}%
    \begingroup
    \lst@basicstyle
    \ifodd\value{lstnumber}%
        \color{#1}%
        \rlap{\hspace*{\lst@numbersep}%
        \color@block{\linewidth}{\ht\strutbox}{\dp\strutbox}%
        }%
    \fi
    \endgroup
}
\makeatother
\newcommand{\breakingperiod}{
\nobreak\hspace{0pt}.\penalty0
}

\ExplSyntaxOn{}

\NewDocumentCommand{\longword}{m}
 {
  \texttt
   {
    \seq_set_split:Nnn \l_michael_lw_seq { . } { #1 }
    \seq_use:Nn \l_michael_lw_seq { \breakingperiod }
   }
 }

\ExplSyntaxOff{}

\lstdefinestyle{CStyle}{
    backgroundcolor=\color{backgroundColour},
    commentstyle=\color{mGreen},
    keywordstyle=\color{magenta},
    numberstyle=\tiny\zebra{lGray},
    stringstyle=\color{mPurple},
    basicstyle=\LSTfont,
    breakatwhitespace=false,
    breaklines=true,
    captionpos=b,    
    escapeinside={\%*}{*)},
    keepspaces=true,
    numbers=left,
    numbersep=5pt,
    showspaces=false,
    showstringspaces=false,
    showtabs=false,
    tabsize=2,
    language=C,
    classoffset=1,
    morekeywords={enter_domain, __writebeforeread, sdrob_call,sdrob_enter,sdrob_malloc,sdrob_exit,sdrob_init,sdrob_destroy,sdrob_free,sdrob_deinit,sdrob_dprotect, sdrad_call,sdrad_enter,sdrad_malloc,sdrad_exit,sdrad_init,sdrad_destroy,sdrad_free,sdrad_deinit,sdrad_dprotect},
    keywordstyle=\color{orange},
    classoffset=2,
    morekeywords={EXECUTION_DOMAIN,ISOLATED_DOMAIN,RETURN_TO_CURRENT,DATA_DOMAIN,SUCCESSFUL_RETURNED,EXECUTION_DOMAIN,NONISOLATED,RETURN_HERE,RETURN_TO_PARENT,NO_HEAP_MERGE,HEAP_MERGE,OK,OK,MALLOC_FAILED,WRITE_ENABLE,READ_ENABLE,ACCESSIBLE,INACCESSIBLE,ACCESSIBLE_DOMAIN,INACCESSIBLE_DOMAIN,udi_t},
    keywordstyle=\color{brown},
    classoffset=0,
}
\crefname{lstlisting}{listing}{listings}
\Crefname{lstlisting}{Listing}{Listings}

\presetkeys{todonotes}{fancyline}{}
\Crefname{section}{\S$\!$}{\S\S$\!$}

\newcounter{tncnt}

\newcounter{mgcnt}

\newcounter{jtcnt}

\newcounter{hecnt}

\newcounter{rscnt}

\renewcommand{\paragraph}{\noindent\textbf}

\def\BibTeX{{\rm B\kern-.05em{\sc i\kern-.025em b}\kern-.08em
    T\kern-.1667em\lower.7ex\hbox{E}\kern-.125emX}}
\newcommand{\Adv}{\gls{adv}\xspace}
\newcommand{\Alloc}{\gls{alloc}\xspace}
\newcommand{\PICASSO}{\acrshort{picasso}\xspace}

\newcommand{\cc}{\glsentrydesc{cc}\xspace}
\newcommand{\ccs}{\glsentrylongpl{cc}\xspace}
\newcommand{\CC}{\Glsentrydesc{cc}\xspace}
\newcommand{\CCs}{\Glsentrylongpl{cc}\xspace}
\newcommand{\cchyph}{\glsentryhyph{cc}\xspace}

\newcommand{\ccproc}{provenance track\xspace}

\newcommand{\ccprocing}{\ccproc{}ing\xspace}

\newcommand{\ccindprocing}{indirect \ccproc{}ing\xspace}

\newcommand{\ccTable}{\Gls{pvt}\xspace}

\newcommand{\cctable}{\gls{pvt}\xspace}
\newcommand{\cctablelong}{\glsentrydesc{pvt}\xspace}

\newcommand{\ccstate}{\ccprocing state\xspace}

\newcommand{\ccactive}{valid\xspace}
\newcommand{\ccinactive}{invalid\xspace}
\newcommand{\ccdeactivate}{invalidate\xspace}
\newcommand{\ccdeactivating}{invalidating\xspace}
\newcommand{\ccdeactivated}{invalidated\xspace}
\newcommand{\ccDeactivated}{Invalidated\xspace}
\newcommand{\ccpid}{\glsentrydesc{pid}\xspace}
\newcommand{\ccpidhyph}{\glsentryhyph{pid}\xspace}
\newcommand{\ccpids}{\glsentrylongpl{pid}\xspace}
\newcommand{\ccshortpid}{identifier\xspace}
\newcommand{\ccshortpids}{identifiers\xspace}

\newcommand{\ccPtlb}{\Gls{ptlb}\xspace}

\newcommand{\caprevoke}{\texttt{caprevoke()}\xspace}
\newcommand{\cherirevokegetshadow}{\texttt{cheri\_revoke\_get\_shadow()}\xspace}
\newcommand{\execve}{\texttt{execve()}\xspace}

\newcommand{\mmap}{\texttt{mmap()}\xspace}
\newcommand{\free}{\texttt{free()}\xspace}

\newcommand{\SPECOverheadRelativeToCheri}{$\approx$~{5\%}\xspace}
\newcommand{\SQLiteOverheadRelativeToCheri}{{8\%}\xspace}

\newcommand{\gRPCOverheadRelativeToCheri}{{3\%}\xspace}

\newcommand{\CheriBSDVersion}{CheriBSD 25.03\xspace}

\ifnotanonymous
\StopCensoring
\fi

\makeatletter
\def\bstctlcite{\@ifnextchar[{\@bstctlcite}{\@bstctlcite[@auxout]}}
\def\@bstctlcite[#1]#2{\@bsphack
  \@for\@citeb:=#2\do{%
    \edef\@citeb{\expandafter\@firstofone\@citeb}%
    \if@filesw\immediate\write\csname #1\endcsname{\string\citation{\@citeb}}\fi}%
  \@esphack}
\makeatother

\makeatletter
\newcommand{\myfnsymbol}[1]{%
  \expandafter\@myfnsymbol\csname c@#1\endcsname
}
\newcommand{\@myfnsymbol}[1]{%
  \ifcase #1
  \or \TextOrMath{\textasteriskcentered}{*}%
  \or \TextOrMath{\textdagger}{\dagger}%
  \or \TextOrMath{$\ddagger$}{\ddagger}%
  \or \TextOrMath{\textsection}{\S}%
  \or \TextOrMath{\textbardbl}{\|}%
  \or \TextOrMath{\P}{\P}%
  \or \TextOrMath{\#}{\#}%
  \or \TextOrMath{$\Delta$}{\Delta}%
  \fi
}
\newcommand{\affiliationA}{\@myfnsymbol{1}}
\newcommand{\affiliationB}{\@myfnsymbol{2}}
\newcommand{\contributedatericsson}{\@myfnsymbol{3}}
\newcommand{\affiliationC}{\@myfnsymbol{4}}
\newcommand{\affiliationD}{\@myfnsymbol{5}}
\newcommand{\affiliationE}{\@myfnsymbol{6}}
\newcommand{\affiliationF}{\@myfnsymbol{7}}
\newcommand{\affiliationG}{\@myfnsymbol{7}}
\makeatother

\begin{document}
\ifarxiv
\bstctlcite{ACMart:BSTcontrol}
\else
\bstctlcite{PlainUrl:BSTcontrol}
\fi

\ifnotarxiv
\title{\Large \bf \PICASSO: Scaling CHERI Use-After-Free Protection\\ to Millions of Allocations using {\color{red}C}{\color{orange}o}{\color{yellow}l}{\color{green}o}{\color{blue}r}{\color{Purple}}e{\color{violet}d} {\color{red}C}{\color{orange}a}{\color{yellow}p}{\color{LimeGreen}a}{\color{green}b}{\color{ForestGreen}i}{\color{Cyan}l}{\color{NavyBlue}i}{\color{blue}t}{\color{Purple}i}{\color{violet}e}{\color{Magenta}s}}
\else
\title{\PICASSO: Scaling CHERI Use-After-Free Protection\\ to Millions of Allocations using {\color{red}C}{\color{orange}o}{\color{yellow}l}{\color{green}o}{\color{blue}r}{\color{Purple}}e{\color{violet}d} {\color{red}C}{\color{orange}a}{\color{yellow}p}{\color{LimeGreen}a}{\color{green}b}{\color{ForestGreen}i}{\color{Cyan}l}{\color{NavyBlue}i}{\color{blue}t}{\color{Purple}i}{\color{violet}e}{\color{Magenta}s}}
\fi
\ifanonymous
\author{
{\rm Your N.\ Here}\\
Your Institution
\and
{\rm Second Name}\\
Second Institution
}

\else
\ifarxiv
\settopmatter{printacmref=false}
\renewcommand\footnotetextcopyrightpermission[1]{}
\pagestyle{plain} 
\renewcommand{\shortauthors}{Gülmez et al.}
\settopmatter{authorsperrow=1}
\fi
\author{Merve G\"{u}lmez\textsuperscript{\affiliationA}, Ruben Sturm\textsuperscript{\affiliationB\contributedatericsson}, Hossam ElAtali\textsuperscript{\affiliationC}, Håkan Englund\textsuperscript{\affiliationA},\\Jonathan Woodruff\textsuperscript{\ \affiliationD}, N. Asokan\textsuperscript{\affiliationC,\affiliationE}, Thomas Nyman\textsuperscript{\affiliationF}\\
\textit{\textsuperscript{\affiliationA}Ericsson Security Research, \textsuperscript{\affiliationB}DistriNet, KU Leuven, \textsuperscript{\affiliationC}University of Waterloo,}\\\textit{ \textsuperscript{\affiliationD}University of Cambridge,\textsuperscript{\affiliationE}KTH Royal Institute of Technology, \textsuperscript{\affiliationF}Ericsson Product Security} \\
\{merve.gulmez,hakan.englund,thomas.nyman\}@ericsson.com, ruben.sturm@kuleuven.be\\hossam.elatali@uwaterloo.ca, jonathan.woodruff@cl.cam.ac.uk, asokan@acm.org}
\fi
\renewcommand{\thefootnote}{\myfnsymbol{footnote}}

\ifarxiv
  \begin{abstract}

\ifarxiv
\small
\fi

While the \acrshort{cheri} \glsdesc{isa} extensions for capabilities enable strong spatial memory safety, \acrshort{cheri} lacks built-in \emph{temporal safety}, particularly for heap allocations.
Prior attempts to augment CHERI with temporal safety
fall short in terms of scalability, memory overhead, and incomplete security guarantees due to periodical sweeps of the system's memory to individually revoke stale capabilities.

We address these limitations by introducing \emph{\ccs} that add a controlled form of \emph{indirection} to \acrshort{cheri}’s capability model.
This enables \emph{\ccprocing} of capabilities to their respective allocations via a hardware-managed \emph{\cctablelong}, allowing bulk retraction of dangling pointers without needing to quarantine freed memory.
\CCs significantly reduce the frequency of capability revocation sweeps while improving security.

We realize \ccs in \acrshort{picasso}, an extension of the CHERI-RISC-V architecture on a speculative out-of-order \acrshort{fpga} softcore (CHERI-Toooba).
We also integrate \cchyph support into the CheriBSD \acrshort{os} and \acrshort{cheri}-enabled Clang/LLVM toolchain.
Our evaluation shows effective mitigation of use-after-free and double-free bugs across all heap-based temporal memory-safety vulnerabilities in NIST Juliet test cases\revise{, real-world \acrshort{cve}s}, only a small performance overhead on SPEC CPU benchmarks (\SPECOverheadRelativeToCheri{} g.m.), less latency, and more consistent performance in long-running SQLite, PostgreSQL, and gRPC workloads compared to prior work.

\end{abstract}
  \maketitle
  \renewcommand{\shortauthors}{Gülmez et al.}
\renewcommand{\shorttitle}{PICASSO}

\else
  \maketitle
  \begin{abstract}

\ifarxiv
\small
\fi

While the \acrshort{cheri} \glsdesc{isa} extensions for capabilities enable strong spatial memory safety, \acrshort{cheri} lacks built-in \emph{temporal safety}, particularly for heap allocations.
Prior attempts to augment CHERI with temporal safety
fall short in terms of scalability, memory overhead, and incomplete security guarantees due to periodical sweeps of the system's memory to individually revoke stale capabilities.

We address these limitations by introducing \emph{\ccs} that add a controlled form of \emph{indirection} to \acrshort{cheri}’s capability model.
This enables \emph{\ccprocing} of capabilities to their respective allocations via a hardware-managed \emph{\cctablelong}, allowing bulk retraction of dangling pointers without needing to quarantine freed memory.
\CCs significantly reduce the frequency of capability revocation sweeps while improving security.

We realize \ccs in \acrshort{picasso}, an extension of the CHERI-RISC-V architecture on a speculative out-of-order \acrshort{fpga} softcore (CHERI-Toooba).
We also integrate \cchyph support into the CheriBSD \acrshort{os} and \acrshort{cheri}-enabled Clang/LLVM toolchain.
Our evaluation shows effective mitigation of use-after-free and double-free bugs across all heap-based temporal memory-safety vulnerabilities in NIST Juliet test cases\revise{, real-world \acrshort{cve}s}, only a small performance overhead on SPEC CPU benchmarks (\SPECOverheadRelativeToCheri{} g.m.), less latency, and more consistent performance in long-running SQLite, PostgreSQL, and gRPC workloads compared to prior work.

\end{abstract}
\fi

\ifnotanonymous
\footnotetext[3]{R.S. contributed to the research while at Ericsson Security Research.}
\fi
\setcounter{footnote}{0}
\renewcommand{\thefootnote}{\arabic{footnote}}

\glsresetall
\section{Introduction}\label{sec:introduction}

\gls{cheri}, an \gls{isa} extensions which replaces conventional pointers with \emph{capabilities}, is a prominent design for \emph{hardware support for memory safety}.
\gls{cheri} has so far been adapted to the MIPS, RISC-V, and Armv8-A \glspl{isa} and \gls{cheri}-enabled processors have been developed by Arm~\cite{Grisenthwaite22}, Microsoft~\cite{Amar23a} and the RISC-V ecosystem.
The baseline \gls{cheri} design inherently provides \emph{spatial safety} \ifnotabridged properties for software running on \gls{cheri}-enabled processors\fi.
\ifnotabridged
To provide \emph{temporal safety} for heap-based allocations, CHERI requires a capability-revocation mechanism, such as Cornucopia~\cite{WesleyFilardo20,Filardo24}, ``\emph{quarantining}'' memory allocators~\cite{Amar23}, or combining \gls{cheri} with additional hardware mechanisms for temporal safety, such as \emph{memory versioning}~\cite{Watson23a}.
We describe each of the approaches in detail in \Cref{sec:background}.
Although these approaches have been deployed in industrial demonstrators and even systems intended for production use, they scale poorly due to:
\begin{inparaenum}[1)]
    \item intermittent latency and performance overhead incurred by memory sweeps looking for capabilities to revoke;
    \item memory overhead of allocations deferred from being reclaimed by relegating them to a ``quarantine'' list until a certain size criteria are reached, and
    \item  an \emph{use-after-reallocation}/\emph{use-after-free} mitigation gap (\Cref{sec:cheri-temporal-safety}) or low probabilistic guarantees of temporal safety due to limited upper bounds for available versions.
\end{inparaenum}
One factor contributing to these gaps is that \gls{cheri} lacks a mechanism that allows access to \emph{sets of capabilities} to be controlled \emph{in bulk}: in \gls{cheri}, individual capabilities must be, by-design, revoked one-by-one by locating them in memory.
\else
Prior attempts to augment \gls{cheri} with temporal safety~\cite{WesleyFilardo20,Filardo24} fall short in terms of scalability, memory overhead, and incomplete security guarantees, such as the existence of an \emph{use-after-reallocation}/\emph{use-after-free} gap (\Cref{sec:cheri-temporal-safety}).
\fi

\noindent
\textbf{This paper and contributions.} We propose \emph{\ccs}, an efficient mechanism for simultaneously \emph{retracting} (not fully revoking)\footnote{Retracting refers to \emph{disabling access} for a \emph{set of capabilities} at once, whereas revocation refers to permanently invalidating specific capabilities.} entire classes of capabilities at once (\Cref{sec:high-level-idea}).
Our insight is that the performance and permission issues in \gls{cheri}'s temporal-safety enforcement---particularly the use-after-reallocation/use-after-free mitigation gap---stem from its  complete removal of \emph{indirection}: central points of control for capability revocation.
\CCs introduce a \emph{limited} form of indirection for \emph{\ccprocing}, using a hardware-managed \emph{\cctable} for efficiently retracting \emph{dangling pointers}.

Using \ccs, we design an improved temporal-safety scheme for \gls{cheri} which \ifnotabridged
simultaneously eliminates both use-after-free risks and the need for quarantining freed memory.
\else
eliminates the \emph{use-after-reallocation}/\emph{use-after-free} gap.
\fi
A key benefit is a significantly reduced frequency required between capability-revocation sweeps compared to prior work (\Cref{sec:cc-temporal-safety}).

To demonstrate the practicality of our approach, we present \PICASSO (\glsentrydesc{picasso}), our realization of \ccs for the CHERI-RISC-V architecture.
\PICASSO is implemented on a \gls{fpga} softcore based on the speculative out-of-order CHERI-Toooba processor and in the QEMU full-system emulator.  We integrate \ccs into the CheriBSD \gls{os} and software stack.
In summary, our contributions are:
\begin{itemize}[topsep=0pt,noitemsep]
    \item A hardware-software co-design for \emph{\ccs} that extends \gls{cheri} with \ccprocing (\Cref{sec:design}).
    \item \PICASSO: a realization of \ccs on the CHERI-RISC-V-QEMU full-system emulator and in an \gls{fpga} softcore based on the speculative out-of-order CHERI-Toooba processor \acrshort{ip} (\Cref{sec:hwimpl}). 
    \item Support for \ccs in the CHERI-enabled Clang/LLVM compiler and CheriBSD \gls{os} (\Cref{sec:swstack}).
    \item Demonstrating \PICASSO successfully detects use-after-free and double-free defects in > 2800 \acrshort{nist} Juliet test cases  without false positives (\Cref{sec:security-eval}).
    \item Showing the efficiency of \PICASSO via the SPEC standard \gls{cpu} benchmark showing small performance impact compared to the baseline CHERI-RISC-V processor (\SPECOverheadRelativeToCheri{} g.m.) and real-world SQLite, PostgreSQL,  and gRPC benchmarks demonstrating less latency and more consistent performance compared to prior work (\gRPCOverheadRelativeToCheri--\SQLiteOverheadRelativeToCheri overhead on average, \Cref{sec:performance-eval}).
\end{itemize}

\revise{An extended version of this conference paper with supplementary material is available at~\cite{Gulmez26}. The open-source PICASSO hardware and software stack is available at \href{https://github.com/orgs/coloredcapabilities}{https://github.com/orgs/coloredcapabilities} and at \cite{Gulmez26b}. }

\section{Background}\label{sec:background}
\subsection{Temporal Safety}\label{sec:temporal-safety}

Temporal safety relates to memory locations containing different data at different points in time during program execution.
For example, after heap memory is freed by the application, accessing the memory \emph{as though the original data is there} leads to undefined behavior. 
This can happen if the program holds a pointer to some memory, frees the memory but keeps the pointer (\emph{dangling pointer}) and later tries to access the memory through the dangling pointer.
The result is a \emph{\gls{uaf}} bug.
If the application reallocates the memory, a subsequent dereference of the dangling pointer leads to a \emph{\gls{uar}} bug.
\gls{uaf} and \gls{uar} bugs are associated with non-deterministic behavior and stem from a confusion of \emph{allocation provenance}: which distinct allocation a particular address to memory represents.
In addition to crashes, these defects can lead to security vulnerabilities.
For example, information leakage can occur if an object containing sensitive data overlaps with the dangling pointer; similarly, arbitrary code execution may be possible if the dangling pointer overlaps with an attacker-controlled address which is inadvertently used in an indirect function call.

\subsection{The \gls{cheri} Capability Architecture}\label{sec:cheri}

\begin{figure}[t]
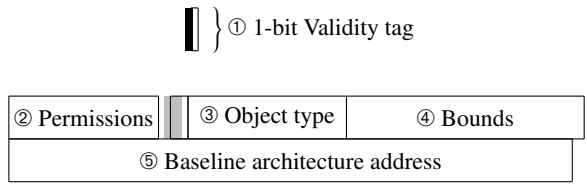

    \begin{bytefield}[bitwidth=0.3em]{1}
        \begin{rightwordgroup}{\dCOne{} 1-bit Validity tag}
            \colorbitbox{black}{1}{}
        \end{rightwordgroup} \\
        \bitbox[]{1}{} 
    \end{bytefield}\\
    \begin{bytefield}[bitwidth=0.37em]{64}
        \bitbox{17}{\dCTwo{} Permissions} & \colorbitbox{lightgray}{2}{} &
        \bitbox{18}{\dCThree{} Object type} & \bitbox{27}{\dCFour{} Bounds} \\
        \bitbox{64}{\dCFive{} Baseline architecture address} \\
    \end{bytefield}
    \caption{In-memory representation of \gls{cheri} capabilities adapted from Watson et al.~\cite{Watson19}}\label{fig:chericap}
\end{figure}

\gls{cheri} is an \gls{isa} extension that augments conventional \glspl{isa} with a capability-based hardware-software co-design for memory protection.
Hardware-supported capabilities enforce \emph{spatial safety} for code or data pointers.
The \gls{cheri} \gls{isa} specification~\cite{Watson23a} defines capability representations in registers and memory, along with instructions to manipulate them safely. 

As shown in \Cref{fig:chericap}, a \gls{cheri} capability is twice the width of a native pointer type---128~bits on 64-bit platforms---and includes one additional \emph{validity-tag} bit \dCOne, stored separately to prevent tampering by  non-capability-aware instructions.
These tags are preserved by valid, capability-aware instructions but are invalidated by unauthorized manipulation or injection of arbitrary capabilities.
In \gls{cheri}-enhanced architectures, general-purpose and address registers are extended to hold full capabilities.
\ifnotabridged For instance, CHERI-RISC-V stores the \gls{pc} and \gls{sp} in \gls{pcc} and \gls{csp} registers, respectively.\fi
Each capability includes:
\begin{itemize}[leftmargin=*, nosep]
    \item \textbf{Permissions} \dCTwo: A bitmask defining allowed operations.
    \item \textbf{Object type (\acrshort{otype})} \dCThree: A signed integer enabling temporary ``sealing'' and ``unsealing'' of capabilities for opaque pointer and fine-grained in-process isolation.
    \item \textbf{Bounds} \dCFour: The valid memory range relative to the baseline architecture address \dCFive, using a compressed encoding~\cite{Woodruff19} to reduce the space taken by a capability at the cost of stricter alignment for larger object allocations.
\end{itemize}
\gls{cheri} enforces spatial safety by associating each allocation with a capability describing its valid range and permissions.
New capabilities are derived from existing ones, while maintaining \emph{monotonicity} that ensures that new capabilities cannot exceed the permissions or bounds of their parent.
Sealed capabilities for compartmentalization and exception handling introduce limited non-monotonicity.
Extensions to \gls{cheri} have also explored sandboxing~\cite{Chisnall17}, initialization safety~\cite{Georges21,Gulmez25a}, safe speculation~\cite{Fuchs23, Fuchs24}, and side-channel resistance\cite{ElAtali25}.

\paragraph{Capability revocation.}
In capability systems, granted authority must be \emph{revocable}, rendering a capability \emph{permanently unusable}.
Historical designs often use \emph{indirection} to address capability revocation\cite{Redell74}.
Capability accesses in such systems occur through protected tables, providing a centralized point from where capabilities used throughout the system can be revoked.
\gls{cheri} takes a different approach, avoiding indirection tables.
Instead, it uses validity tags \dCOne stored either in a dedicated memory region, or distributed throughout program memory and \gls{cpu} register files.
A capability's validity can thus be determined without implicit indirection on memory access, eliminating the performance bottlenecks associated with a centralized protection table.
This design choice optimizes \emph{memory accesses} over \emph{efficient revocation}.
Because \gls{cheri} capabilities do not explicitly track their provenance, revoking a capability (and all its derivatives) necessitates finding their in-memory and register-file representations through exhaustive search and invalidating the associated validity tags.

\paragraph{Capability sealing.}
A capability is ``sealed'' when its \gls{otype}~\dCThree is not \texttt{-1} (all ones).
A sealed capability is \emph{temporarily} rendered unusable until it is ``unsealed''. 
In \gls{cheri} \gls{isa}v9, sealing is primarily used to protect control-flow transfers between isolated, \gls{cheri}-enforced compartments.
\ifnotabridged
Transitions between compartments use the \texttt{CInvoke} instruction which atomically unseals and jumps to a pair of code and data capabilities whose \glspl{otype} match.
Like revocation, a capability's \gls{otype} and sealing state is tracked by its in-memory representation.
The \texttt{CSeal} and \texttt{CUnseal} instructions change a capability's sealing state.
They operate on a data capability as operand, and require a code capability with the appropriate seal or unseal permission for the given \gls{otype}.
\fi

\paragraph{\gls{cheri}-enabled software stack.}
CheriBSD is a \gls{cheri}-enhanced version of the FreeBSD \gls{os}, supporting CHERI-RISC-V both in emulation and hardware.
It serves as a fully functional prototype demonstrating \gls{cheri} integration in a conventional \gls{os}.
The CheriBSD kernel and userspace can be built in ``\emph{pure-capability}'' \gls{cheri} C/C++, where all conventional memory pointers are replaced with capabilities by the CHERI-LLVM compiler.

\begin{figure}[t]
        \centering
        \includegraphics[width=0.9\linewidth]{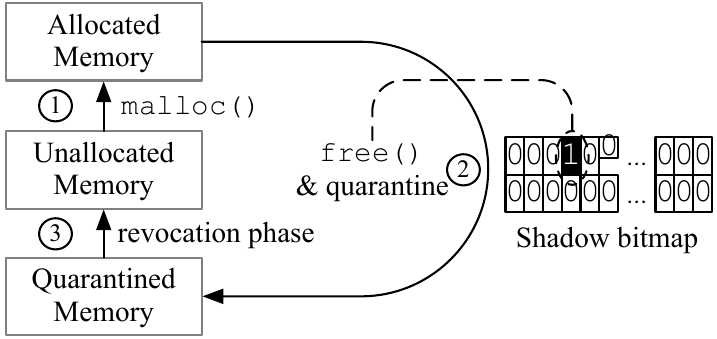}
        \caption{Memory allocation lifecycle in Cornucopia.}
        \label{fig:corn}
    \end{figure}

\subsection{\gls{cheri} Temporal Safety}\label{sec:cheri-temporal-safety}

\paragraph{Cornucopia.}
\ifabridged
\emph{Cornucopia}~\cite{Filardo24} enforces deterministic C/C++ temporal safety for heap allocations. It is integrated into CheriBSD via a \emph{shim layer}, known as the \gls{mrs}, which sits on top of the underlying BSD libc memory allocator, and a kernel-resident, concurrent revocation service that periodically performs \emph{sweeping revocation}---that is, periodically scans memory to revoke capabilities referencing freed memory.
Freeing memory in Cornucopia's memory lifecycle (\Cref{fig:corn}) involves three separate steps:
\begin{enumerate}[nolistsep]
    \item [\dCOne] Freed memory is placed into a \emph{quarantine} buffer.
    \item [\dCTwo] Quarantined regions are marked at word granularity in a \emph{shadow bitmap} managed by the \gls{mrs}.
    \item [\dCThree] Once the quarantine buffer exceeds a user-defined threshold, a revocation phase is triggered to find and invalidate stale capabilities referencing the quarantined memory.
\end{enumerate}

To prevent \gls{uar} exploits, memory is not reused until revocation completes. However, this introduces several drawbacks:
\begin{inparaenum}[1)]
    \item \textbf{Dangling pointers remain valid during quarantine}: while quarantining can reduce exploitability it also \emph{obscures} \gls{uaf} defects.
    \item \textbf{Increased memory overhead} due to delayed reuse of freed blocks.
    \item \textbf{Performance trade-offs}: more frequent recycling reduces memory footprint and increases the effectiveness of the mitigation at the cost of higher performance overhead. Immediate revocation on \free (\glsdesc{rof}) is, however, prohibitively expensive (\Cref{appendix:eval}).
    \item \textbf{Stop-the-world phase}: despite concurrent sweeping, Cornucopia halts all threads at the end to scan register files, recently modified pages, and kernel-held capabilities. This disrupts execution and aborts active system calls.
\end{inparaenum}
Cornucopia trusts the \gls{mrs} and allocator to maintain the quarantine buffer and bitmap.
These structures are protected by an internal API, requiring the caller to hold a capability with the \texttt{CHERI\_PERM\_SW\_VMEM} permission.
Allocators can derive such capabilities from a higher-level “super” capability (e.g., via \mmap~\cite{Bramley23}), but the permission is stripped before returning capabilities to untrusted userspace. Cornucupia is also employed by the internal libc allocator and the thread-local storage memory management area, following a similar approach. However, it holds its own guaranteed memory and revocation mechanism.
\else
As discussed in \Cref{sec:introduction}, to provide temporal safety for heap-based allocations, \gls{cheri} requires a capability revocation mechanism, such as Cornucopia~\cite{Filardo24}.
Cornucopia is capability revocation system for \gls{cheri} that implements deterministic C/C++ temporal safety for standard heap allocations.
Cornucopia is integrated into CheriBSD via a \emph{shim layer}, known as the \gls{mrs}, which sits on top of the underlying BSD libc memory allocator.
It also provides a kernel-resident concurrent revocation service which periodically performs \emph{sweeping revocation}---that is, periodically scanning \emph{all} application memory to identify and remove capabilities to freed memory.
The Cornucopia allocation life cycle is illustrated in \Cref{fig:corn}.
When memory is freed under Cornucopia, it is retained in a \emph{quarantine buffer} \dCOne, and the corresponding region is marked (at word granularity) as quarantined in a \emph{shadow bitmap} \dCTwo maintained by the Cornucopia shim.
Periodically, as the quarantine buffer grows to a (user-configurable) size, Cornucopia initiates a revocation phase \dCThree.
During this phase, Cornucopia's revocation service scans the process’s memory for any remaining capabilities referencing the freed memory region.
If such capabilities are found, their tags are cleared to revoke them. 

Cornucopia ensures that memory is not reallocated until sweeping revocation completes, preventing \gls{uar} exploits. However, it has several limitations, 
firstly,  the quarantine buffer still holds valid memory, allowing programs to retain and potentially use dangling pointers.
While this reduces the likelihood of \gls{uaf} exploits, it obscures underlying programming errors.
Additionally, the use of a quarantine buffer and shadow bitmap increases memory usage, which may be unsuitable for some applications.
The size of the shadow bitmap is relative to the total size of the address space and the size of a capability's in-memory representation.
The quarantine buffers effectively delays the deallocation operations, preventing freed blocks to be immediately available for reuse.
Since such freed blocks cannot be reallocated before the exit quarantine, they still effectively contribute to the application's memory footprint.
Blocks held in quarantine will be eventually be recycled once certain size criteria are reached.
This is essentially a delayed freelist which can help mitigate some \gls{uaf} situations but is fairly costly in terms of performance and memory footprint.
A lower recycling threshold reduces
memory footprint and increases the effectiveness of the mitigation, at the cost of higher overhead. Cornucopia can also be configured to initiate revocation immediately when memory if freed, bypassing the quarantine buffer, but this comes at severe performance cost.
Third, although Cornucopia’s sweeping revocation runs mostly concurrently, it requires a final \emph{stop-the-world} phase to scan register files, memory pages that received capability writes during the concurrent sweeps, and capabilities held by the kernel on behalf of userspace programs. This phase is intrusive, as it halts all application threads and aborts any in-progress system calls.

In the Cornucopia system model the \gls{mrs} and allocator are trusted to maintain the shadow bitmap and quarantine buffer list in userspace.
An internal \gls{api} protects these data structures from unauthorized modification by requiring calling code to possess a capability to the allocation with the \texttt{CHERI\_PERM\_SW\_VMEM} permission.
While allocator can derive such a capability from a ``super'' capability to the underlying memory region it holds as a result of an \texttt{mmap} call~\cite{Bramley23}, the permission is removed before a capability to a new allocation is returned to untrusted userspace. Cornucupia is also employed by the internal libc allocator and the thread-local storage memory management area, following a similar approach. However, it defines its own guaranteed memory and its own revocation mechanism.
\fi

\begin{figure}[t]
        \centering
        \includegraphics[width=0.9\linewidth]{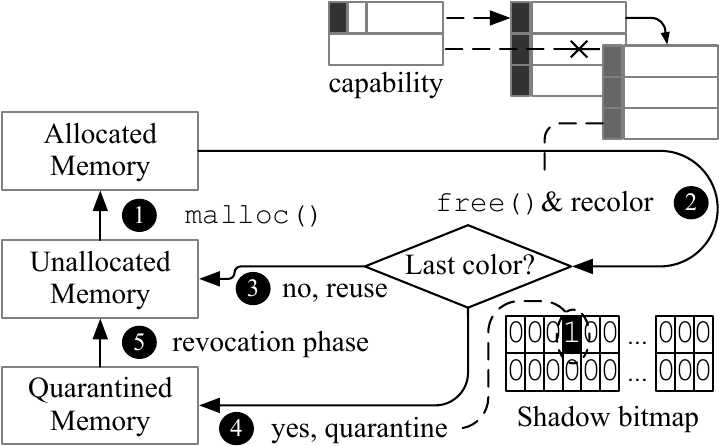}
        \caption{Allocation lifecycle in \gls{cheri} + Memory Tagging.}
        \label{fig:cherimte}
\end{figure}

\paragraph{Cornucopia Reloaded.}
\ifabridged
Filardo et al.~\cite{Filardo24} identify Cornucopia’s long \emph{stop-the-world} pauses as its main limitation.
Their \emph{Cornucopia Reloaded} improves revocation concurrency by introducing \emph{capability load barriers}, which associate each \gls{pte} with a \emph{revocation epoch}. A new \emph{epoch counter} in \gls{cpu} control registers tracks completed atomic revocation sweeps.
If a \gls{pte}’s epoch differs from the counter, any capability load from that page traps, allowing the revoker to complete revocation before the \gls{cpu} processes the load.

Because sweeps are atomic, pages with recent capability writes cannot retain stale capabilities, eliminating the need for global pauses.
Applications thus avoid major stop-the-world events.
However, within an active epoch, programs can still hold, copy, or dereference pointers to memory freed in that same epoch.
Only allocations freed in earlier epochs are guaranteed to be revoked.
Consequently, memory added to quarantine cannot be reused until at least a full epoch has passed.
As with the original Cornucopia, this leads to de-allocation delays and increased memory usage.

\else
Filardo et al.~\cite{Filardo24} identify its impractical “stop-the-world” pause times as the most pressing limitation of Cornucopia.
Filardo et al.'s \emph{Cornucopia Reloaded} improves the concurrency of capability revocation through \emph{capability load barriers} that associate each \gls{pte} with an \emph{revocation epoch}.
An \emph{epoch counter} is introduced into the \gls{cpu} control registers, that counts the number of completed (atomic) revocation sweeps.
If the \gls{pte} revocation epoch does not match the value of the epoch counter, any capability load from the corresponding memory page will trap.
This allows the concurrently executing revoker to catch up, revoking any capabilities pointing to quarantined pages in the page before the loads are processed by the \gls{cpu}.
As revocation sweeps under Cornucopia Reloaded are atomic, memory pages that received capability writes no longer run the risk of holding capabilities revoked under that epoch.
As a result, under Cornucopia Reloaded applications no longer experience significant revocation-induced stop-the-world periods.
However, within a given epoch, the application might yet hold, copy, and dereference pointers to memory that has been freed within that epoch.
Only frees from prior epochs are guaranteed to be seen as already revoked during revocation.
In other words, memory added to quarantine in one epoch cannot be removed from quarantine until at least one complete epoch has passed -- i.e., the epoch counter has been increased by two.
Cornucopia Reloaded thus suffers from the same quarantine-induced de-allocation delays and increase in memory footprint as  original Cornucopia.
\fi

\revise{Sweeping revocation inherited from Cornucopia remains the most contested aspect of
\gls{cheri}\cite{Rebert24}. Alternatives involve accelerating \gls{cheri} temporal-safety
enforcement by combining Cornucopia with approaches not relying on sweeps.}

\paragraph{Composing \gls{cheri} and Memory Versioning.}
\ifabridged
Filardo~\cite{Filardo24,Filardo24a} proposes combining \gls{cheri} with \emph{memory versioning} to close the \gls{uar}/\gls{uaf} mitigation gap.
A variant of this approach is included in CHERI ISAv9~\cite[Appendix C.5]{Watson23a}.
As shown in~\Cref{fig:cherimte}, both the pointer and its allocated memory region are assigned a matching version ID (or “color”)~\dOne.
On free, the memory is recolored to invalidate stale pointers~\dTwo, and the freed color becomes available for reuse~\dThree.
This blocks dangling pointer use.
However, the number of colors is limited, as version IDs are stored in dedicated \emph{memory tags}.
Due to performance constraints, architectures like Arm’s \gls{mte} restrict tags to 4 bits, allowing only 16 unique IDs.
Reusing colors risks collisions between valid and stale pointers, yielding only probabilistic \gls{uaf} detection.
To prevent this, once all version IDs are used~\dFour, freed memory is quarantined, and the system falls back to Cornucopia-style sweeping revocation~\dFive.
\else
Filardo~\cite{Filardo24,Filardo24a} proposes combining \gls{cheri} with \emph{memory versioning} to close the \gls{uar}/\gls{uaf} mitigation gap in the aforementioned Cornucopia designs.
A variant of such a composition has also made its way into the \gls{cheri} ISA-9 specification~\cite{Watson23a}.
In it, illustrated in~\Cref{fig:cherimte}, both the pointer and the corresponding memory region are assigned a matching version identifier (or "color") when memory is allocated~\dOne.
On free, the memory region is "recolored" to invalidate stale references~\dTwo and the corresponding color is available to use for recoloring the memory regions~\dThree.  
This method effectively detects and blocks the use of dangling pointers.
However, the number of available colors is finite and severely limited due to the need to carry the color bits as dedicated \emph{memory tags}.
Memory tagging exhibits poor performance as the number of tag bits increase. 
Thus schemes such as the Arm architecture's \gls{mte} limit the number of tag bits to 4.
A 4-bit versioning scheme provides only 16 unique identifiers.
Reusing a color can lead to collisions between new and stale pointers, which results in only probabilistic guarantees for detecting \gls{uaf} errors.
To avoid this, once all version identifiers have been exhausted \dFour, the freed memory is quarantined and the system falls back to Cornucopia’s capability sweeping mechanism \dFive, as described above.
\fi
A practical implementation for this design concept is not yet available.
\section{System and Threat Model}\label{sec:systemmodel}\label{sec:threatmodel}

\paragraph{System Model.}
We assume the same system model as \gls{cheri}~\cite{WesleyFilardo20} and that its spatial-safety properties hold.
We rely on the underlying \Alloc satisfying assumptions:

\begin{enumerate}[label=\textbf{A\arabic*},labelsep*=0pt,itemsep=2pt]
    \item:~\Alloc is part of the trusted runtime.\label{asm:alloc-is-trusted}
    \item:~\Alloc maintains correct bounds on capabilities it issues.\label{asm:correct-bounds} 
    \item:~\Alloc can discern malformed input to \texttt{free()}.\label{asm:malformed-free}
\end{enumerate}

\paragraph{Threat Model.}
We assume an \Adv with the ability to manipulate program behavior through malicious input.
Due to flaws in the victim program, such as faulty input validation, \Adv can influence it to invoke arbitrary sequences of \texttt{malloc()} and \texttt{free()} calls (or C++ \texttt{new} and \texttt{delete} statements).
Thanks to \gls{cheri}'s architectural guarantees, \Adv is unable to produce arbitrary, counterfeit capabilities.
Nor is \Adv capable of directly tampering with \Alloc's data structures, control-data, or code beyond interaction through the \texttt{malloc()} and \texttt{free()} family of functions (\ref{asm:alloc-is-trusted}).
This is consistent with the threat model employed by prior work~\cite{WesleyFilardo20, Filardo24}.

\gls{cheri} capabilities are architecturally guaranteed to be \emph{monotonically nonincreasing} (\Cref{sec:cheri}); the permissions and bounds set on a capability can be \emph{decreased} by software (by deriving a new, more narrowly bounded capability), but cannot be \emph{increased} beyond the parent capability permissions and bounds.
Thus, \ref{asm:correct-bounds} implies that capabilities pointing to the heap fall within the bounds of (once) heap-allocated objects.
\ref{asm:malformed-free} in turn ensures that capabilities not pointing to beginning of legitimate heap-allocated objects are rejected by \Alloc.
Invalid capabilities, as well as capabilities for non-heap-allocated data, are rejected by the \gls{mrs} (\Cref{sec:swstack}).

The exploitation of uninitialized reads of heap-allocated memory objects are out of scope, but can be efficiently prevented using previously proposed \gls{cheri} extensions~\cite{Gulmez25a} or compiler-based uninitialized memory sanitization~\cite{Chow05,Milburn17,Joly20}.

While temporal-safety concerns are also valid for stack-based allocations (c.f. \emph{use-after-return}~\cite{MITRE24}), they are easier to protect~\cite{Dang17}, significantly rarer compared to heap-based use-after-free~\cite{Younan15}, and according to Lee et al. \emph{``unlikely to be exploitable''}~\cite{Lee15}.
Compiler-based \emph{escape analysis}  can verify that references to stack variables do not outlive the enclosing stack frame and is widely-deployed in modern C and C++ toolchains~\cite{GCCTeam25,LLVMTeam25}.
In fact, mitigations dating back a decade expect that run-time temporal-safety mitigation are only needed to dynamically allocated heap memory~\cite{Lee15,Younan15,Dang17,WesleyFilardo20,Filardo24}.
Consequently, following prior work~\cite{WesleyFilardo20, Filardo24}, we consider stack-based use-after-return out of scope.
\revise{Additionally, similar to prominent prior implementations of \gls{cheri} on application-class
processors, our threat model excludes transient-execution and other micro-architectural
attacks and side-channels~\cite{Watson23b}. We discuss the gap with respect to safe speculation in
\Cref{sec:futurework}.}

\section{Goals and Challenges}\label{sec:goals-and-challenges}

Our goals are:
\begin{enumerate}[label=\textbf{G\arabic*},labelsep*=0pt,itemsep=2pt]
    \item: Deterministically close the~\emph{\gls{uaf}/\gls{uar} gap}\label{goal:close-uaf-gap}.
    \item: Allow the freed memory to be \emph{immediately reused}\label{goal:no-quarantine}.
    \item: Reduce the \emph{frequency of \gls{cheri} revocation sweeps}\label{goal:reduce-sweeps}.
\end{enumerate}

\paragraph{Deterministically closing the \gls{uaf}/\gls{uar} gap.}
Designs that compose \gls{cheri} with memory versioning (\Cref{sec:cheri-temporal-safety}) can only achieve \emph{probabilistic} \gls{uaf} mitigation if deployed without quarantining due to their limited tag space---as low as 4 bits.
In contrast, we aim to achieve \emph{deterministic} prevention of \gls{uaf}.
The inherent challenge is scaling the ability to efficiently track the \emph{provenance}---which allocation a particular address to memory represents---to \emph{millions} of allocations.

\paragraph{Immediate reuse of freed memory.}
A common source of memory overhead for each scheme discussed in \Cref{sec:cheri-temporal-safety} is the quarantine buffer.
Quarantining memory delays its availability for reuse until a revocation sweep occurs.
This is necessary, as the \gls{cheri} design lacks a mechanism for \emph{retracting} capabilities, that is, disabling capabilities whose provenance can be traced to particular allocation until their eventual revocation.
Quarantining instead ensures \emph{provenance validity} of capabilities by \emph{artificially extending} the lifetime of the associated allocations beyond the \texttt{free()} call that de-allocates them until a time when all capabilities to that allocation have been revoked.
Even though this is deferred until certain size criteria are reached, it may be costly in terms of performance and memory footprint for the application.
For this reason, production-grade memory allocators, such as LLVM's Scudo, disable quarantining by default~\cite{LLVMTeam25a}.
Therefore we target immediate reuse of freed memory without quarantine.

\paragraph{Reducing the frequency of revocation sweeps.}
Recall from \Cref{sec:cheri} that due to \gls{cheri} prioritizing the efficiency of memory accesses over the efficiency of revocation, the revocation sweep is a ``\emph{necessary evil}'' that is inherent to the \gls{cheri} design.
Our goal is not to fundamentally alter the core tenets of \gls{cheri}.
Rather, we want to significantly reduce the frequency of the necessary sweeps (without a quarantine buffer \labelcref{goal:no-quarantine}).
The challenge of how to achieve reduced revocation frequency goes hand-in-hand with scaling the number of allocation provenances for capabilities that can be \emph{simultaneously} tracked. By increasing the number of allocations that can simultaneously co-exist in a system while still being discernible from each other, the system can operate without risking overlapping capabilities with different allocation provenance being confused with one another and necessitating revocation of stale capabilities.
 
\section{Colored Capability Design}\label{sec:design}
\glsreset{pvt}

Recall from \Cref{sec:cheri} that \gls{cheri}'s design disregards revocation implemented via table-based lookups or indirection on pointer operations in favor of validity bits stored in tagged memory.
However, as discussed in \Cref{sec:cheri-temporal-safety},  \Cref{sec:goals-and-challenges}, and shown in \Cref{sec:performance-eval}, revoking all capabilities for an allocation at the moment that allocation is freed is prohibitively expensive in terms of run-time overhead in \gls{cheri}.
To achieve \labelcref{goal:close-uaf-gap,goal:no-quarantine,goal:reduce-sweeps} outlined in \Cref{sec:goals-and-challenges} we aim to decouple the \emph{validity of a capability's provenance}---the allocation the capability represents---from the \emph{validity of the capability itself}.
This requires introducing a form of \emph{limited indirection} to the \gls{cheri} architecture.
While many historical capability systems, including the initial design by Dennis and Van Horn\cite{Dennis66}, use table-based indirection to manage all capability metadata, \gls{cheri} explicitly has avoided such indirection as a design goal~\cite[2.3.16]{Watson23a}.
The challenge we must overcome is introducing the necessary indirection in a manner which limits its impact only on the subset of capabilities whose provenances needs to be tracked, and reduce the run-time impact of table-based lookups by minimizing the information needed to validate provenance.

In this section we present the high-level idea behind the \cc design (\Cref{sec:high-level-idea}), how we apply \ccs for temporal-safety protection (\Cref{sec:cc-temporal-safety}), and describe the overall system architecture (\Cref{sec:system-arch}).

\subsection{High-Level Idea}\label{sec:high-level-idea}
\CCs introduce a \emph{limited} form of indirection to improve temporal safety.
\revise{Whereas historical indirection designs centralize all capability metadata~\cite{Dennis66,Redell74}, we demonstrate that 1-bit of indirect metadata for a set of \ccs is sufficient for accelerating temporal-safety enforcement.}
This allows \emph{all} capabilities in a set referencing a given heap allocation to be \emph{retracted} simultaneously. Retracted capabilities are disabled so that attempts to dereference them result in a \gls{cheri} fault (c.f. \emph{capability sealing}, \Cref{sec:cheri}).
Retracted capabilities need to \emph{eventually} be fully revoked, but do \textbf{not pose a risk for \gls{uaf}} in the interim, thus \textbf{removing the need for quarantining freed memory}.
Since \gls{cheri}'s direct revocation aligns with the typical execution style of C-language programs--where pointers freely propagate through memory without enforced (capability-to-capability~\cite{Redell74}) indirection--we instead target a mechanism for \emph{\ccindprocing}.
Recall from \Cref{sec:background} that \gls{cheri}'s sealing encodes both the sealing state and the object type directly in the capability (the presence of a non-default object type doubles as both the indicator of a capability's sealing state and the conditions for its unsealing).
In contrast, \ccs \emph{decouple} the capability's \emph{state} from its \emph{type}: the \emph{\ccpid} remains embedded in the capability, but the \ccstate is expressed by a \emph{\gls{pvb}} stored separately in a \emph{\gls{pvt}}, indexed by the \ccpid.
\revise{\CCs only used to track heap allocations; non-heap capabilities are unaffected.}

\CCs derive their name from the use of ``color'' to distinguish allocation provenance in memory-versioning schemes (\Cref{sec:cheri-temporal-safety}). We use \emph{color} to refer to the \texttt{otype} of a capability while it represents a \ccpid.
\revise{We reserve a portion of \gls{cheri}'s object type for conventional sealing.}
Unlike memory versioning, we only assign a color to \ccs, not the associated memory allocation.
Instead, each entry in the \gls{pvt} holds a single \gls{pvb} indicating whether \emph{all} capabilities with that \ccpid are \ccactive (the corresponding allocation is active). The \ccpid in the capability determines if the capability is a \cc or not.
Hardware enforces provenance-validity checks for any load or store using a \cc by looking up the corresponding \gls{pvb} in the \cctable.
Any memory access using a \cc, where the \gls{pvt} indicates that the color (\ccpid) is \ccinactive, faults.

\begin{figure}
    \centering
    \includegraphics[width=0.8\linewidth]{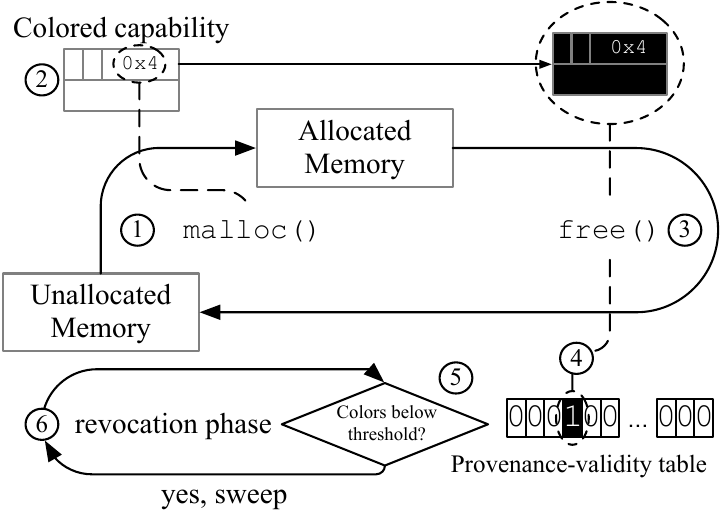}
    \caption{\CCs allocation lifecycle.}
    \label{fig:cc}
\end{figure}

\subsection{Temporal Safety via Colored Capabilities}\label{sec:cc-temporal-safety}
Using \ccs and \ccprocing{}, we design the temporal memory safety scheme illustrated in \Cref{fig:cc}. 
On allocation \dCOne, a unique \ccpid, from a finite pool, is assigned to the capability \dCTwo, and its corresponding \gls{pvb} in the \cctable is marked \emph{\ccactive}.
On de-allocation~\dCThree, this bit is marked \emph{\ccinactive}~\dCFour.
As explained earlier, all load or stores via colored capabilities are subject to a hardware check against the \gls{pvt}.
Once a \ccpid has been \ccdeactivated, the hardware ensures that all capabilities with that \ccpid cannot be dereferenced.
This eliminates the need for a quarantine buffer and shadow memory to track freed regions, allowing freed memory to be safely reused immediately.

The \ccpid pool supports over two million entries per process with 128-bit capabilities (\Cref{sec:implementation}).
In long-running programs, this pool may eventually be exhausted.
To be able to retract and reuse previously assigned identifiers, \ccs still need a mechanism to identify and revoke such retracted capabilities.
Once the identifier pool is depleted up to a pre-determined threshold~\dCFive, a revocation sweep is needed~\dCSix.
Colored capability revocation is based on Cornucopia Reloaded's improved revocation sweep using capability load barriers and inherits the reduced stop-the-world periods, but does not require capability sweeps to be atomic. As \ccprocing already guarantees that capabilities pointing to freed memory cannot be referenced, even the in-kernel scanning for capabilities can occur asynchronously. A revocation pass for a given \ccpid has to simply complete before that \ccshortpid is returned to the pool.

\begin{figure}
    \centering
    \includegraphics[width=0.9\linewidth]{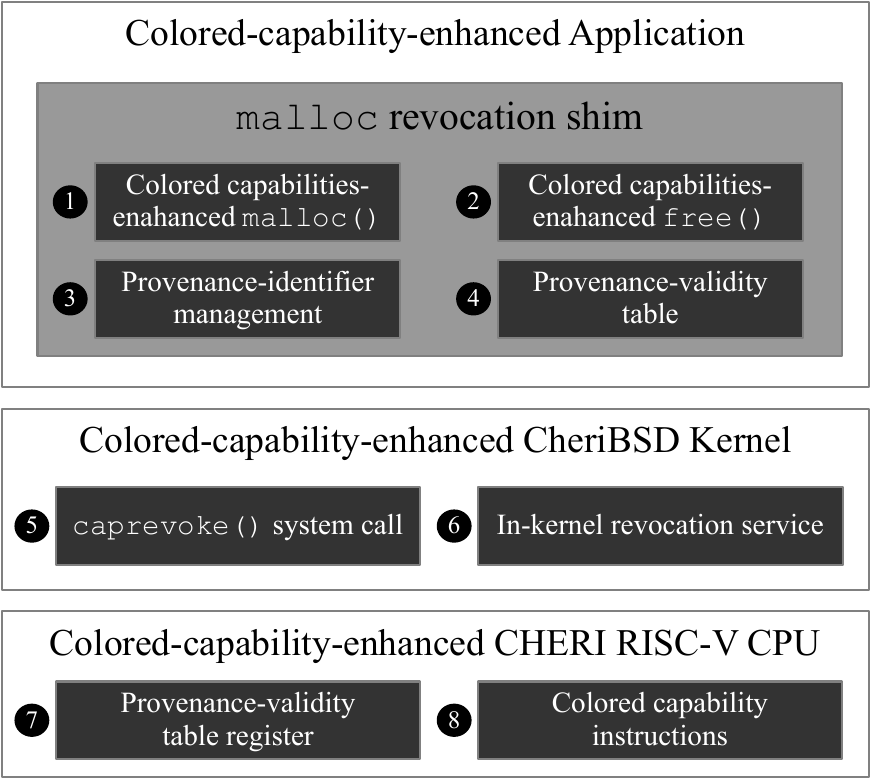}
    \caption{\CCs overview }
    \label{fig:cc-overview}
\end{figure}

\subsection{System Architecture}\label{sec:system-arch}

\Cref{fig:cc-overview} illustrates the high-level system architecture for a \cc{}-enhanced \gls{cpu} and \gls{os}.
At the application-level, \ccs integrate with the regular \texttt{malloc}-family of functions and \texttt{free()} through enhanced versions of these memory management \glspl{api}~\dOne,~\dTwo integrated via the \glsdesc{mrs} (\acrshort{mrs}, \Cref{sec:cheri-temporal-safety}).
For brevity, we refer in \Cref{fig:cc} and subsequent figures simply to \cc-enhanced \texttt{malloc()} instead of the whole \texttt{malloc}-family of functions.
When a new \cc is issued by \texttt{malloc()}, or a memory object is freed by passing a \cc to \texttt{free()}, the allocator must keep track of assigned and still unassigned \ccpids.
The \emph{\ccpidhyph management} \dThree within the \gls{mrs} is responsible for assigning each allocation a unique \ccpid which is retained by any capabilities derived from the issued \cc{}.
Recall from \Cref{sec:threatmodel} that thanks to \ref{asm:correct-bounds} and monotonically non-increasing bounds (\Cref{sec:cheri}), any such derived capabilities are guaranteed to point within the bounds of the allocated object.
When an allocation is freed, the \texttt{free()} implementation marks the corresponding \ccpid\ifnotabridged{} (which it learns from its argument \cc)\fi as \ccdeactivated via the \gls{pvt}~\dFour.
This has the effect of simultaneously \ccdeactivating \emph{all} \ccs pointing to the the relevant allocation since they share the corresponding \ccpid.

\ccDeactivated \ccpids are kept out of rotation by the \ccpidhyph management until all corresponding capabilities have been revoked.
This occurs when the number of unused IDs fall below a predetermined threshold, at which point any \texttt{malloc()} or \texttt{free()} call engages the kernel's revocation service~\dSix by making a \texttt{caprevoke()} system call~\dFive.
Other system software not using \ccs do not require modification (apart from adaptation to CHERI).

The \cc-enhanced \gls{cpu} core(s) is augmented via a \emph{\gls{pvtr}}~\dSeven which holds the virtual start address of the \cctable available to the \cc-enhanced instructions~\dEight.
Load (\texttt{l[bhwd]}) and store (\texttt{s[bhwd]}) instructions, when given a \cc as operand, lookup its corresponding \gls{pvb} in the \gls{pvt}.

\glsreset{pvtr}
\glsreset{pvb}

\begin{figure}[t]
    \centering
    \includegraphics[width=\linewidth]{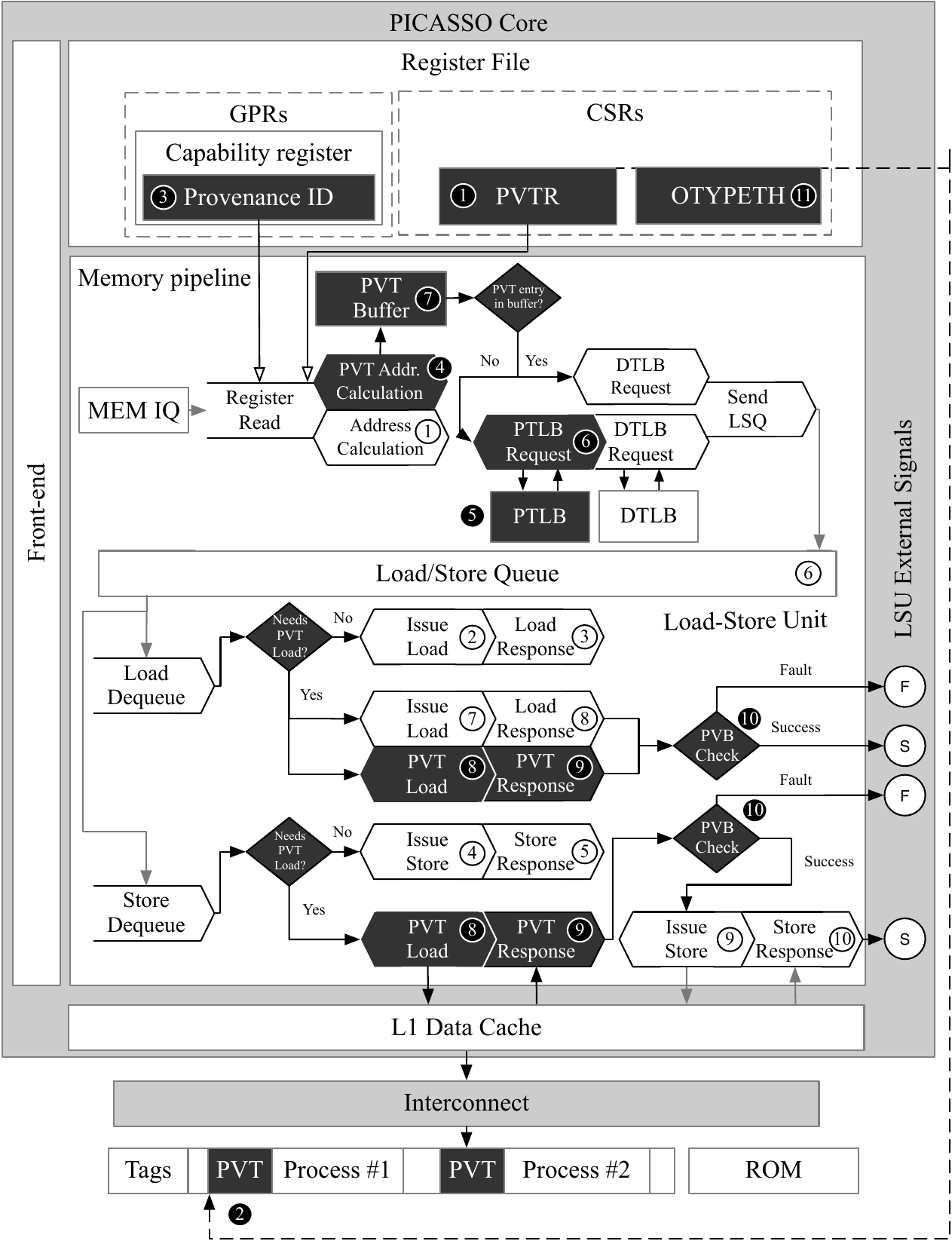}
    \caption{\PICASSO hardware architecture.}\label{fig:impl_verif}
\end{figure}

\glsreset{picasso}
\section{Implementation}\label{sec:implementation}

In this section, we present \gls{picasso}, our implementation of  \ccs for the CHERI-RISC-V architecture. We
implemented two versions of \gls{picasso}: 1) a QEMU-system-CHERI128 full-system emulator, and 2) a softcore based on the CHERI-Toooba processor \gls{ip}. For brevity, we focus on the Toooba implementation in the following sections, while the QEMU implementation is detailed in \Cref{appendix:qemu}.

\subsection{\gls{picasso} Extension for CHERI-RISC-V}\label{sec:hwimpl}
To add colored capability architectural support to the CHERI-Toooba core, we modified 34 source files, totaling approximately 1,700 lines of changes spanning the build system, core \gls{cpu}, \gls{isa}, memory pipeline, and CHERI capability logic. The most significant changes affect loads and stores.

\paragraph{\gls{isa} extension.}
\Cref{fig:impl_verif} illustrates our changes to the baseline CHERI-Toooba processor.
A new \gls{csr} called the \gls{pvtr}~\dOne stores the base virtual address of the \cctable~\dTwo. 
Access to the \gls{pvtr} is restricted to kernel mode, ensuring only privileged code can control the \cctable location.
A new \texttt{ccsettype} instruction assigns \ccpids~\dThree to capabilities.
Recall from \Cref{sec:cc-temporal-safety} that Cornucopia requires allocator capabilities to hold the \texttt{CHERI\_PERM\_SW\_VMEM} software-defined permission.
Similarly, \texttt{ccsettype} requires this as a hardware-defined permission to ensure that only the trusted allocator can modify \ccpids.
Loads and stores via capabilities with a \ccpid (\ccs) implicitly read the corresponding 1-bit \gls{pvt} entry, the \gls{pvb}, at \texttt{[PVTBase + ProvenanceID]} and fault if the \gls{pvb} indicates the provenance is invalid, i.e., the corresponding capabilities have been retracted.

\paragraph{Accessing the \gls{pvt}.}
The \gls{cheri}-Toooba memory pipeline consists of four main stages: \emph{Register Read}, \emph{Address Calculation}~\dCOne, \emph{Address Translation}, \emph{Issuing Load/Store}~\dCTwo,~\dCFour, and \emph{Load/Store Response}~\dCThree,~\dCFive.
\PICASSO extends the memory pipeline to distinguish between regular and \ccs.
For \ccs, the \ccpid is extracted from the source capability register~\dThree, while the \gls{pvt} base address is read from the \gls{pvtr}~\dOne. Address calculation~\dFour for the \gls{pvt} entry is performed concurrently with normal address calculation~\dCOne.

\paragraph{\gls{pvt} address translation.}
To translate the physical address for a \gls{pvb}, \PICASSO adds a dedicated \emph{\gls{ptlb}}~\dFive to the memory pipeline.
Accessing the \gls{ptlb} introduces an additional pipeline stage~\dSix with a one-cycle latency.
\gls{pvt} address translation feeds into the \gls{lsq}~\dCSix. Rather than adding separate \gls{lsq} entries for implicit \gls{pvt} loads, each load or store entry is augmented to track its associated \gls{pvt} load, preserving the existing queue structure.
The \gls{ptlb} is only used for address translation in implicit \gls{pvt} loads; writes to the \gls{pvt} use the standard store instructions and memory hierarchy (and require a corresponding capability with privileges to modify the \gls{pvt}).

\paragraph{\gls{pvt} buffer.} To avoid \gls{pvt} address translation and access for all \cc loads and stores, \PICASSO introduces a small \gls{pvt} buffer~\dSeven that caches recently accessed 128-bit \gls{pvt} words with valid~\gls{pvb}s.
The buffer is indexed directly by the virtual \gls{pvb} address, allowing it to bypass all of \gls{ptlb} translation~\dSix, \gls{pvt} load~\dEight, response~\dNine, and \gls{pvb} check~\dTen.
To maintain coherence, all buffer entries are invalidated on \texttt{fence} instructions; correctness relies on fences accompanying page-table updates and writes to the \gls{pvt}. Each 128-bit word in the \gls{pvt} buffer can contain 128 \gls{pvb}s.
The buffer is looked up combinationally to avoid adding latency to the common case and is therefore kept small: only 64 words, 4-way set-associative.
Despite its size, it can buffer \gls{pvb}s for 8192 objects, and achieves high hit rates in practice (\Cref{appendix:eval}).

\paragraph{Loads with \ccs.}
For \cc loads the data request and the \gls{pvt} load to L1 data cache~\dCSeven are issued concurrently by the \gls{lsu}. Although the cache services only one access per cycle, the two requests are independent and their latencies can overlap. A load can commit only after both responses~\dCEight~\dNine arrive, and \gls{pvb} check~\dTen succeeds.
On mis-speculation, the \gls{lsu} waits for outstanding responses before freeing the corresponding \gls{lsq} entry.

\paragraph{Stores with \ccs.}
For stores, the \gls{pvt} load is issued early~\dEight, after address translation, but the store itself~\dCNine is delayed until the \gls{pvt} response~\dNine and the \gls{pvb} check~\dTen passes. This ordering prevents use-after-free writes.
If a store is squashed before issue, its \gls{lsq} entry must remain allocated until the \gls{pvt} response. This requires additional waiting logic analogous to handling mis-speculated loads.

\paragraph{Supporting millions of allocations.} In CHERI’s 128-bit capability representation 18 bits are reserved for the \gls{otype} which we reuse for \ccpids, allowing for $2^{18}$ distinct allocation provenances to be tracked. 
In our current design, applications benefit from a much larger \ccpid space, which extends the period between needed invocations of the revocation mechanism. We extend the \gls{otype} field by adding one reserved bit and two unused software-defined permission bits, resulting in a total of $2^{21}$ allocations. 

\paragraph{Compatibility with \gls{cheri} sealing.}\label{sec:sealin_coexistence}
Recall from \Cref{sec:background} that \gls{cheri} hardware prevents manipulation of any sealed capability. However, \ccs disallow de-referencing of capabilities with an invalid \gls{pvb}. To support both the sealing mechanism and \ccs \PICASSO provides a way to detect which feature is in use whenever an \gls{otype} or \ccpid is set in a capability. \PICASSO adds an additional CSR, \gls{otth}~\dEleven, that determines if the capability is interpreted as a normal, sealed or \cc according to the following rule: 

{\footnotesize
\[ 
    f(\text{\gls{otype}})=
   \begin{cases}
   
    \text{unsealed capability}, & \text{if \gls{otype}}=-1 \\
    \text{\cc},& \text{if }0<\text{\gls{otype}}<\text{\gls{otth}} \\
    \text{sealed capability}, &  \text{if \gls{otype}} \geq \text{\gls{otth}} 
  \end{cases}.
\]
}

\revise{This ensures that \PICASSO is fully compatible with \gls{cheri} \gls{isa}v9 sealing and compartmentalization~\cite{Watson23a}.}

\ifnotabridged
\begin{figure}[t!]
    \centering
    \includegraphics[width=\linewidth]{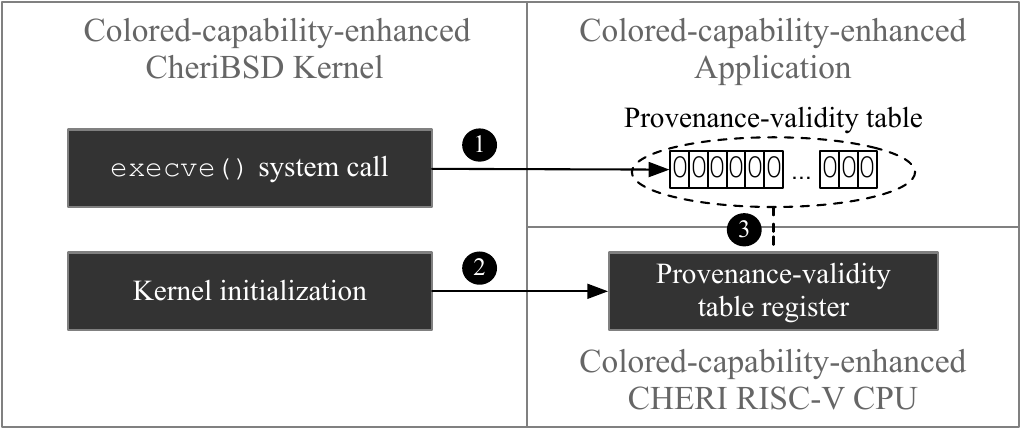}
    \caption{\gls{pvt} initialization at process creation.}\label{fig:cc-init}
\end{figure}
\fi

\subsection{\gls{picasso} Software Stack}\label{sec:swstack}
We integrated \ccs into \CheriBSDVersion, resulting in $\approx$3 500 changed lines of code over 57 distinct files.

\paragraph{Provenance Validity Table.} 
To efficiently track which \ccpids are \ccactive, we map a \cctable into each process's address space during the \execve system call\ifnotabridged~\dOne (see \Cref{fig:cc-init}\fi. The \ccTable is placed directly under the stack, at a fixed virtual address determined when the kernel initializes based on the address space size. The virtual address of the \cctable is stored in the \gls{pvtr}\ifnotabridged~\dTwo\fi, enabling the hardware to reference the \ccTable during load or store instructions\ifnotabridged~\dThree\fi. By using a fixed virtual address across all processes, we eliminate the need to update the \gls{pvtr} during context switches.
Cornucopia introduces the \cherirevokegetshadow system call to provide a capability for its shadow bitmap to userspace; we augment this with a new flag that allows the \cchyph-enhanced \gls{mrs} to request a capability to the \cctable instead.

\paragraph{\Glsentryhyph{pid} management.}\label{cc:pidmanagment} The \gls{mrs} tracks which \ccpids are valid, which are in use, and which can be assigned to new capabilities.
This requires it to keep track of $2^{21}$ ($\approx$~2 million) \ccpids in order to efficiently find and claim the first available one. When the number of unclaimed \ccpids reaches a user-defined threshold, less than 1\% by default, the \gls{mrs} releases the \ccpids associated with freed memory. The number of reclaimed IDs is program-dependent; however, a single sweep can release up to 99\% of reserved \ccpids.
Completely exhausting the \ccpids pool prevents further allocations until IDs can be reclaimed.

We use the BSD \gls{unr} allocator, \texttt{alloc\_unr}, a mechanism used in the FreeBSD kernel for efficient allocation of integer identifiers by utilizing a doubly linked list of heap-allocated \texttt{unr} objects. It stores a compressed representation of \ccpids, keeping track of consecutive sequences of identifiers as runs instead of individual numbers. We optimize \texttt{alloc\_unr} to batch \ccpids that are to be released and make corresponding changes to the \gls{unr} data structure over a single iteration. For further details our modified implementation, refer to Appendix~\ref{appendix:unr}. The BSD \gls{unr} is a kernel-space facility. We adapt our design for userspace to avoid issuing system calls to manage \glspl{unr}. 

\begin{figure}[t!]
    \includegraphics[width=0.9\linewidth]{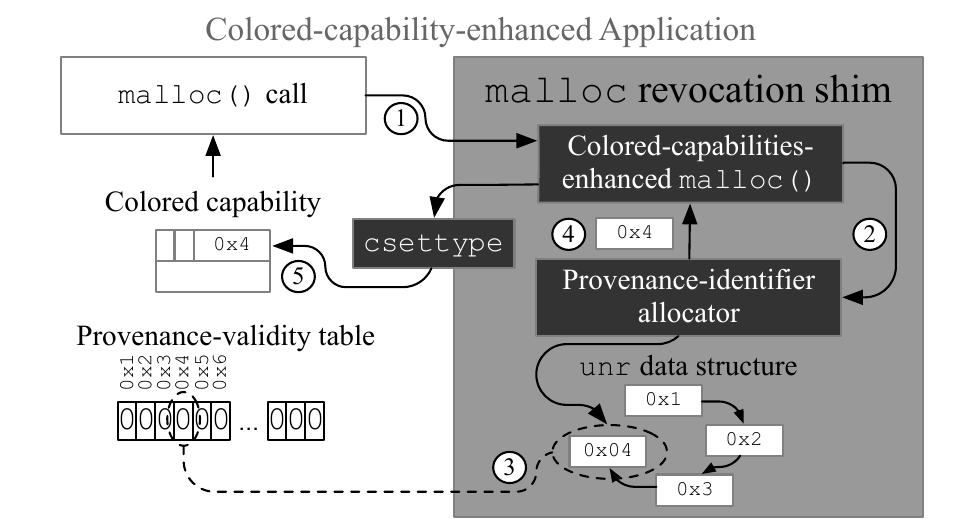}
    \caption{Allocation using \ccs.}\label{fig:cc-alloc}

\end{figure}

\paragraph{Userspace memory allocator.} We modify the CheriBSD \gls{mrs} to support our allocation-\ccprocing mechanism similar to ~\cite{WesleyFilardo20, Filardo24}.
\Cref{fig:cc-alloc} shows how we assign \ccpids to newly allocated memory objects. Any \texttt{malloc()} call first invokes \gls{mrs}~\dCOne. After the memory is allocated, the \gls{unr} allocator~\dCTwo claims an available \ccpid~\dCThree and assigns it to the newly created capability via \texttt{ccsettype}~\dCFive. This mechanism is also integrated into the internal libc and thread‑local storage allocators.

\begin{figure}[t!]
    \includegraphics[width=0.9\linewidth]{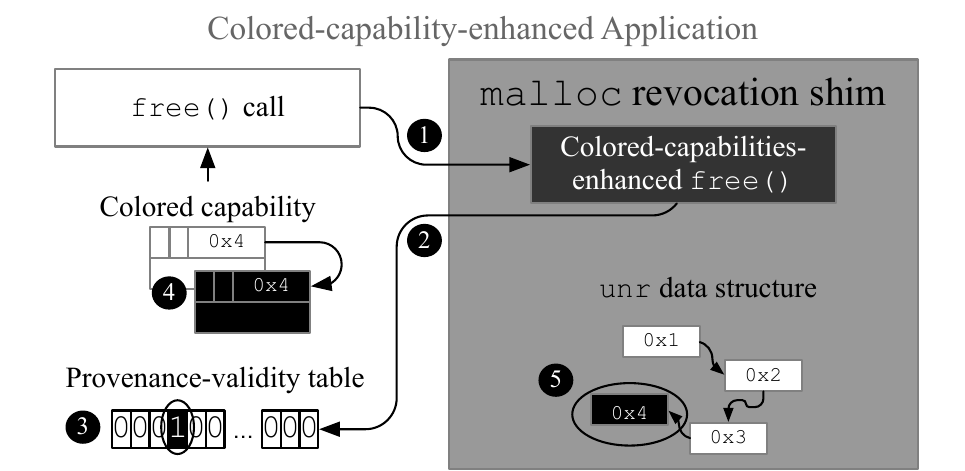}
    \caption{Free using \ccs.}\label{fig:cc-free}
\end{figure}

Figure~\ref{fig:cc-free} shows our approach to \ccdeactivate \ccpids when memory is freed. When a \texttt{free()} call is made, it invokes \gls{mrs}~\dOne, similar to \texttt{malloc()} calls. \emph{\textbf{Unlike Cornucopia, we do not quarantine the memory.}} Instead, we read the \ccpid from the capability~\dTwo and \ccdeactivate the corresponding \gls{pvb} in the \cctable~\dThree. Prior to invalidation, the \gls{mrs} checks whether the \gls{pvb} is already \ccdeactivated. This allows detection of \gls{df} violations regardless of the underlying allocator.
After that, the memory is freed~\dFour and ready to be reused. However, the \ccpid remains reserved by the \gls{unr} to prevent capabilities to a new memory object from using an \ccdeactivated \ccpid~\dFive.
By default, the CheriBSD \gls{mrs} rejects invalid capabilities (\labelcref{asm:malformed-free} \Cref{sec:systemmodel});
our extension validates any capability passed to free has a \ccpid, ensuring that capabilities pointing to non-heap-allocated data are rejected.


\begin{figure}[t!]
    \centering
    \includegraphics[width=0.8\linewidth]{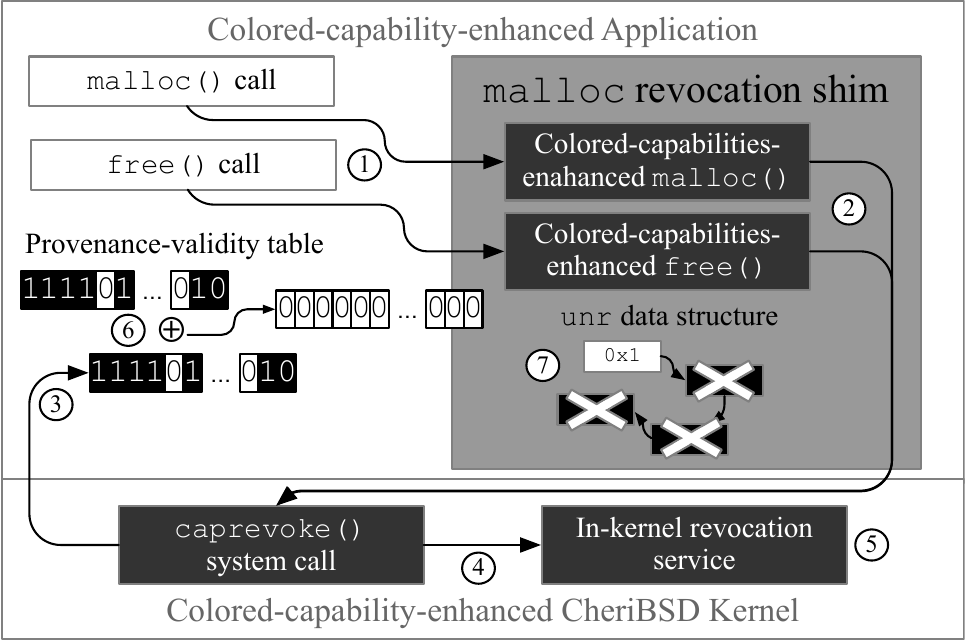}
    \caption{\CC revocation.}\label{fig:cc-revocation}
\end{figure}

\paragraph{Revocation.} While \PICASSO removes the need for a quarantine buffer, the finite number of \ccpids still requires periodic revocation in long‑running programs to reclaim \ccdeactivated identifiers for reuse in the \gls{unr}.
\Cref{fig:cc-revocation} shows how we implement this revocation step to release \ccdeactivated \ccpids. During each memory allocation, the \gls{mrs} monitors the population of reserved \ccpids against a predefined threshold \dCOne. Exceeding this limit triggers a \caprevoke system call \dCTwo to initiate a kernel-level revocation sweep. The kernel then creates a snapshot of the current \cctable~\dCThree to establish a definitive set of \ccpids that are \ccdeactivated at that moment. Any \ccpids that become \ccdeactivated after this snapshot are excluded from the current revocation cycle and will remain in their \ccdeactivated state upon its completion. The kernel subsequently spawns a kernel thread to asynchronously revoke all \ccdeactivated capabilities~\dCFour. We employ the same revocation method~\dCFive as described in Cornucopia Reloaded\cite{Filardo24}, with the exception that, rather than using a shadow bitmap, we inspect the copy of the \cctable to determine whether the \ccpid of the capability is marked as \ccdeactivated. Any subsequent malloc call will then poll the kernel to determine if the \caprevoke is finished. Once done, we clear the corresponding bits in the \cctable that represent the \ccpid values involved in the revocation~\dCSix. Then perform a batch release of all \ccdeactivated \ccpids within the \gls{unr}~\dCSeven (\Cref{appendix:unr}).

\paragraph{\ccPtlb page fault handling.}
\ifabridged
\ccTable pages can be evicted from physical memory to swap.
De-referencing heap objects via \ccs can trigger page faults accessing the corresponding \ccpid entry. User-space page faults are handled as usual, but \ccs passed to system calls can  trigger \ccpid check in kernel code.
As the kernel does not expect memory accesses in privileged code to fault, we extend the CheriBSD page-fault handler to support faults for the \cctable even in privileged mode.
\else
Just like any other memory, it is possible for the \cctable (or individual pages of the \cctable) to leave physical memory. Any attempt to dereference a \cc to a heap object could then trigger a page fault if the part of the \cctable that corresponds with that \cc \ccpid is not in physical memory. This is not a problem if the dereference happens while running in unprivileged mode. However, if it happens while running in privileged mode, it causes the entire operating system to panic and shut down, as the kernel expects all of its memory accesses outside of specific functions to be valid. This can happen if a \cc is provided as an argument for a system call. To solve this problem we extend the page fault handler to handle page fault for the \cctable even if they happen in privileged mode.
\fi

\section{Evaluation}\label{sec:evaluation}

\begin{table*}[t!]
    \centering

    \caption{Area and power costs on VCU118 @ 25MHz expressed in number of LUTs and registers, and Watts respectively.}\label{tab:hwcost}
    \resizebox{\textwidth}{!}{
    \begin{tabular}{r rcc rcc rcc rcc}\toprule
    & logic      & \multicolumn{2}{c}{$\Delta (\%)$} & memory & \multicolumn{2}{c}{$\Delta (\%)$} 
    & registers  & \multicolumn{2}{c}{$\Delta (\%)$} & power  & \multicolumn{2}{c}{$\Delta (\%)$} \\ \midrule
    CHERI-Toooba          & {688096} & \multicolumn{2}{c}{-}
                               & 20113 & \multicolumn{2}{c}{-}
                               & {419300} & \multicolumn{2}{c}{{-}}
                               & {6.471} & \multicolumn{2}{c}{-} \\
    \PICASSO CHERI-Toooba without PVT buffer & {709869} & \multicolumn{2}{c}{21773(+3.16)}
                               & {20079} & \multicolumn{2}{c}{-34(-0.17)}
                               & {436574} & \multicolumn{2}{c}{17274(+4.12)}
                               & {6.478} & \multicolumn{2}{c}{{0.01}} \\
    \PICASSO CHERI-Toooba with PVT buffer & {720347} & \multicolumn{2}{c}{32251(+4.69)}
                               & {21611} & \multicolumn{2}{c}{1498(+7.65)}
                               & {445129} & \multicolumn{2}{c}{25829(+6.16)}
                               & {6.515} & \multicolumn{2}{c}{{0.04}} \\
    \bottomrule
    \end{tabular}}
\end{table*}

We evaluate \PICASSO across four dimensions: functionality, security, performance, and hardware area cost.
For the performance evaluation we use the BESSPIN-GFE security evaluation platform, which provides out-of-the-box support for the CHERI-Toooba softcore. We replace the standard core with our \PICASSO CHERI-Toooba, an extension of the CHERI-RISC-V Toooba FPGA softcore\ifarxiv\linebreak\fi(RV64ACDFIMSUxCHERI) built on the open-source Bluespec RISC-V 64-bit Toooba core. 
We synthesize the \gls{soc} at 25MHz (default for BESSPIN-GFE) targeting the Xilinx Virtex UltraScale+ VCU118 FPGA. Functional and security evaluations are conducted using the hardware platform and a colored-capability–enhanced \texttt{QEMU-system-CHERI128} full-system emulator.
We answer the research questions:
\begin{enumerate}[label=\textbf{RQ\arabic*},labelsep*=0pt]
    \item:~{Can \PICASSO reliably detect \gls{uaf} conditions?}\label{rq:uaf-detection}
    \item:~{How does \PICASSO impact the baseline processor in terms of hardware cost and performance?}\label{rq:non-cc-performance}
    \item:~{What is the performance and memory overhead of enforcing temporal-safety with \PICASSO?}\label{rq:cc-performance}
    \item:~{To what extent does \PICASSO reduce the frequency of CHERI revocation sweeps?}\label{rq:cc-revoke}
\end{enumerate}

 \subsection{Security Evaluation}\label{sec:security-eval}
To answer \labelcref{rq:uaf-detection}, we evaluate \gls{picasso} using version 1.3 of the U.S. \acrshort{nist} \acrshort{sard} Juliet Test Suite~\cite{Boland12}. Juliet contains thousands of C/C++ test cases illustrating common programming flaws that can lead to memory vulnerabilities. The tests are organized by \gls{cwe} categories, with “bad” cases containing known flaws and corresponding “good” cases representing the patched versions. We focus on CWE-416 (Use After Free) and CWE-415 (Double Free), encompassing 1385 vulnerable (“bad”) and as many fixed (``good'') C and C++ test cases. \gls{picasso} detects all bad cases without false positives.
Cornucopia fails to detect any bad CWE-416 cases in its default configuration, but achieves full success rate with \glsdesc{rof} (\acrshort{rof}, \Cref{sec:cheri-temporal-safety}). \acrshort{rof}, however, comes with an impractical performance cost (\Cref{appendix:eval}). Cornucopia detects all CWE-415 bad cases.

\revise{We additionally confirm PICASSO`s efficacy in preventing real-world CVEs in BZip2 (CVE-2016-3189), libiberty (CVE-2016-4487), \texttt{nasm} (CVE-2017-10686, CVE-2019-8343), libzip (CVE-2019-17582), NGINX (CVE-2020-24346), LibreDWG (CVE-2022-35164), Lua (CVE-2019-6706), and other issues reported in \texttt{mjs} (issues 73 and 78) and \texttt{yasm} (issue 91) reproduced from the test suites in~\cite{Nguyen20,Ahn24}.}

\begin{figure*}[t!]
     \centering
     \begin{subfigure}[b]{0.45\linewidth}
         \centering
         \includegraphics[width=\linewidth]{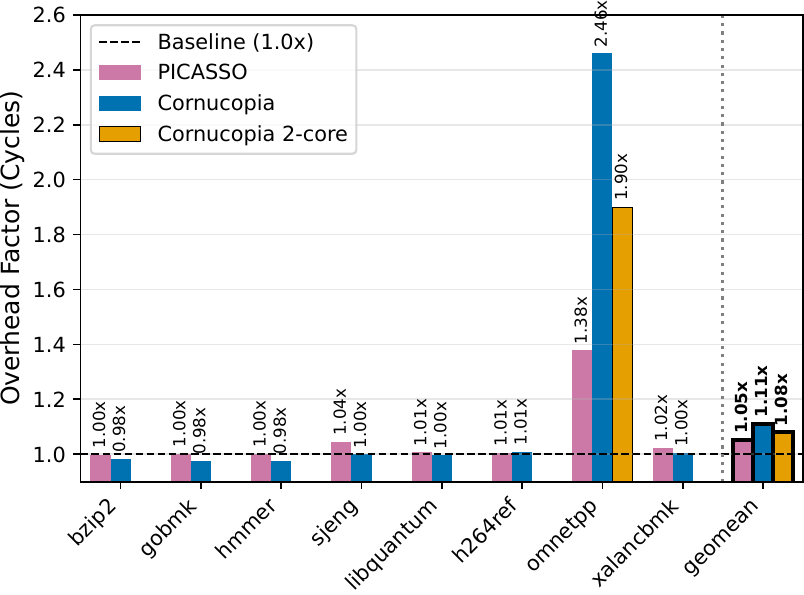}
         \caption{Normalized \gls{cpu} cycles. We omit bars for the 2-core configuration when Cornucopia's revocation less frequently triggers.}\label{fig:spec-cycles}
     \end{subfigure}
     \hfill
     \begin{subfigure}[b]{0.45\linewidth}
         \centering
         \includegraphics[width=\linewidth]{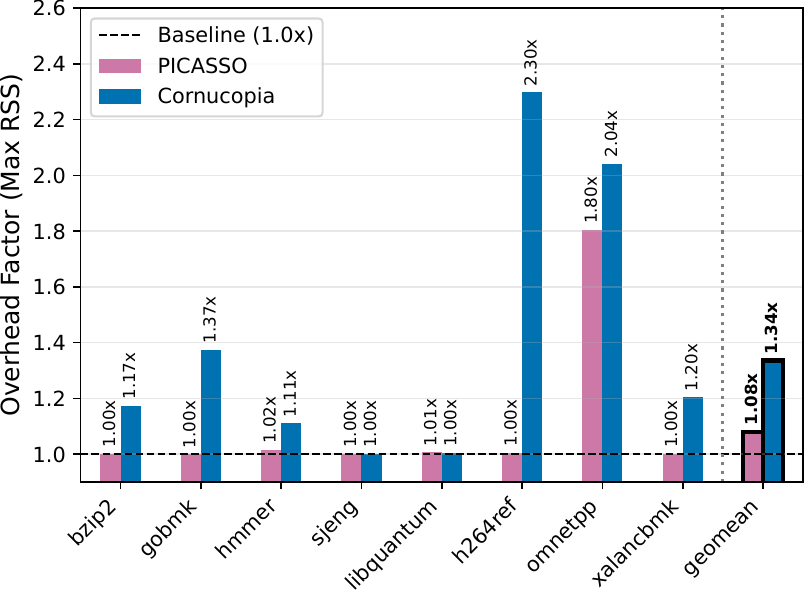}
         \caption{Normalized memory overhead measured as maximum \gls{rss}.}\label{fig:spec-memory}
     \end{subfigure}
     \caption{SPEC CPU 2006 performance overhead for \PICASSO ($1.05\times$ g.m.) and Cornucopia ($1.11\times$ g.m. in single-core, and $1.08\times$ g.m. in 2-core configuration), and memory overhead of 
\PICASSO ($1.08\times$ g.m.) and Cornucopia ($1.34\times$ g.m.).}
     \label{fig:spec-comparison}
\end{figure*}

\subsection{Performance Evaluation}\label{sec:performance-eval}

\paragraph{Impact on baseline CHERI-Toooba processor.} To answer \labelcref{rq:non-cc-performance}, we synthesize \PICASSO CHERI-Toooba for the VCU118 \gls{fpga}. We also evaluated the design without the PVT buffer to isolate the logic overhead introduced by the optimization. \Cref{tab:hwcost} shows power and resource utilization reported by Xilinx Vivado 2019.1. Without the \gls{pvt} buffer, the overheads are minimal ($\approx$3\% in logic and $\approx$4\% in registers) compared to unmodified CHERI-Toooba. Incorporating the PVT buffer increases the overheads to $\approx$4.7\% in logic $\approx$8\% in memory and $\approx$6\% in registers, a moderate impact justified by the performance tradeoff of the \gls{pvt} buffer (\Cref{appendix:eval}).

We investigated the performance overhead of our memory pipeline changes when not using the colored capabilities features caused by the additional pipeline latency introduced by \gls{ptlb} lookup (\dSix in \Cref{fig:impl_verif}, \Cref{sec:hwimpl}).
This overhead is not fundamental to the approach, but is nevertheless an artifact of the current prototype.
EEMBC CoreMark~\cite{EEMBC24} incurred between 3.67\% and 6.07\% overhead depending on compilation mode on the VCU118, and MiBench incurred 1.32\% overhead on average in simulation (see \Cref{tab:coremark,tab:mibench} in \Cref{appendix:eval}).
Overheads in larger benchmarks were not as pronounced.

\paragraph{Performance and memory overhead of enforcing temporal safety.} To answer \labelcref{rq:cc-performance}, we measure the overhead of \PICASSO relative to the \gls{cheri}-enabled memory allocator in \CheriBSDVersion, which is based on the Cornucopia Reloaded design. 
Unlike the allocator described by Filardo et al.~\cite{Filardo24}, the CheriBSD implementation performs the \caprevoke system call asynchronously by spawning a new kernel thread that handles the revocation sweep. We utilize the Cornucopia configuration, where the quarantine limit is set to one-fourth of the total allocated memory.  For brevity, we refer to the \CheriBSDVersion allocator implementation as ``Cornucopia''. 

We use a single-core \gls{cpu} configuration, our results include the full cost of revocation: preempting the application by the revocation thread results in stop-the-world behavior.
On multicore systems, this overhead could be reduced by overlapping the revocation sweep with application execution.
The overhead we report is therefore an overestimate but enables a conservative assessment of the total system overhead introduced by the mechanism.

We instrument \caprevoke to collect revocation counts,
measure peak memory overhead using \texttt{time -l} in CheriBSD, which reports the maximum resident set size during execution, and collect hardware performance metrics using CheriBSD’s \texttt{minimal\_counts\_stats} tool.
We evaluate \PICASSO's performance impact across the following benchmarks:

\paragraph{SPEC CPU2006 INT.} We run the subset of the SPEC CPU2006 INT benchmarks~\cite{spec06} that execute on the baseline CHERI system without code modifications~\cite{Rugg23}. Similar to prior work~\cite{Rugg23}, we use the \emph{train} input set, as even these inputs per configuration require over 24 hours to complete on \gls{fpga}. \Cref{fig:spec-comparison} shows the overall cycle and memory normalized overhead, for the CheriBSD allocator on the unmodified CHERI-Toooba core (Cornucopia) and for \PICASSO. 
In 7 out of 8 INT tests, the number of allocations is too few to trigger revocation in \PICASSO, however, it triggers in most cases for Cornucopia. In four tests, Cornucopia achieves better performance than the baseline since it postpones memory deallocation and instead places freed objects into the quarantine buffer, which involves less bookkeeping than reclaiming the memory. 
For omnetpp, \PICASSO and Cornucopia trigger the revocation mechanism 76 and 2792 times, respectively, resulting in a worst-case overhead of 38\% and 268\%.
To understand the impact of parallelizing revocation, we obtain measurements for the SPEC CPU 2006 INT benchmarks on an unmodified CHERI-Toooba in a 2-core configuration from \censor{the Cambridge Computer Laboratory}.
Parallelizing revocation reduces the Cornucopia overhead to 90\% relative to our baseline for omnetpp but has no effect in cases where revocation less frequently triggers.
\PICASSO in a single-core configuration still outperforms Cornucopia in the 2-core configuration. \PICASSO shows negligible memory overhead in most cases. However, unlike Cornucopia, whose memory consumption peaks as the quarantine buffer grows to its maximum size, \PICASSO's memory usage increases during revocation due to handling two \gls{pvt} tables and the memory consumed by the \gls{unr} allocator (\Cref{sec:swstack}). We observe a worst-case memory overhead in \PICASSO of 80\% during revocation in omnetpp,
lower than Cornucopia's in all benchmarks.
To quantify the impact on DRAM, we collect hardware counters that characterize DRAM traffic. \PICASSO increases DRAM traffic by 21\% (g.m.), whereas Cornucopia increases it by 65\% (g.m.) (see \Cref{fig:cc-specdram} in \Cref{appendix:eval}).

\begin{figure*}[t]
    \centering
    \includegraphics[width=\linewidth]{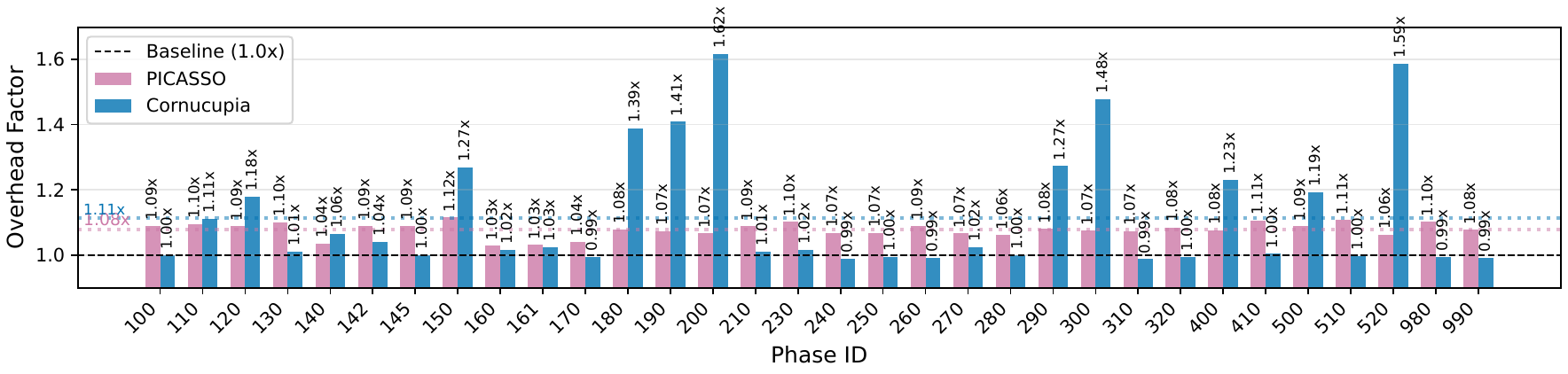}
    \caption{Normalized Performance overhead of \PICASSO ($1.08\times$ avg., $1.12\times$ max.) and Cornucopia ($1.12\times$ avg., $1.62\times$ max.) across SQLite speedtest1 operational phases. Phase description and raw data are available in \ifarxiv\Cref{tab:sqlitephase} in \Cref{appendix:eval}\fi  \ifnotarxiv \revise{Table 8 in Appendix C of \cite{Gulmez26}}\fi.}\label{fig:cc-sqlite-phase}
\end{figure*}

\paragraph{SQLite.} We further evaluate \PICASSO using SQLite 3.22.0~\cite{sqlite} with its built-in speedtest1 benchmark, which consists of multiple phases of database operations such as create, insert, reorder, and delete. A complete list of benchmark phase descriptions can be found in \ifarxiv\Cref{tab:sqlitephase} in \Cref{appendix:eval}\else \revise{Table 8 in Appendix C of \cite{Gulmez26}}\fi. \Cref{fig:cc-sqlite-phase} illustrates the performance overhead across these distinct operational phases.
Overall, \PICASSO incurs a $8\%$ performance overhead on average, while Cornucopia incurs a $11\%$ average performance overhead and Cornucopia exhibits a $99\%$ memory overhead, whereas \PICASSO incurs no observable memory overhead.
Unlike Cornucopia, which suffers performance spikes up to 62\% during revocation, \PICASSO maintains a consistent profile with a peak overhead of only 12\%. In the SQLite \texttt{speedtest1} benchmark, Cornucopia triggers 272 revocations while \PICASSO triggers zero.

\begin{figure}[t]
    \centering
    \includegraphics[scale=0.55]{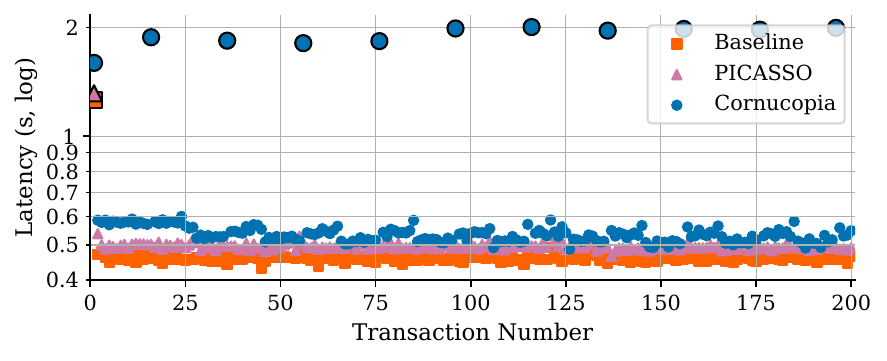}
    \caption{Latency  for 200 sample \texttt{pgbench} transactions. Markers denote individual response times. Initial latency spike at $t=1$ is due to connection establishment.}\label{fig:cc-pgbench}
\end{figure}

\paragraph{PostgreSQL \texttt{pgbench}.} 
To evaluate \PICASSO impact on latency we run the PostgreSQL server\cite{postgresql, cheripostgress} on the CHERI-Toooba core with \texttt{pgbench} executed on a separate host connected over the local network. We configure a scale factor of 10 and execute $10^3$ transactions with a single client. To reduce I/O overhead and ensure single-threaded execution suitable for the single-core CHERI-Toooba processor, we disable fsync, synchronous commit, full page writes, statistics collection, and parallel workers. As illustrated in \Cref{fig:cc-pgbench}, which plots individual transaction latencies, Cornucopia exhibits periodic latency spikes occurring approximately every 20 transactions (60 allocations per transaction). Each latency spike represents a tail-latency penalty, with p99 latency increasing by over 300\%, due to the non-concurrent revocation sweep in a single-core environment (see \Cref{fig:cc-pgbench} in \Cref{appendix:eval}). Overall, \PICASSO degrades \gls{tps} by 6.2\%, compared to 23.1\% for Cornucopia. 
We measure that \PICASSO requires revocation only once every $66\times10^3$ transactions (vs. once every 20 in Cornucopia).
For a production-speed core similar to Arm Morello (2.5GHz), we estimate this corresponds to one revocation every 2.5 min vs. one every 71.4 ms for Cornucopia \cite{Filardo24}.

\begin{figure}[t]
    \centering
    \includegraphics[scale=0.54]{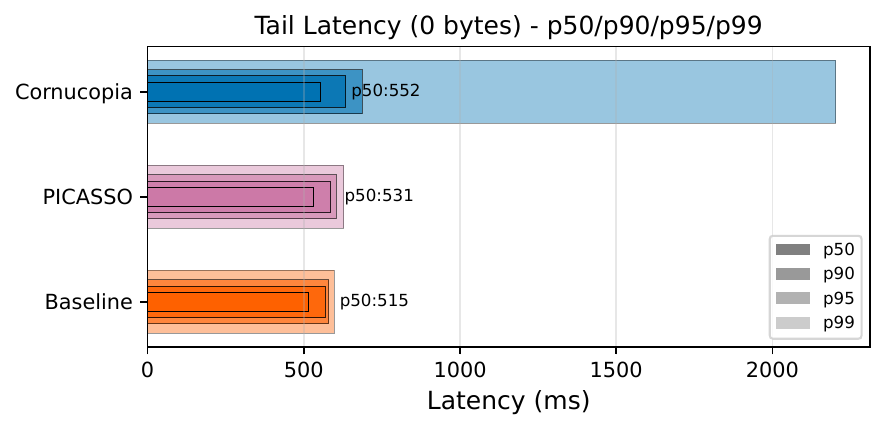}
    \caption{gRPC QPS latency percentile. Raw data are available in \Cref{tab:grpc-benchmark} in \Cref{appendix:eval}.}\label{fig:cc-qps}
\end{figure}

\paragraph{gRPC QPS.} We run a gRPC 1.54.2 QPS\cite{Tiller25}, client-server workload. Similar to pgbench setup, we run the server on the FPGA, and run the clients on a separate host.  To isolate and measure only gRPC server overhead, we use a custom gRPC scenario: transport security is disabled, each of the client and server is a single process, both client and server are synchronous and apply latency-focused tuning to a minimal stack, the client opens 1 gRPC channel with 8 outstanding messages. The client sends messages to the server and awaits responses, while measuring throughput and latency percentiles.  Each run lasts 120 seconds with an initial 15‑second warmup, and each configuration is executed twice with 0‑byte request sizes. \Cref{fig:cc-qps} shows the different latency percentiles. For p50 latency, \PICASSO incurs a 3\% overhead, while Cornucopia incurs a 7\% overhead. At p99, \PICASSO incurs only a 5\% overhead, whereas Cornucopia exhibits a $3.7\times$ increase due to revocation. Overall, \PICASSO introduces a 3\% \gls{qps} degradation, while Cornucopia incurs a 15.6\% degradation.
To further evaluate when \PICASSO triggers revocation, we run the same benchmark for 8 hours, more than 8 million allocations, resulting in a total of four revocations. These four revocations over 8 hours do not affect any latency percentile, even up to p999. 
In comparison, Cornucopia triggers seven revocations within just 120 seconds.

\revise{\paragraph{Performance and memory overhead analysis}. \PICASSO overhead is bounded by the revocation frequency, determined by the ratio of allocations to frees. For example, omnetpp revocations reclaim $\approx$80\% of provenance
identifiers on average. As the number of allocations increase, revocation counts scale proportionally but given a comparable reclamation rate of \ccpids the per-revocation cost remains similar. \PICASSO can fall back to Cornucopia non-provenance-tracked allocations if provenance
identifiers are exhausted. As revocation rate scales with available \ccpids, \PICASSO’s advantage widens relative to Cornucopia under allocation/free pressure as Cornucopia’s revocation frequency depends on the amount of quarantined memory. 
PVT memory overhead is 256 KiB per process (\Cref{sec:hwimpl}, \Cref{sec:swstack}) independent of the amount of heap memory consumed since PVBs are associated with capabilities, not allocations.
Fragmentation of \ccpids also contribute to \PICASSO's memory overhead due to memory consumed by the \texttt{unr} allocator.}

\paragraph{Reduction in revocations.}\label{sec:revocation-eval}
To answer \labelcref{rq:cc-revoke}, we collect statistics on the number of allocations and revocations as discussed for each individual benchmark above. As shown in \Cref{tab:revokecount}, \PICASSO does not trigger revocation in the majority of benchmarks. Our further experiment with long-running PostgreSQL and gRPC shows that \PICASSO reduces the revocation time significantly without increasing the latency percentile even with p999.   This demonstrates that \PICASSO eliminates frequent revocation sweeps for long-running software, significantly reducing overall system load and overhead.


\begin{table}[h]
    \centering
    \caption{Allocations and revocations for \Cref{sec:performance-eval} benchmarks.}\label{tab:alloc_revoke_counts}
    \resizebox{\linewidth}{!}{
    \begin{tabular}{l r r r}
        \toprule
        & & \multicolumn{2}{c}{\textbf{\# Revocations}} \\
        \cmidrule(lr){3-4}
        \textbf{Benchmark} & \textbf{\# Allocations} & \textbf{Cornucopia} & \textbf{PICASSO} \\
        \midrule
        bzip2  & 87 & 3 & 0 \\
        gobmk  & 121361 & 17 & 0 \\
        hmmer       & 170126 & 16 & 0 \\
        sjeng      & 5  & 0 & 0\\
        libquantum  & 108   & 1  & 0 \\
        h264ref   & 38270 & 12 & 0 \\
        omnetpp & 129620400 & 2792 & 76 \\
        xalancbmk      & 1057345  & 2  & 0\\
        SQLite      & 224917  & 272  & 0\\
       \toprule
        \textbf{Benchmark} & \textbf{\# Transactions} & \textbf{Cornucopia} & \textbf{\PICASSO} \\    
        \midrule
        PostgreSQL  & $\approx$66000 & $\approx3300$ & 1\\
        gRPC  & $\approx$100000 & $\approx$ 400 & 1\\
        \bottomrule
    \end{tabular}}
    \label{tab:revokecount}
\end{table}

\section{Related Work}\label{sec:relatedwork}
Research on \gls{uaf} broadly follows two strategies: \emph{detection}, which identifies violations during program execution, and \emph{prevention}, which blocks exploitation of freed memory.

\paragraph{Detection.}
``\emph{Lock-and-key}'' schemes~\cite{Duck18, Li22, Nagarakatte10, Caballero12, Cho22} pair each pointer with a ``key'' and each memory object with a ``lock''.
When an object is freed, its lock is invalidated.
At dereference, the key and lock are compared to detect temporal violations.
These methods resemble \PICASSO's provenance-based detection but require complex static data-flow analysis and run-time instrumentation of pointer dereferences, leading to high run-time overhead (e.g., CETS~\cite{Nagarakatte10} incurs 48\%).
PTAuth~\cite{Farkhani21} applies a similar scheme using Arm pointer authentication~\cite{QualcommProductSecurity17} by  creating an authentication between a pointer and its object using a random ID, but suffers from costly base-address lookups for interior pointers (pointers to the middle of objects).
DangNull~\cite{Lee15} automatically nullifies all pointers to freed objects by tracing pointer references through run-time instrumenting, at a steep 80\% run-time overhead.
In contrast, \PICASSO is a hardware feature that detects \gls{uaf} without compiler transformations or pointer analysis. 
Another class of detection relies on \emph{page-permission–based} schemes~\cite{Dang17, Kouwe17}.
In these schemes each heap object is allocated on a separate virtual page.
When an object is freed, access to the corresponding page is revoked.
Any dangling dereference immediately triggers a hardware fault, avoiding the need to instrument dereference sites. 

\emph{Tag-based memory versioning} assigns matching tags to pointer and objects and reassigns new tags to objects when they are freed.
This is probabilistic due to tag collisions, especially with small tag spaces (e.g., ARM MTE’s 4-bit tags).
Combining \gls{cheri} with memory versioning (\Cref{sec:cheri-temporal-safety}) mitigates the ``reallocation gap” by reassigning tags before quarantining freed memory.
However, with 4-bit versioning, they can track only 16 distinct allocations before memory tags collide.
In contrast, \PICASSO supports over two million unique allocation-\ccpids without relying on memory tagging thus enabling better scaling, deterministic protection.

\paragraph{Prevention.} Prevention approaches avoid reusing memory until all references to freed regions are eliminated~\cite{Kouwe17, Ainsworth20, Liu18, Erdos22, Park24}, or reduce the likelihood of successful exploitation by delaying reuse of freed blocks~\cite{Silvestro17, Berger06, Novark10, Akritidis10, Liu19, Silvestro18}. Another strategy employs virtual address–based allocators~\cite{Wickman21, Ahn24}, exploiting the abundant 64-bit address space to ensure each allocation receives a unique virtual address and is never reused at the same location. These techniques do not detect developer mistakes, but can block \gls{uaf}-based attacks outright. \revise{These schemes also assign unique identifiers to every allocation to prevent temporal aliasing. This makes their memory overhead scale with application memory usage. In contrast, \PICASSO's main source of memory overhead stems from the \gls{pvt}; a fixed cost per process since \ccpids are associated with capabilities and \glspl{pvb} describes whether the allocation which provenance the corresponding capabilities relate is still is valid; they do not enumerate unique allocations.}

\paragraph{Temporal-safety solutions for \gls{cheri}.}
Apart from quarantine-based approaches \ifnotarxiv (\Cref{sec:cheri-temporal-safety}) \else (Cornucopia, Cornucopia Reloaded, and composing \gls{cheri} and memory versioning, detailed in \Cref{sec:cheri-temporal-safety})\fi, temporal-safety solutions for \gls{cheri} also include the CHERIoT \gls{rtos}~\cite{Amar23,Amar23a}.
CHERIoT is geared towards embedded systems without virtual memory that use a ``flat'' physical memory space shared between programs (tasks) on the device.
These systems avoid heap allocation and preallocate all needed memory, requiring total memory equal to the sum of all component's  peak memory usage.
C++ \texttt{new} and \texttt{delete} operators are likely to still use the \texttt{malloc}-family of functions for dynamic allocations.
CHERIoT enables safe reuse of memory across \gls{rtos} tasks. All allocations/frees require an allocator capability tied to a specific quota, and pointers can only be freed using the same capability that allocated them, preventing unauthorized releases.
A \emph{revocation bitmap} (1-bit per 8 bytes heap, similar to Cornucopia's shadow bitmap \Cref{sec:cheri-temporal-safety}) tracks freed objects.
When capabilities load from memory to registers, a \emph{load filter} checks the revocation bit associated capability's base and clears the validity tag if the capability points to freed memory, preventing dangling pointers.
Freed objects are quarantined while periodic revocation sweeps to invalidate any lingering stale capabilities.
Because the load filter prevents new invalid pointers from forming, this sweep can be done safely in the background.
Unlike \Cref{sec:cheri-temporal-safety} approaches, CHERIoT deterministically prevents de-referencing dangling pointers.
However, this requires the hardware-managed load filter check every capability load against a bitmap of \emph{freed memory}, making the CHERIoT approach only suitable for resource-constrained devices with limited memory (and manageable revocation bitmap sizes).
Further, unlike \PICASSO, CHERIoT still relies on quarantining memory.

\paragraph{In Summary.} PICASSO addresses the \gls{uaf}/\gls{uar} gap, can extend the revocation period without requiring any memory to be quarantined, and can detect the \gls{uaf} bugs deterministically, without any compiler instrumentation and pointer analysis.

\section{Discussion and Future Work}\label{sec:discussion}
\ifnotabridged
This section reflects on the relevance of \ccs with respect to past and current capability systems and outlines directions for future work.
\fi

\ifnotabridged
\begin{table} \footnotesize
\caption{Estimates for \ccpid space size and corresponding memory consumption for different encodings.}
\label{tab:object-type-sizes}
\begin{tabular}{lclr}
\toprule
Encoding & \# Bits & \# Colors & \gls{pvt} size\\
\midrule
\gls{cheri} \gls{isa}v9 & 18 &  $2^{18}$ & 32~KiB\\
RISC-V Zcheri & 18 & $2^{18}$ & 32~KiB\\
\gls{cheri}-Toooba (\PICASSO) & 21 & $2^{21}$ & 256~KiB\\
\gls{cheri}-Toooba w/ pointer masking & 21+16 & $2^{37}$ & 16~GiB\\
\bottomrule
\end{tabular}
\end{table}
\fi

\subsection{Historical and Current Relevance}

\paragraph{Rethinking indirection.}
\ifnotabridged
In their 2017 ACM Turing Award lecture, John Hennessy and David Patterson
\else
Hennessy and Patterson
\fi~\cite{Hennessy18} reflected on how early security concepts from the 1970's, including protection domains and rings, as seen in systems like Multics~\cite{Saltzer74}, and  even capabilities~\cite{Levy84}, ended up underused, partly due to a lack of \emph{architectural techniques} necessary to make them sufficiently fast and performant.
Today, with the resurgence of capabilities through \gls{cheri}, it is timely to revisit these historical ideas using modern design tools.

As noted in \Cref{sec:cheri}, \gls{cheri} avoids table-based indirection for revocation, opting instead for tagged memory to track capability validity.
\gls{cheri} ISAv9~\cite[Appendix C.7.2]{Watson23a} discusses indirect capabilities as an experimental feature, favoring limited tagging for performance reasons.
With \emph{\ccs} and \PICASSO, we argue that rejecting indirection outright also lead to suboptimal designs.
While indirection on \emph{every} capability access is costly, our approach shows that \emph{selective} indirection, applied only where beneficial, combined with microarchitectural optimizations, can minimize overhead.

\ifnotabridged
We hope this work inspires the \gls{cheri} and broader hardware security communities to re-explore past ideas through the lens of modern \glsdesc{eda}, turning once-theoretical concepts into practical, performant systems.
\fi

\revise{\paragraph{Applicability to other processors.} \PICASSO targets application processors whereas embedded systems are better served by CHERIoT’s load filter~\cite{Amar23,Amar23a}. CHERIoT's revocation bitmap, however, scales poorly to virtual-memory systems since its size depends on the total available memory whereas PICASSO’s \gls{pvt} scales with numbers of \ccpids reserved from  \gls{otype}. \PICASSO’s \gls{isa}-level design (\gls{pvtr}, \texttt{ccsettype}, \gls{pvb} check) can be applied to any core. Our implementation of \ccs on CHERI-Toooba  with the microarchitectural \gls{pvt} buffer and \gls{ptlb} optimizations demonstrate that our design can  be applicable to superscalar, out-of-order CHERI processors.}

\paragraph{Relevancy for \gls{cheri} standardization.}
\gls{cheri} is in a \ifarxiv\linebreak\fi pre-commercialization phase and is being standardized by the \gls{cheri} Task Group within RISC-V International, under the proposed ``Y'' extension (called ``Zcheri'' during development).
\ifabridged
\gls{rv64y} is receiving additional extensions~\cite{Aird25,Aird25a} enabling temporal-safety enforcement similar to Cornucopia Reloaded.
\else
A key addition for temporal-safety enforcement is the ``Svucrg'' extension~\cite{Aird25} for \gls{rv64y}, which introduces a \gls{crg} bit in each \gls{pte}, along with a corresponding \gls{ucrg} bit in \gls{cpu} status registers.
This mechanism mirrors Cornucopia Reloaded's per-\gls{pte} revocation epochs.
\gls{crg} interacts with the \gls{cw} bit (from the ``Svy'' extension~\cite{Aird25a}) which controls whether capabilities are permitted to be written or read from a page.
When the \gls{pte}'s \gls{cw} and \gls{u} bits are set and the \gls{crg} bit indicates the current generation of the virtual memory page with regards to the ongoing capability revocation cycle.
In this mode, capability stores are allowed but capability reads are controlled by whether the \gls{crg} bit equals \gls{ucrg}.
If \gls{crg} $\neq$ \gls{ucrg}, capability loads to that page trigger a \gls{cheri} load-page-fault, preventing the use of stale capabilities.
This removes the need for revocation sweeps to pre-mark all \glspl{pte} unreadable; toggling the \gls{ucrg} globally is sufficient.
Without \glsentrydescplural{crg}, a concurrent capability revocation sweep must begin by visiting all \glspl{pte} to mark them unreadable.
In addition, a page-granularity capability write control eliminates many pages from the sweep that are known to not contain capabilities.
The \gls{cw} bit also allows the sweep to skip pages that cannot contain valid capabilities.
\gls{cw} is unset, a capability load is executed, the implementation clears the capability tag bit of the capability read from the virtual page.
\fi
Unlike \gls{cheri} \gls{isa}v9, which uses five bits for  capability load/store barrier controls~\cite[4.3.12]{Watson23a}, \ifnotabridged, including whether capabilities operations trap, raise fault, or strip capability tags,\fi
\gls{rv64y} simplifies the encoding due to limited available free \gls{pte} bits.
This, however, requires changes to the revocation state machine and may degrade Zcheri revocation performance compared to \gls{cheri} \gls{isa}v9.

\PICASSO is orthogonal to optimizing the revocation sweep; it reduces how often revocation is needed.
Furthermore, \PICASSO does not require changes to the \gls{pte}.
Consequently, \PICASSO-style improvements are potentially interesting to future \gls{cheri} standardization as a complement for improving temporal-safety enforcement performance.
We believe \ccs are compatible with the currently available work-in-progress drafts of the Zcheri specification, although \gls{otype} has not yet been ratified as a Zcheri feature.
\ifabridged
As fully compatible \glspl{cpu}
\else
As Svucrg-enabled \glspl{cpu} 
\fi
design are not yet readily available we leave integrating \PICASSO into Zcheri as future work.

\subsection{\revise{Limitations and} Future Work}\label{sec:futurework}

\revise{\paragraph{Safe speculative execution.} It is possible for \Gls{pvt} checks that miss the \gls{pvt} buffer to be bypassed in transient execution in our current Toooba implementation, as the \gls{pvt} and data loads are issued in parallel.
Stores are delayed until after the check and cannot be bypassed (see in \Cref{sec:hwimpl}).
Future work could explore the security/performance tradeoff of adapting defensive specification, as in Capability Speculation Contracts~\cite{Fuchs23, Fuchs24}, or integration with state-of-the-art taint tracking~\cite{Yu19a, ElAtali25} to selectively delay loads until after PVT checks.}

\revise{\paragraph{PVT coherence on multi-core.} \Gls{pvt} coherence in core-private caches is maintained by the cache coherence protocol. Furthermore, free() applies fence instructions to discard \glspl{pvb} cached in the \gls{pvt} buffer (\Cref{sec:hwimpl}). In multi-core scenarios, our current implementation would require all cores to perform a fence before a \gls{pvt} update is guaranteed to be seen by all cores.
Future work would add a hardware mechanism to broadcast invalidated colors to other cores to selectively ``shootdown'' modified entries.}

\paragraph{Software-Design Optimization.} We observe that high fragmentation within the \ccpids space causes memory bloat in the \texttt{unr} allocator. To address this, future work could explore alternative allocators, such as those used in GCC~\cite{Carlini25} or the Linux ID allocator~\cite{Wilcox25}. Another approach involves the direct use of a bitmap accelerated by RISC-V hardware extensions such as Bitmanip~\cite{Shanbhogue24} or Vector~\cite{Zabrocki24} for efficient provenance identifier searching. Additionally, the current revocation process requires a full copy of the \gls{pvt}. Implementing a Copy-on-Write (CoW) strategy would significantly reduce this memory overhead and avoid unnecessary duplication during the revocation cycle.

\paragraph{Dynamically-adjustable \gls{pvt}.}
The favorable performance trade-off of \ccs depends on how many \ccpids can be encoded in the capability format.
\ifnotabridged
\Cref{tab:object-type-sizes} shows the bit budget and corresponding \ccshortpid counts for different \gls{cheri} variants.
\fi
In \Cref{sec:implementation} and \Cref{sec:evaluation}, we describe \PICASSO with a 21-bit \ccpid on \gls{cheri}-Toooba.
A larger \ccshortpid space allows more allocations between revocation sweeps, further improving overall system performance.
However, it also has costs:
\begin{inparaenum}[1)]
  \item It consumes finite bits from the capability representation, competing with other \gls{cheri} extensions,
  \item It increases memory usage for the \cctable and \gls{unr} storage,
  \item It raises revocation overhead due to more \gls{unr} entries.
\end{inparaenum}
For example, \gls{cheri}-RISC-V with pointer masking\ifnotabridged (``Zjpm'' extension~\cite{Zabrocki24})\else~\cite{Zabrocki24}\fi{} could support up to $2^{37}$ \ccpids, but would require a 16~GiB \gls{pvt}---impractical for most use cases.
To address this, we envision allocating a small \gls{pvt} at process startup, which can grow dynamically when more \ccshortpids are needed.
This is especially beneficial for short-lived programs which use only a fraction of available \ccshortpids.
Smaller initial \glspl{pvt} reduce memory footprint without impacting performance.

\section{Conclusion}\label{sec:conclusion}

We introduce \emph{\ccs}, a novel hardware-software co-design that enhances \gls{cheri}’s temporal memory safety through efficient, \emph{bulk retraction} of capabilities.
By adding limited indirection to \gls{cheri}’s capability model, \ccs addresses key shortcomings of existing revocation-based approaches--notably, mitigating \gls{uaf} vulnerabilities without requiring quarantining memory management.
Our implementation, \PICASSO, demonstrates the practicality of this approach by integrating \ccs into CHERI-RISC-V, the CheriBSD \gls{os}, and the LLVM toolchain.
Evaluation on thousands of \acrshort{nist} Juliet test cases, \revise{real-world CVEs}, SPEC CPU benchmarks, and latency-sensitive real-world benchmarks confirm that colored capabilities improve temporal safety with reasonable performance overhead ($3\%-8\%$ overhead on average).
\ifnotarxiv
\input{sections/ethicalconsiderations.tex}
\input{sections/openscience.tex}
\fi
\ifnotanonymous
\section*{Acknowledgments}
\revise{We thank our anonymous Shepherd and our collegues at Ericsson: Santeri Paavolainen, Sini Ruohomaa, and Zoltán Turányi for their feedback on this manuscript.
We are equally grateful to the University of Cambridge Computer Laboratory Security Group members, particularly Alfredo Mazzinghi, Peter Rugg, Jessica Clarke, Robert Watson, and Simon Moore.}

\revise{This work is supported in part by the Wallenberg Visiting Professor Program, the Natural Sciences and Engineering Research Council of Canada (grant number RGPIN-2026- 04826), and the Government of Ontario (RE011-038). Views expressed in the paper are those of
the authors and do not necessarily reflect the position of the
funders.}
\fi
\ifreviewer
\input{sections/usenix.tex}
\fi
\ifarxiv
\bibliographystyle{ACM-Reference-Format}
\else
\bibliographystyle{plainurl}
\fi
\bibliography{main, local}


\begin{thebibliography}{72}


\ifx \showCODEN    \undefined \def \showCODEN     #1{\unskip}     \fi
\ifx \showISBNx    \undefined \def \showISBNx     #1{\unskip}     \fi
\ifx \showISBNxiii \undefined \def \showISBNxiii  #1{\unskip}     \fi
\ifx \showISSN     \undefined \def \showISSN      #1{\unskip}     \fi
\ifx \showLCCN     \undefined \def \showLCCN      #1{\unskip}     \fi
\ifx \shownote     \undefined \def \shownote      #1{#1}          \fi
\ifx \showarticletitle \undefined \def \showarticletitle #1{#1}   \fi
\ifx \showURL      \undefined \def \showURL       {\relax}        \fi
\providecommand\bibfield[2]{#2}
\providecommand\bibinfo[2]{#2}
\providecommand\natexlab[1]{#1}
\providecommand\showeprint[2][]{arXiv:#2}

\bibitem[pos({[n.\,d.]})]%
        {postgresql}
 \bibinfo{year}{[n.\,d.]}\natexlab{}.
\newblock \bibinfo{title}{{{PostgreSQL}}}.
\newblock
\urldef\tempurl%
\url{https:/postgresql.org/}
\showURL{%
\tempurl}


\bibitem[spe({[n.\,d.]})]%
        {spec06}
 \bibinfo{year}{[n.\,d.]}\natexlab{}.
\newblock \bibinfo{title}{{{SPEC CPU}} 2006}.
\newblock
\urldef\tempurl%
\url{https://spec.org/cpu2006/}
\showURL{%
\tempurl}


\bibitem[sql({[n.\,d.]})]%
        {sqlite}
 \bibinfo{year}{[n.\,d.]}\natexlab{}.
\newblock \bibinfo{title}{{{SQLite Home Page}}}.
\newblock
\urldef\tempurl%
\url{https://sqlite.org/}
\showURL{%
\tempurl}


\bibitem[Hen(2018)]%
        {Hennessy18}
 \bibinfo{year}{2018}\natexlab{}.
\newblock \bibinfo{title}{John {{Hennessy}} and {{David Patterson}} 2017 {{ACM A}}.{{M}}. {{Turing Award Lecture}}}.
\newblock
\urldef\tempurl%
\url{https://www.youtube.com/watch?v=3LVeEjsn8Ts}
\showURL{%
\tempurl}


\bibitem[Ahn et~al\mbox{.}(2024)]%
        {Ahn24}
\bibfield{author}{\bibinfo{person}{Junho Ahn}, {and} others.} \bibinfo{year}{2024}\natexlab{}.
\newblock \showarticletitle{{{BUDAlloc}}: Defeating Use-after-Free Bugs by Decoupling Virtual Address Management from Kernel}. In \bibinfo{booktitle}{\emph{{{USENIX Security}} '24}}. \bibinfo{publisher}{{USENIX}}, \bibinfo{pages}{181--197}.
\newblock
\showISBNx{978-1-939133-44-1}
\urldef\tempurl%
\url{https://www.usenix.org/conference/usenixsecurity24/presentation/ahn}
\showURL{%
\tempurl}


\bibitem[Ainsworth and Jones(2020)]%
        {Ainsworth20}
\bibfield{author}{\bibinfo{person}{Sam Ainsworth} {and} \bibinfo{person}{Timothy~M. Jones}.} \bibinfo{year}{2020}\natexlab{}.
\newblock \showarticletitle{{{MarkUs}}: {{Drop-in}} Use-after-Free Prevention for Low-Level Languages}. In \bibinfo{booktitle}{\emph{S\&{{P}} '20}}. \bibinfo{publisher}{IEEE}, \bibinfo{pages}{578--591}.
\newblock
\showISSN{2375-1207}
\href{https://doi.org/10.1109/SP40000.2020.00058}{doi:\nolinkurl{10.1109/SP40000.2020.00058}}


\bibitem[Aird et~al\mbox{.}(2025b)]%
        {Aird25a}
\bibfield{author}{\bibinfo{person}{Thomas Aird}, {and} others.} \bibinfo{year}{2025}\natexlab{b}.
\newblock \bibinfo{title}{"{{Svucrg}}" {{Extension}}, {{Version}} 1.0 for {{RV64Y}} ({{Stable}} State)}.
\newblock
\urldef\tempurl%
\url{https://riscv.github.io/riscv-cheri/#section_cheri_priv_crg_ext}
\showURL{%
\tempurl}


\bibitem[Aird et~al\mbox{.}(2025a)]%
        {Aird25}
\bibfield{author}{\bibinfo{person}{Thomas Aird}, {and} others.} \bibinfo{year}{2025}\natexlab{a}.
\newblock \bibinfo{title}{"{{Svy}}" {{Extension}}, {{Version}} 1.0 for {{RV64Y}} ({{Stable}} State)}.
\newblock
\urldef\tempurl%
\url{https://riscv.github.io/riscv-cheri/#section_priv_cheri_vmem}
\showURL{%
\tempurl}


\bibitem[Akritidis(2010)]%
        {Akritidis10}
\bibfield{author}{\bibinfo{person}{Periklis Akritidis}.} \bibinfo{year}{2010}\natexlab{}.
\newblock \showarticletitle{Cling: {{A}} Memory Allocator to Mitigate Dangling Pointers}. In \bibinfo{booktitle}{\emph{{{USENIX Security}} '10}}. \bibinfo{publisher}{{USENIX}}, \bibinfo{pages}{12}.
\newblock
\urldef\tempurl%
\url{https://www.usenix.org/conference/usenixsecurity10/cling-memory-allocator-mitigate-dangling-pointers}
\showURL{%
\tempurl}


\bibitem[Amar et~al\mbox{.}(2023a)]%
        {Amar23}
\bibfield{author}{\bibinfo{person}{Saar Amar}, {and} others.} \bibinfo{year}{2023}\natexlab{a}.
\newblock \bibinfo{booktitle}{\emph{{{CHERIoT}}: {{Rethinking}} Security for Low-Cost Embedded Systems}}.
\newblock \bibinfo{type}{Technical {{Report}}} MSR-TR-2023-6. \bibinfo{institution}{Microsoft}.
\newblock
\urldef\tempurl%
\url{https://www.microsoft.com/en-us/research/wp-content/uploads/2023/02/cheriot-63e11a4f1e629.pdf}
\showURL{%
\tempurl}


\bibitem[Amar et~al\mbox{.}(2023b)]%
        {Amar23a}
\bibfield{author}{\bibinfo{person}{Saar Amar}, {and} others.} \bibinfo{year}{2023}\natexlab{b}.
\newblock \showarticletitle{{{CHERIoT}}: {{Complete Memory Safety}} for {{Embedded Devices}}}. In \bibinfo{booktitle}{\emph{{{MICRO}} '23}}. \bibinfo{publisher}{ACM}, \bibinfo{pages}{641--653}.
\newblock
\showISBNx{979-8-4007-0329-4}
\href{https://doi.org/10.1145/3613424.3614266}{doi:\nolinkurl{10.1145/3613424.3614266}}


\bibitem[Berger and Zorn(2006)]%
        {Berger06}
\bibfield{author}{\bibinfo{person}{Emery~D. Berger} {and} \bibinfo{person}{Benjamin~G. Zorn}.} \bibinfo{year}{2006}\natexlab{}.
\newblock \showarticletitle{{{DieHard}}: Probabilistic Memory Safety for Unsafe Languages}.
\newblock \bibinfo{journal}{\emph{ACM SIGPLAN Notices}} \bibinfo{volume}{41}, \bibinfo{number}{6} (\bibinfo{date}{June} \bibinfo{year}{2006}), \bibinfo{pages}{158--168}.
\newblock
\showISSN{0362-1340}
\href{https://doi.org/10.1145/1133255.1134000}{doi:\nolinkurl{10.1145/1133255.1134000}}


\bibitem[Boland and Black(2012)]%
        {Boland12}
\bibfield{author}{\bibinfo{person}{Tim Boland} {and} \bibinfo{person}{Paul~E. Black}.} \bibinfo{year}{2012}\natexlab{}.
\newblock \showarticletitle{Juliet 1.1 {{C}}/{{C}}++ and {{Java Test Suite}}}.
\newblock \bibinfo{journal}{\emph{Computer}} \bibinfo{volume}{45}, \bibinfo{number}{10} (\bibinfo{date}{Oct.} \bibinfo{year}{2012}), \bibinfo{pages}{88--90}.
\newblock
\showISSN{0018-9162}
\href{https://doi.org/10.1109/MC.2012.345}{doi:\nolinkurl{10.1109/MC.2012.345}}


\bibitem[Bramley et~al\mbox{.}(2023)]%
        {Bramley23}
\bibfield{author}{\bibinfo{person}{Jacob Bramley}, {and} others.} \bibinfo{year}{2023}\natexlab{}.
\newblock \showarticletitle{Picking a {{CHERI Allocator}}: {{Security}} and {{Performance Considerations}}}. In \bibinfo{booktitle}{\emph{{{ISMM}} 2023}}. \bibinfo{publisher}{{ACM}}, \bibinfo{pages}{111--123}.
\newblock
\showISBNx{979-8-4007-0179-5}
\href{https://doi.org/10.1145/3591195.3595278}{doi:\nolinkurl{10.1145/3591195.3595278}}


\bibitem[Caballero et~al\mbox{.}(2012)]%
        {Caballero12}
\bibfield{author}{\bibinfo{person}{Juan Caballero}, {and} others.} \bibinfo{year}{2012}\natexlab{}.
\newblock \showarticletitle{Undangle: Early Detection of Dangling Pointers in Use-after-Free and Double-Free Vulnerabilities}. In \bibinfo{booktitle}{\emph{{{ISSTA}} '12}}. \bibinfo{publisher}{{ACM}}, \bibinfo{pages}{133--143}.
\newblock
\showISBNx{978-1-4503-1454-1}
\href{https://doi.org/10.1145/2338965.2336769}{doi:\nolinkurl{10.1145/2338965.2336769}}


\bibitem[Carlini et~al\mbox{.}(2025)]%
        {Carlini25}
\bibfield{author}{\bibinfo{person}{Paolo Carlini}, {and} others.} \bibinfo{year}{2025}\natexlab{}.
\newblock \bibinfo{title}{The Bitmap\_allocator}.
\newblock
\urldef\tempurl%
\url{https://gcc.gnu.org/onlinedocs/gcc-15.2.0/libstdc++/manual/manual/bitmap_allocator.html}
\showURL{%
\tempurl}


\bibitem[Chisnall et~al\mbox{.}(2017)]%
        {Chisnall17}
\bibfield{author}{\bibinfo{person}{David Chisnall}, {and} others.} \bibinfo{year}{2017}\natexlab{}.
\newblock \showarticletitle{{{CHERI JNI}}: {{Sinking}} the {{Java Security Model}} into the {{C}}}. In \bibinfo{booktitle}{\emph{{{ASPLOS}} '17}}. \bibinfo{publisher}{{ACM}}, \bibinfo{pages}{569--583}.
\newblock
\showISBNx{978-1-4503-4465-4}
\href{https://doi.org/10.1145/3037697.3037725}{doi:\nolinkurl{10.1145/3037697.3037725}}


\bibitem[Cho et~al\mbox{.}(2022)]%
        {Cho22}
\bibfield{author}{\bibinfo{person}{Haehyun Cho}, {and} others.} \bibinfo{year}{2022}\natexlab{}.
\newblock \showarticletitle{{{ViK}}: Practical Mitigation of Temporal Memory Safety Violations through Object {{ID}} Inspection}. In \bibinfo{booktitle}{\emph{{{ASPLOS}} '22}}. \bibinfo{publisher}{{ACM}}, \bibinfo{pages}{271--284}.
\newblock
\showISBNx{978-1-4503-9205-1}
\href{https://doi.org/10.1145/3503222.3507780}{doi:\nolinkurl{10.1145/3503222.3507780}}


\bibitem[Chow et~al\mbox{.}(2005)]%
        {Chow05}
\bibfield{author}{\bibinfo{person}{Jim Chow}, {and} others.} \bibinfo{year}{2005}\natexlab{}.
\newblock \showarticletitle{Shredding {{Your Garbage}}: {{Reducing Data Lifetime Through Secure Deallocation}}}. In \bibinfo{booktitle}{\emph{{{USENIX Security}} '15}}. \bibinfo{publisher}{{USENIX}}.
\newblock
\urldef\tempurl%
\url{https://www.usenix.org/conference/14th-usenix-security-symposium/shredding-your-garbage-reducing-data-lifetime-through}
\showURL{%
\tempurl}


\bibitem[CTSRD(2019)]%
        {cheripostgress}
\bibfield{author}{\bibinfo{person}{CTSRD}.} \bibinfo{year}{2019}\natexlab{}.
\newblock \bibinfo{title}{{{CTSRD-CHERI}}/Postgres}.
\newblock
\urldef\tempurl%
\url{https://github.com/CTSRD-CHERI/postgres}
\showURL{%
\tempurl}


\bibitem[Dang et~al\mbox{.}(2017)]%
        {Dang17}
\bibfield{author}{\bibinfo{person}{Thurston H.~Y. Dang}, \bibinfo{person}{Petros Maniatis}, {and} \bibinfo{person}{David Wagner}.} \bibinfo{year}{2017}\natexlab{}.
\newblock \showarticletitle{Oscar: A Practical Page-Permissions-Based Scheme for Thwarting Dangling Pointers}. In \bibinfo{booktitle}{\emph{{{USENIX Security}} '17}}. \bibinfo{publisher}{{USENIX}}, \bibinfo{pages}{815--832}.
\newblock
\showISBNx{978-1-931971-40-9}
\urldef\tempurl%
\url{https://www.usenix.org/conference/usenixsecurity17/technical-sessions/presentation/dang}
\showURL{%
\tempurl}


\bibitem[Dennis and Van~Horn(1966)]%
        {Dennis66}
\bibfield{author}{\bibinfo{person}{Jack~B. Dennis} {and} \bibinfo{person}{Earl~C. Van~Horn}.} \bibinfo{year}{1966}\natexlab{}.
\newblock \showarticletitle{Programming Semantics for Multiprogrammed Computations}.
\newblock \bibinfo{journal}{\emph{Commun. ACM}} \bibinfo{volume}{9}, \bibinfo{number}{3} (\bibinfo{date}{March} \bibinfo{year}{1966}), \bibinfo{pages}{143--155}.
\newblock
\showISSN{0001-0782}
\href{https://doi.org/10.1145/365230.365252}{doi:\nolinkurl{10.1145/365230.365252}}


\bibitem[Duck and Yap(2018)]%
        {Duck18}
\bibfield{author}{\bibinfo{person}{Gregory~J. Duck} {and} \bibinfo{person}{Roland H.~C. Yap}.} \bibinfo{year}{2018}\natexlab{}.
\newblock \showarticletitle{{{EffectiveSan}}: Type and Memory Error Detection Using Dynamically Typed {{C}}/{{C}}++}. In \bibinfo{booktitle}{\emph{{{PLDI}} '18}}. \bibinfo{publisher}{ACM}, \bibinfo{pages}{181--195}.
\newblock
\showISBNx{978-1-4503-5698-5}
\href{https://doi.org/10.1145/3192366.3192388}{doi:\nolinkurl{10.1145/3192366.3192388}}


\bibitem[EEMBC(2018)]%
        {EEMBC24}
\bibfield{author}{\bibinfo{person}{EEMBC}.} \bibinfo{year}{2018}\natexlab{}.
\newblock \bibinfo{title}{Eembc/Coremark}.
\newblock \bibinfo{howpublished}{Embedded Microprocessor Benchmark Consortium}.
\newblock
\urldef\tempurl%
\url{https://github.com/eembc/coremark}
\showURL{%
\tempurl}


\bibitem[ElAtali et~al\mbox{.}(2025)]%
        {ElAtali25}
\bibfield{author}{\bibinfo{person}{Hossam ElAtali}, {and} others.} \bibinfo{year}{2025}\natexlab{}.
\newblock \showarticletitle{{{BLACKOUT}}: {{Data-Oblivious Computation}} with {{Blinded Capabilities}}}. In \bibinfo{booktitle}{\emph{{{CCS}} '25}}. \bibinfo{publisher}{{ACM}}, \bibinfo{pages}{2039--2053}.
\newblock
\showISBNx{979-8-4007-1525-9}
\href{https://doi.org/10.1145/3719027.3765169}{doi:\nolinkurl{10.1145/3719027.3765169}}


\bibitem[Erd{\H o}s et~al\mbox{.}(2022)]%
        {Erdos22}
\bibfield{author}{\bibinfo{person}{M{\'a}rton Erd{\H o}s}, \bibinfo{person}{Sam Ainsworth}, {and} \bibinfo{person}{Timothy~M. Jones}.} \bibinfo{year}{2022}\natexlab{}.
\newblock \showarticletitle{{{MineSweeper}}: A ``Clean Sweep'' for Drop-in Use-after-Free Prevention}. In \bibinfo{booktitle}{\emph{{{ASPLOS}} '22}}. \bibinfo{publisher}{{ACM}}, \bibinfo{pages}{212--225}.
\newblock
\showISBNx{978-1-4503-9205-1}
\href{https://doi.org/10.1145/3503222.3507712}{doi:\nolinkurl{10.1145/3503222.3507712}}


\bibitem[Farkhani et~al\mbox{.}(2021)]%
        {Farkhani21}
\bibfield{author}{\bibinfo{person}{Reza~Mirzazade Farkhani}, \bibinfo{person}{Mansour Ahmadi}, {and} \bibinfo{person}{Long Lu}.} \bibinfo{year}{2021}\natexlab{}.
\newblock \showarticletitle{{{PTAuth}}: {{Temporal Memory Safety}} via {{Robust Points-to Authentication}}}. In \bibinfo{booktitle}{\emph{{{USENIX Security}} '21}}. \bibinfo{publisher}{{USENIX}}, \bibinfo{pages}{1037--1054}.
\newblock
\showISBNx{978-1-939133-24-3}
\urldef\tempurl%
\url{https://www.usenix.org/conference/usenixsecurity21/presentation/mirzazade}
\showURL{%
\tempurl}


\bibitem[Filardo(2024)]%
        {Filardo24a}
\bibfield{author}{\bibinfo{person}{Nathaniel Filardo}.} \bibinfo{year}{2024}\natexlab{}.
\newblock \bibinfo{title}{For Discussion: {{MSR}}'s {{CHERI}}+{{MTE}} Composition}.
\newblock
\urldef\tempurl%
\url{https://github.com/riscv/riscv-cheri/issues/340}
\showURL{%
\tempurl}


\bibitem[Filardo et~al\mbox{.}(2024)]%
        {Filardo24}
\bibfield{author}{\bibinfo{person}{Nathaniel~Wesley Filardo}, {and} others.} \bibinfo{year}{2024}\natexlab{}.
\newblock \showarticletitle{Cornucopia {{Reloaded}}: {{Load Barriers}} for {{CHERI Heap Temporal Safety}}}. In \bibinfo{booktitle}{\emph{{{ASPLOS}} '24}}. \bibinfo{publisher}{ACM}, \bibinfo{pages}{251--268}.
\newblock
\showISBNx{979-8-4007-0385-0}
\href{https://doi.org/10.1145/3620665.3640416}{doi:\nolinkurl{10.1145/3620665.3640416}}


\bibitem[Fuchs et~al\mbox{.}(2024)]%
        {Fuchs24}
\bibfield{author}{\bibinfo{person}{Franz~A. Fuchs}, {and} others.} \bibinfo{year}{2024}\natexlab{}.
\newblock \showarticletitle{Safe {{Speculation}} for {{CHERI}}}. In \bibinfo{booktitle}{\emph{{{ICCD}} '24}}. \bibinfo{publisher}{IEEE}, \bibinfo{pages}{364--372}.
\newblock
\showISBNx{979-8-3503-8040-8}
\showISSN{2576-6996}
\href{https://doi.org/10.1109/ICCD63220.2024.00063}{doi:\nolinkurl{10.1109/ICCD63220.2024.00063}}


\bibitem[Fuchs et~al\mbox{.}(2023)]%
        {Fuchs23}
\bibfield{author}{\bibinfo{person}{Franz~A Fuchs}, {and} others.} \bibinfo{year}{2023}\natexlab{}.
\newblock \showarticletitle{Architectural {{Contracts}} for {{Safe Speculation}}}. In \bibinfo{booktitle}{\emph{{{ICCD}} '23}}. \bibinfo{publisher}{{IEEE}}, \bibinfo{pages}{578--586}.
\newblock
\href{https://doi.org/10.1109/ICCD58817.2023.00093}{doi:\nolinkurl{10.1109/ICCD58817.2023.00093}}


\bibitem[Georges et~al\mbox{.}(2021)]%
        {Georges21}
\bibfield{author}{\bibinfo{person}{A{\"i}na~Linn Georges}, {and} others.} \bibinfo{year}{2021}\natexlab{}.
\newblock \showarticletitle{Efficient and Provable Local Capability Revocation Using Uninitialized Capabilities}. In \bibinfo{booktitle}{\emph{{{POPL}} '21}}. \bibinfo{pages}{6:1--6:30}.
\newblock
\href{https://doi.org/10.1145/3434287}{doi:\nolinkurl{10.1145/3434287}}


\bibitem[Grisenthwaite(2022)]%
        {Grisenthwaite22}
\bibfield{author}{\bibinfo{person}{Richard Grisenthwaite}.} \bibinfo{year}{2022}\natexlab{}.
\newblock \showarticletitle{Arm {{Morello Evaluation Platform}} -{{Validating CHERI-based Security}} in a {{High-performance System}}}. In \bibinfo{booktitle}{\emph{{{HCS}} '22}}. \bibinfo{publisher}{IEEE}, \bibinfo{pages}{1--22}.
\newblock
\showISSN{2573-2048}
\href{https://doi.org/10.1109/HCS55958.2022.9895591}{doi:\nolinkurl{10.1109/HCS55958.2022.9895591}}


\bibitem[G{\"u}lmez et~al\mbox{.}(2025)]%
        {Gulmez25a}
\bibfield{author}{\bibinfo{person}{Merve G{\"u}lmez}, {and} others.} \bibinfo{year}{2025}\natexlab{}.
\newblock \showarticletitle{Mon {{CH\'ERI}}: {{Mitigating Uninitialized Memory Access}} with {{Conditional Capabilities}}}. In \bibinfo{booktitle}{\emph{S\&{{P}} '15}}. \bibinfo{publisher}{{IEEE}}, \bibinfo{pages}{829--847}.
\newblock
\showISBNx{979-8-3315-2236-0}
\showISSN{2375-1207}
\href{https://doi.org/10.1109/SP61157.2025.00133}{doi:\nolinkurl{10.1109/SP61157.2025.00133}}


\bibitem[G{\"u}lmez et~al\mbox{.}(2026)]%
        {Gulmez26}
\bibfield{author}{\bibinfo{person}{Merve G{\"u}lmez}, {and} others.} \bibinfo{year}{2026}\natexlab{}.
\newblock \bibinfo{title}{{{PICASSO}}: {{Scaling CHERI Use-After-Free Protection}} to {{Millions}} of {{Allocations}} Using {{Colored Capabilities}}}.
\newblock
\showeprint[arxiv]{2602.09131}~[cs.CR]
\href{https://doi.org/10.48550/arXiv.2602.09131}{doi:\nolinkurl{10.48550/arXiv.2602.09131}}


\bibitem[Gülmez et~al\mbox{.}(2026)]%
        {Gulmez26b}
\bibfield{author}{\bibinfo{person}{Merve Gülmez} {et~al\mbox{.}}} \bibinfo{year}{2026}\natexlab{}.
\newblock \bibinfo{booktitle}{\emph{{PICASSO} Artifact}}.
\newblock
\href{https://doi.org/10.5281/zenodo.20353117}{doi:\nolinkurl{10.5281/zenodo.20353117}}


\bibitem[Joly et~al\mbox{.}(2020)]%
        {Joly20}
\bibfield{author}{\bibinfo{person}{Nicolas Joly}, \bibinfo{person}{Saif ElSherei}, {and} \bibinfo{person}{Saar Amar}.} \bibinfo{year}{2020}\natexlab{}.
\newblock \bibinfo{booktitle}{\emph{Security {{Analysis}} of {{CHERI ISA}}}}.
\newblock \bibinfo{type}{{T}echnical {R}eport}. \bibinfo{institution}{Microsoft Security Response Center}. \bibinfo{pages}{42} pages.
\newblock
\urldef\tempurl%
\url{https://msrc.microsoft.com/blog/2020/10/security-analysis-of-cheri-isa/}
\showURL{%
\tempurl}


\bibitem[Lee et~al\mbox{.}(2015)]%
        {Lee15}
\bibfield{author}{\bibinfo{person}{Byoungyoung Lee}, {and} others.} \bibinfo{year}{2015}\natexlab{}.
\newblock \showarticletitle{Preventing {{Use-after-free}} with {{Dangling Pointers Nullification}}}. In \bibinfo{booktitle}{\emph{{{NDSS}} '15}}. \bibinfo{publisher}{Internet Society}.
\newblock
\showISBNx{978-1-891562-38-9}
\href{https://doi.org/10.14722/ndss.2015.23238}{doi:\nolinkurl{10.14722/ndss.2015.23238}}


\bibitem[Levy(1984)]%
        {Levy84}
\bibfield{author}{\bibinfo{person}{Henry~M. Levy}.} \bibinfo{year}{1984}\natexlab{}.
\newblock \bibinfo{booktitle}{\emph{Capability-{{Based Computer Systems}}}}.
\newblock \bibinfo{publisher}{Butterworth-Heinemann}.
\newblock
\showISBNx{978-0-932376-22-0}
\urldef\tempurl%
\url{https://homes.cs.washington.edu/~levy/capabook/}
\showURL{%
\tempurl}


\bibitem[Li et~al\mbox{.}(2022)]%
        {Li22}
\bibfield{author}{\bibinfo{person}{Yuan Li}, {and} others.} \bibinfo{year}{2022}\natexlab{}.
\newblock \showarticletitle{{{PACMem}}: {{Enforcing Spatial}} and {{Temporal Memory Safety}} via {{ARM Pointer Authentication}}}. In \bibinfo{booktitle}{\emph{{{CCS}} '22}}. \bibinfo{publisher}{{ACM}}, \bibinfo{pages}{1901--1915}.
\newblock
\showISBNx{978-1-4503-9450-5}
\href{https://doi.org/10.1145/3548606.3560598}{doi:\nolinkurl{10.1145/3548606.3560598}}


\bibitem[Liu et~al\mbox{.}(2019)]%
        {Liu19}
\bibfield{author}{\bibinfo{person}{Beichen Liu}, \bibinfo{person}{Pierre Olivier}, {and} \bibinfo{person}{Binoy Ravindran}.} \bibinfo{year}{2019}\natexlab{}.
\newblock \showarticletitle{{{SlimGuard}}: {{A Secure}} and {{Memory-Efficient Heap Allocator}}}. In \bibinfo{booktitle}{\emph{Middleware '19}}. \bibinfo{publisher}{{ACM}}, \bibinfo{pages}{1--13}.
\newblock
\showISBNx{978-1-4503-7009-7}
\href{https://doi.org/10.1145/3361525.3361532}{doi:\nolinkurl{10.1145/3361525.3361532}}


\bibitem[Liu et~al\mbox{.}(2018)]%
        {Liu18}
\bibfield{author}{\bibinfo{person}{Daiping Liu}, \bibinfo{person}{Mingwei Zhang}, {and} \bibinfo{person}{Haining Wang}.} \bibinfo{year}{2018}\natexlab{}.
\newblock \showarticletitle{A {{Robust}} and {{Efficient Defense}} against {{Use-after-Free Exploits}} via {{Concurrent Pointer Sweeping}}}. In \bibinfo{booktitle}{\emph{{{CCS}} '18}}. \bibinfo{publisher}{{ACM}}, \bibinfo{pages}{1635--1648}.
\newblock
\showISBNx{978-1-4503-5693-0}
\href{https://doi.org/10.1145/3243734.3243826}{doi:\nolinkurl{10.1145/3243734.3243826}}


\bibitem[Milburn et~al\mbox{.}(2017)]%
        {Milburn17}
\bibfield{author}{\bibinfo{person}{Alyssa Milburn}, \bibinfo{person}{Herbert Bos}, {and} \bibinfo{person}{Cristiano Giuffrida}.} \bibinfo{year}{2017}\natexlab{}.
\newblock \showarticletitle{{{SafeInit}}: {{Comprehensive}} and {{Practical Mitigation}} of {{Uninitialized Read Vulnerabilities}}}. \bibinfo{publisher}{Internet Society}.
\newblock
\showISBNx{978-1-891562-46-4}
\href{https://doi.org/10.14722/ndss.2017.23183}{doi:\nolinkurl{10.14722/ndss.2017.23183}}


\bibitem[MITRE(2024)]%
        {MITRE24}
\bibfield{author}{\bibinfo{person}{MITRE}.} \bibinfo{year}{2024}\natexlab{}.
\newblock \bibinfo{title}{{{CWE-562}}: {{Return}} of {{Stack Variable Address}} (4.17)}.
\newblock
\urldef\tempurl%
\url{https://cwe.mitre.org/data/definitions/562.html}
\showURL{%
\tempurl}


\bibitem[Nagarakatte et~al\mbox{.}(2010)]%
        {Nagarakatte10}
\bibfield{author}{\bibinfo{person}{Santosh Nagarakatte}, {and} others.} \bibinfo{year}{2010}\natexlab{}.
\newblock \showarticletitle{{{CETS}}: Compiler Enforced Temporal Safety for {{C}}}.
\newblock \bibinfo{journal}{\emph{SIGPLAN Not.}} \bibinfo{volume}{45}, \bibinfo{number}{8} (\bibinfo{date}{June} \bibinfo{year}{2010}), \bibinfo{pages}{31--40}.
\newblock
\showISSN{0362-1340}
\href{https://doi.org/10.1145/1837855.1806657}{doi:\nolinkurl{10.1145/1837855.1806657}}


\bibitem[Nguyen et~al\mbox{.}(2020)]%
        {Nguyen20}
\bibfield{author}{\bibinfo{person}{Manh-Dung Nguyen}, {and} others.} \bibinfo{year}{2020}\natexlab{}.
\newblock \showarticletitle{Binary-Level {{Directed Fuzzing}} for {{Use-After-Free Vulnerabilities}}}. \bibinfo{pages}{47--62}.
\newblock
\showISBNx{978-1-939133-18-2}
\urldef\tempurl%
\url{https://www.usenix.org/conference/raid2020/presentation/nguyen}
\showURL{%
\tempurl}


\bibitem[Novark and Berger(2010)]%
        {Novark10}
\bibfield{author}{\bibinfo{person}{Gene Novark} {and} \bibinfo{person}{Emery~D. Berger}.} \bibinfo{year}{2010}\natexlab{}.
\newblock \showarticletitle{{{DieHarder}}: Securing the Heap}. In \bibinfo{booktitle}{\emph{{{CCS}} '10}}. \bibinfo{publisher}{ACM}, \bibinfo{pages}{573--584}.
\newblock
\showISBNx{978-1-4503-0245-6}
\href{https://doi.org/10.1145/1866307.1866371}{doi:\nolinkurl{10.1145/1866307.1866371}}


\bibitem[Park and Moon(2024)]%
        {Park24}
\bibfield{author}{\bibinfo{person}{Chanyoung Park} {and} \bibinfo{person}{Hyungon Moon}.} \bibinfo{year}{2024}\natexlab{}.
\newblock \showarticletitle{Efficient {{Use-After-Free Prevention}} with {{Opportunistic Page-Level Sweeping}}}. \bibinfo{publisher}{Internet Society}.
\newblock
\showISBNx{978-1-891562-93-8}
\href{https://doi.org/10.14722/ndss.2024.24804}{doi:\nolinkurl{10.14722/ndss.2024.24804}}


\bibitem[Qualcomm(2017)]%
        {QualcommProductSecurity17}
\bibfield{author}{\bibinfo{person}{Qualcomm}.} \bibinfo{year}{2017}\natexlab{}.
\newblock \bibinfo{booktitle}{\emph{Pointer {{Authentication}} on {{ARMv8}}.3: {{Design}} and {{Analysis}} of the {{New Software Security Instructions}}}}.
\newblock \bibinfo{type}{Whitepaper}. \bibinfo{institution}{Qualcomm}. \bibinfo{pages}{12} pages.
\newblock
\urldef\tempurl%
\url{https://www.qualcomm.com/content/dam/qcomm-martech/dm-assets/documents/pointer-auth-v7.pdf}
\showURL{%
\tempurl}


\bibitem[Rebert and Kern(2024)]%
        {Rebert24}
\bibfield{author}{\bibinfo{person}{Alex Rebert} {and} \bibinfo{person}{Christoph Kern}.} \bibinfo{year}{2024}\natexlab{}.
\newblock \bibinfo{booktitle}{\emph{Secure by {{Design}}: {{Google}}'s {{Perspective}} on {{Memory Safety}}}}.
\newblock \bibinfo{type}{Technical {{Report}}}. \bibinfo{institution}{Google}. \bibinfo{pages}{13} pages.
\newblock
\urldef\tempurl%
\url{https://research.google/pubs/secure-by-design-googles-perspective-on-memory-safety/}
\showURL{%
\tempurl}


\bibitem[Redell(1974)]%
        {Redell74}
\bibfield{author}{\bibinfo{person}{David~D. Redell}.} \bibinfo{year}{1974}\natexlab{}.
\newblock \bibinfo{booktitle}{\emph{Naming and {{Protection}} in {{Extendible Operating Systems}}}}.
\newblock \bibinfo{type}{Technical {{Report}}} MAC TR-140. \bibinfo{institution}{Massachussets Institute of Technology}. \bibinfo{pages}{169} pages.
\newblock
\urldef\tempurl%
\url{https://apps.dtic.mil/sti/tr/pdf/ADA001721.pdf}
\showURL{%
\tempurl}


\bibitem[Rugg(2023)]%
        {Rugg23}
\bibfield{author}{\bibinfo{person}{Peter~David Rugg}.} \bibinfo{year}{2023}\natexlab{}.
\newblock \bibinfo{booktitle}{\emph{Efficient Spatial and Temporal Safety for Microcontrollers and Application-Class Processors}}.
\newblock \bibinfo{type}{Technical {{Report}}} UCAM-CL-TR-984. \bibinfo{institution}{University of Cambridge}. \bibinfo{pages}{198} pages.
\newblock
\showISSN{1476-2986}
\urldef\tempurl%
\url{https://www.cl.cam.ac.uk/techreports/UCAM-CL-TR-984.pdf}
\showURL{%
\tempurl}


\bibitem[Saltzer(1974)]%
        {Saltzer74}
\bibfield{author}{\bibinfo{person}{Jerome~H. Saltzer}.} \bibinfo{year}{1974}\natexlab{}.
\newblock \showarticletitle{Protection and the {{Control}} of {{Information Sharing}} in {{Multics}}}.
\newblock \bibinfo{journal}{\emph{Commun. ACM}} \bibinfo{volume}{17}, \bibinfo{number}{7} (\bibinfo{date}{July} \bibinfo{year}{1974}), \bibinfo{pages}{388--402}.
\newblock
\showISSN{0001-0782}
\href{https://doi.org/10.1145/361011.361067}{doi:\nolinkurl{10.1145/361011.361067}}


\bibitem[Shanbhogue(2024)]%
        {Shanbhogue24}
\bibfield{author}{\bibinfo{person}{Ved Shanbhogue}.} \bibinfo{year}{2024}\natexlab{}.
\newblock \bibinfo{title}{"{{B}}" {{Extension}} for {{Bit Manipulation}}, {{Version}} 1.0.0 :: {{RISC-V Ratified Specifications Library}}}.
\newblock
\urldef\tempurl%
\url{https://docs.riscv.org/reference/isa/unpriv/b-st-ext.html}
\showURL{%
\tempurl}


\bibitem[Silvestro et~al\mbox{.}(2017)]%
        {Silvestro17}
\bibfield{author}{\bibinfo{person}{Sam Silvestro}, {and} others.} \bibinfo{year}{2017}\natexlab{}.
\newblock \showarticletitle{{{FreeGuard}}: {{A Faster Secure Heap Allocator}}}. In \bibinfo{booktitle}{\emph{{{CCS}} '17}}. \bibinfo{publisher}{{ACM}}, \bibinfo{pages}{2389--2403}.
\newblock
\showISBNx{978-1-4503-4946-8}
\href{https://doi.org/10.1145/3133956.3133957}{doi:\nolinkurl{10.1145/3133956.3133957}}


\bibitem[Silvestro et~al\mbox{.}(2018)]%
        {Silvestro18}
\bibfield{author}{\bibinfo{person}{Sam Silvestro}, {and} others.} \bibinfo{year}{2018}\natexlab{}.
\newblock \showarticletitle{Guarder: A Tunable Secure Allocator}. In \bibinfo{booktitle}{\emph{{{USENIX Security}} '18}}. \bibinfo{publisher}{{USENIX}}, \bibinfo{pages}{117--133}.
\newblock
\showISBNx{978-1-931971-46-1}
\urldef\tempurl%
\url{https://www.usenix.org/conference/usenixsecurity18/presentation/silvestro}
\showURL{%
\tempurl}


\bibitem[Team(2025a)]%
        {GCCTeam25}
\bibfield{author}{\bibinfo{person}{GCC Team}.} \bibinfo{year}{2025}\natexlab{a}.
\newblock \bibinfo{title}{Options to {{Request}} or {{Suppress Warnings}}: -{{Wno-return-local-addr}}}.
\newblock
\urldef\tempurl%
\url{https://gcc.gnu.org/onlinedocs/gcc-15.2.0/gcc/Warning-Options.html#index-Wno-return-local-addr}
\showURL{%
\tempurl}


\bibitem[Team(2025b)]%
        {LLVMTeam25a}
\bibfield{author}{\bibinfo{person}{LLVM Team}.} \bibinfo{year}{2025}\natexlab{b}.
\newblock \bibinfo{title}{Diagnostic Flags in {{Clang}}}.
\newblock
\urldef\tempurl%
\url{https://releases.llvm.org/20.1.0/tools/clang/docs/DiagnosticsReference.html#wreturn-stack-address}
\showURL{%
\tempurl}


\bibitem[Team(2025c)]%
        {LLVMTeam25}
\bibfield{author}{\bibinfo{person}{LLVM Team}.} \bibinfo{year}{2025}\natexlab{c}.
\newblock \bibinfo{title}{Scudo {{Hardened Allocator}}}.
\newblock
\urldef\tempurl%
\url{https://releases.llvm.org/20.1.0/docs/ScudoHardenedAllocator.html}
\showURL{%
\tempurl}


\bibitem[Tiller et~al\mbox{.}(2025)]%
        {Tiller25}
\bibfield{author}{\bibinfo{person}{Craig Tiller}, {and} others.} \bibinfo{year}{2025}\natexlab{}.
\newblock \bibinfo{title}{{{gRPC}}}.
\newblock
\urldef\tempurl%
\url{https://github.com/grpc/grpc/tree/v1.54.2}
\showURL{%
\tempurl}


\bibitem[van~der Kouwe et~al\mbox{.}(2017)]%
        {Kouwe17}
\bibfield{author}{\bibinfo{person}{Erik van~der Kouwe}, \bibinfo{person}{Vinod Nigade}, {and} \bibinfo{person}{Cristiano Giuffrida}.} \bibinfo{year}{2017}\natexlab{}.
\newblock \showarticletitle{{{DangSan}}: {{Scalable Use-after-free Detection}}}. In \bibinfo{booktitle}{\emph{{{EuroSys}} '17}}. \bibinfo{publisher}{{ACM}}, \bibinfo{pages}{405--419}.
\newblock
\showISBNx{978-1-4503-4938-3}
\href{https://doi.org/10.1145/3064176.3064211}{doi:\nolinkurl{10.1145/3064176.3064211}}


\bibitem[Watson et~al\mbox{.}(2023a)]%
        {Watson23b}
\bibfield{author}{\bibinfo{person}{Robert N~M Watson}, {and} others.} \bibinfo{year}{2023}\natexlab{a}.
\newblock \bibinfo{booktitle}{\emph{Arm {{Morello Programme}}: {{Architectural}} Security Goals and Known Limitations}}.
\newblock \bibinfo{type}{Technical {{Report}}} UCAM-CL-TR-982. \bibinfo{institution}{University of Cambridge}. \bibinfo{pages}{8} pages.
\newblock
\showISSN{1476-2986}
\urldef\tempurl%
\url{https://www.cl.cam.ac.uk/techreports/UCAM-CL-TR-982.pdf}
\showURL{%
\tempurl}


\bibitem[Watson et~al\mbox{.}(2023b)]%
        {Watson23a}
\bibfield{author}{\bibinfo{person}{Robert N.~M. Watson}, {and} others.} \bibinfo{year}{2023}\natexlab{b}.
\newblock \bibinfo{booktitle}{\emph{Capability {{Hardware Enhanced RISC Instructions}}: {{CHERI Instruction-Set Architecture}} ({{Version}} 9)}}.
\newblock \bibinfo{type}{Technical {{Report}}} UCAM-CL-TR-987. \bibinfo{institution}{University of Cambridge}. \bibinfo{pages}{523 pages} pages.
\newblock
\showISSN{1476-2986}
\href{https://doi.org/10.48456/TR-987}{doi:\nolinkurl{10.48456/TR-987}}


\bibitem[Watson et~al\mbox{.}(2019)]%
        {Watson19}
\bibfield{author}{\bibinfo{person}{Robert N~M Watson}, {and} others.} \bibinfo{year}{2019}\natexlab{}.
\newblock \bibinfo{booktitle}{\emph{Capability {{Hardware Enhanced RISC Instructions}}: {{CHERI Instruction-Set Architecture}} ({{Version}} 7)}}.
\newblock \bibinfo{type}{Technical {{Report}}} UCAM-CL-TR-927. \bibinfo{institution}{University of Cambridge}. \bibinfo{pages}{496 pages} pages.
\newblock
\showISSN{1476-2986}
\href{https://doi.org/10.48456/TR-927}{doi:\nolinkurl{10.48456/TR-927}}


\bibitem[Wesley~Filardo et~al\mbox{.}(2020)]%
        {WesleyFilardo20}
\bibfield{author}{\bibinfo{person}{Nathaniel Wesley~Filardo}, {and} others.} \bibinfo{year}{2020}\natexlab{}.
\newblock \showarticletitle{Cornucopia: {{Temporal Safety}} for {{CHERI Heaps}}}. In \bibinfo{booktitle}{\emph{S\&{{P}} '20}}. \bibinfo{publisher}{IEEE}, \bibinfo{pages}{608--625}.
\newblock
\showISSN{2375-1207}
\href{https://doi.org/10.1109/SP40000.2020.00098}{doi:\nolinkurl{10.1109/SP40000.2020.00098}}


\bibitem[Wickman et~al\mbox{.}(2021)]%
        {Wickman21}
\bibfield{author}{\bibinfo{person}{Brian Wickman}, {and} others.} \bibinfo{year}{2021}\natexlab{}.
\newblock \showarticletitle{Preventing {{Use-After-Free Attacks}} with {{Fast Forward Allocation}}}. In \bibinfo{booktitle}{\emph{{{USENIX Security Symposium}} '21}}. \bibinfo{publisher}{{USENIX}}, \bibinfo{pages}{2453--2470}.
\newblock
\urldef\tempurl%
\url{https://www.usenix.org/conference/usenixsecurity21/presentation/wickman}
\showURL{%
\tempurl}


\bibitem[Wilcox(2025)]%
        {Wilcox25}
\bibfield{author}{\bibinfo{person}{Matthew Wilcox}.} \bibinfo{year}{2025}\natexlab{}.
\newblock \bibinfo{title}{{{ID Allocation}}}.
\newblock
\urldef\tempurl%
\url{https://www.kernel.org/doc/html/v6.18/core-api/idr.html#id-allocation}
\showURL{%
\tempurl}


\bibitem[Woodruff et~al\mbox{.}(2019)]%
        {Woodruff19}
\bibfield{author}{\bibinfo{person}{Jonathan Woodruff}, {and} others.} \bibinfo{year}{2019}\natexlab{}.
\newblock \showarticletitle{{{CHERI Concentrate}}: {{Practical Compressed Capabilities}}}.
\newblock \bibinfo{journal}{\emph{IEEE Trans. Comput.}} \bibinfo{volume}{68}, \bibinfo{number}{10} (\bibinfo{date}{Oct.} \bibinfo{year}{2019}), \bibinfo{pages}{1455--1469}.
\newblock
\showISSN{1557-9956}
\href{https://doi.org/10.1109/TC.2019.2914037}{doi:\nolinkurl{10.1109/TC.2019.2914037}}


\bibitem[Younan(2015)]%
        {Younan15}
\bibfield{author}{\bibinfo{person}{Yves Younan}.} \bibinfo{year}{2015}\natexlab{}.
\newblock \showarticletitle{{{FreeSentry}}: {{Protecting Against Use-After-Free Vulnerabilities Due}} to {{Dangling Pointers}}}. In \bibinfo{booktitle}{\emph{{{NDSS}} '15}}. \bibinfo{publisher}{Internet Society}.
\newblock
\showISBNx{978-1-891562-38-9}
\href{https://doi.org/10.14722/ndss.2015.23190}{doi:\nolinkurl{10.14722/ndss.2015.23190}}


\bibitem[Yu et~al\mbox{.}(2019)]%
        {Yu19a}
\bibfield{author}{\bibinfo{person}{Jiyong Yu}, {and} others.} \bibinfo{year}{2019}\natexlab{}.
\newblock \showarticletitle{Speculative {{Taint Tracking}} ({{STT}}): {{A Comprehensive Protection}} for {{Speculatively Accessed Data}}}. In \bibinfo{booktitle}{\emph{{{MICRO-52}}}}. \bibinfo{publisher}{{ACM}}, \bibinfo{pages}{954--968}.
\newblock
\showISBNx{978-1-4503-6938-1}
\href{https://doi.org/10.1145/3352460.3358274}{doi:\nolinkurl{10.1145/3352460.3358274}}


\bibitem[Zabrocki et~al\mbox{.}(2024)]%
        {Zabrocki24}
\bibfield{author}{\bibinfo{person}{Adam Zabrocki}, {and} others.} \bibinfo{year}{2024}\natexlab{}.
\newblock \bibinfo{title}{{{RISC-V Pointer Masking}}, {{Version}} 1.0 ({{Ratified}})}.
\newblock
\urldef\tempurl%
\url{https://raw.githubusercontent.com/riscv/riscv-j-extension/master/zjpm-spec.pdf}
\showURL{%
\tempurl}


\end{thebibliography}
\appendix
\section{QEMU Implementation}\label{appendix:qemu}
We implemented a functional model of PICASSO in QEMU-system-CHERI128 to enable kernel development and functional validation prior to hardware development. The QEMU implementation mirrors the hardware ISA extensions, including the \gls{pvtr} and the   \texttt{ccsettype} instruction, as well as the compressed capability format modifications, the expanded 21-bit object type field identical to the hardware implementation. The primary architectural difference lies in the \gls{pvt} access model. Whereas Toooba has its own \gls{tlb} and buffer, QEMU performs \gls{pvt} lookups equentially during capability validation. When a load or store uses a  \ccpids, the emulator computes the \gls{pvb} address and reads the corresponding \gls{pvt} word from emulated memory before the instruction completes. To optimize the common case, QEMU first probes its software TLB for a cached translation. If found, the \gls{pvt} word is read directly from host memory, avoiding full address translation. On a TLB miss, the emulator falls back to complete guest address translation with proper exception handling.

\section{Optimized \gls{unr} allocator}\label{appendix:unr}
As described in section~\ref{cc:pidmanagment}, we utilize the \gls{unr} from FreeBSD to manage both claimed and available \ccpids. This approach enables us to store the state of all \ccpids in a compressed form, implemented as a linked list of \texttt{unr} objects. These objects can represent either a \texttt{run} or a \texttt{bitmap}. A \texttt{run} is a series of claimed or available numbers represented by a type (claimed or available) and a length. For example, 50 consecutive allocated numbers can be represented in an efficient way using: \{state = ``claimed'', len = 50\}. To save memory, multiple small \texttt{runs} can be combined into one \texttt{bitmap} representation of a fixed size (in our case 512 bits), for example: \{state = ``claimed'', len = 3\} + \{state = ``available'', len = 2\} + \{state = ``claimed'', len = 3\} = ``11100111''. These bitmaps are formed where they replace at least three \texttt{runs}, because this is the point where the bitmap takes up less memory then the runs it replaced.

The approach of leveraging runs of claimed and available \ccpids offers two advantages. Firstly, it optimizes memory usage by representing consecutive claimed or available IDs as a single run rather than tracking each ID individually. This compression is effective because, when claiming the first available \ccpids sequentially, long runs of claimed IDs naturally form. Secondly, finding the first available number is efficient because the sequence of claimed numbers will always start with a consecutive run from the beginning. This ensures that the first unclaimed number immediately follows this claimed run.

However, releasing numbers incurs significant overhead, which becomes problematic when releasing more than 1 million \ccpids simultaneously. This overhead arises from the three steps involved in releasing an ID. First, the system iterates over the doubly linked list to find the \gls{unr} objects that contain the to be released ID. Second, it restructures the data to indicate that the number is now available by resizing the current and/or neighboring \gls{unr} objects. Finally, it optimizes memory usage by searching for any new bitmaps that can be formed. 

To optimize the batch‐releasing process, we eliminate the overhead of the first and last steps by releasing IDs in sorted order and scanning the allocator exactly once. As we traverse the allocator, we free each number on the fly, avoiding a separate search for every release. Simultaneously, we look for opportunities to form a ``first good bitmap'' rather than waiting to compute the absolute optimal bitmaps. Whenever the current layout yields any memory savings by switching from runs to a bitmap, we convert immediately, even if it’s not the maximal compression. This approach reduces memory usage with minimal additional overhead.

\section{Complementary Evaluation}\label{appendix:eval}

\begin{figure*}[t]
    \centering
    \includegraphics[scale=0.420]{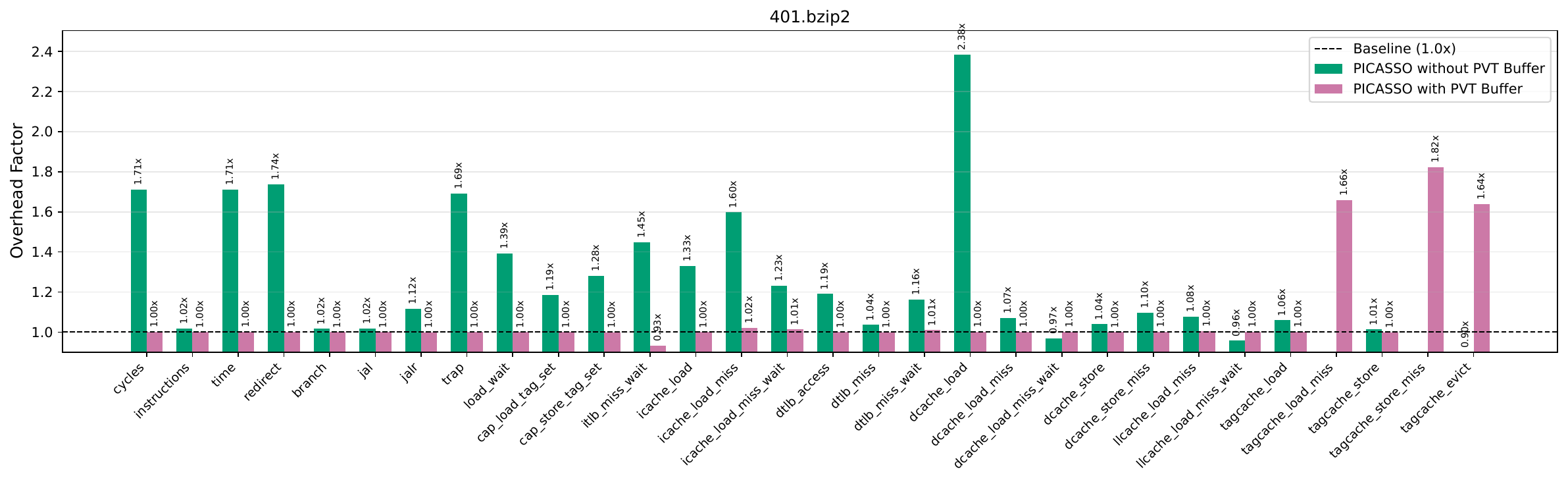}
    \caption{SPEC CPU 2006 bzip performance overhead for \PICASSO without \gls{pvt} buffer and \PICASSO with \gls{pvt} buffer.}\label{fig:cc-bzip}
\end{figure*}

\paragraph{\PICASSO \gls{pvt} Buffer Optimization.} 
We evaluate the performance impact of the \gls{pvt} buffer using the SPEC CPU2006 \texttt{bzip2} benchmark. While bzip2 includes three distinct inputs for the \emph{train} workload, we utilize a single representative input for this evaluation. The workload performs 29 allocations, executing compression and decompression operations on each. \Cref{fig:cc-bzip} illustrates the normalized hardware performance counter overheads, comparing \PICASSO without the \gls{pvt} buffer against the optimized design, both normalized to the baseline. Without the \gls{pvt} buffer, \PICASSO incurs a $1.7\times$ cycle overhead, primarily due to a $2.39\times$ increase in data cache loads. This is due to the frequent memory accesses required to verify allocation-provenance validity (\glspl{pvb}). In contrast, with the \gls{pvt} buffer enabled, the overhead drops to $1.002\times$. The buffer effectively eliminates the performance penalty of \gls{pvt} lookups by caching recently accessed \glspl{pvb}.

\paragraph{\PICASSO Comparison with Cornucopia-\gls{rof} on SQLite.}: To address the \gls{uaf}, Cornucopia can be configured in a \acrlong{rof} mode, which triggers a global revocation sweep upon every \texttt{free} operation. From a security standpoint, \PICASSO and Cornucopia-\gls{rof} provide equivalent protection against \gls{uaf}/\gls{uar}. However, a key mechanical distinction remains: \PICASSO does not perform a full memory scan upon deallocation; instead, it retracts the capability, deferring the computationally expensive global sweep until the \ccpids space is exhausted. We evaluate Cornucopia-\gls{rof} using the SQLite \texttt{speedtest1} benchmark under the same configuration described in \Cref{sec:performance-eval}. \Cref{fig:sqllitesystemmetric} shows the cycles, instruction and memory overhead for Cornucopia, Cornucopia-\gls{rof} and \PICASSO. Our results demonstrate that while Cornucopia-RoF reduces memory overhead from $1.99\times$ to $1.1\times$ (by eliminating the quarantine buffer) compare to Cornucopia, this comes at a prohibitive performance cost of $6.10\times$ overhead relative to the baseline. 
In contrast, \PICASSO achieves equivalent security guarantees without the penalty of frequent memory scans. 
\ifnotarxiv
\captionof{table}{PICASSO overhead on MiBench}\label{tab:mibench}
\centering
\resizebox{.4\textwidth}{!}{
\begin{tabular}{lrrr}
\toprule
Benchmark & Baseline & PICASSO & Overhead (\%) \\
\midrule
adpcm\_decode & 3325012 & 3380972 & 1.68\% \\
adpcm\_encode & 3048620 & 3083658 & 1.15\% \\
aes & 74548 & 75133 & 0.78\% \\
bitcount & 2632864 & 2632886 & 0.00\% \\
blowfish & 1393763 & 1465512 & 5.15\% \\
crc & 14435 & 14509 & 0.51\% \\
dijkstra & 1427512 & 1432030 & 0.32\% \\
limits & 4434 & 4436 & 0.05\% \\
patricia & 1610777 & 1632902 & 1.37\% \\
picojpeg & 2038299 & 2074049 & 1.75\% \\
qsort & 514081 & 513281 & -0.16\% \\
randmath & 38221 & 38601 & 0.99\% \\
rc4 & 67938 & 68072 & 0.20\% \\
rsa & 49450 & 52004 & 5.16\% \\
sha & 1464582 & 1474223 & 0.66\% \\
\midrule
Average & & & 1.31\% \\
\bottomrule
\end{tabular}}

\begin{figure}[h!]
 \begin{minipage}{.48\textwidth}
    \centering
    \includegraphics[width=\linewidth]{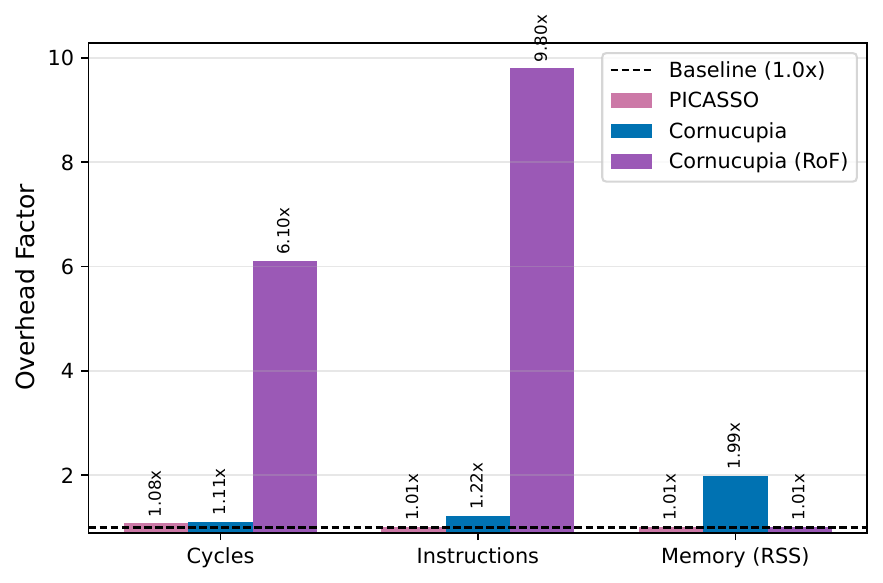}
    \caption{Normalized cycles, instructions, and memory overhead for SQLite speedtest1 for Cornucopia, Cornucopia-RoF and \PICASSO.}\label{fig:sqllitesystemmetric}
 \end{minipage}
\end{figure}

    \begin{minipage}{.5\textwidth}
    \centering
    \includegraphics[scale=0.48]{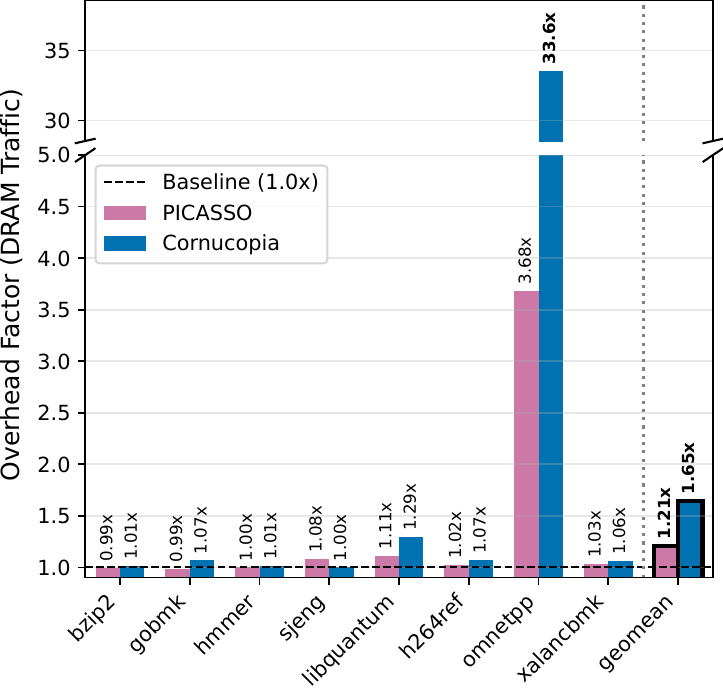}
    \captionof{figure}{Approximation of DRAM traffic overhead.}\label{fig:cc-specdram}
    \end{minipage}

 \fi
\onecolumn

\ifarxiv
\begin{figure*}[t!]
    \centering
    \begin{minipage}{.55\textwidth}
    \includegraphics[scale=0.50]{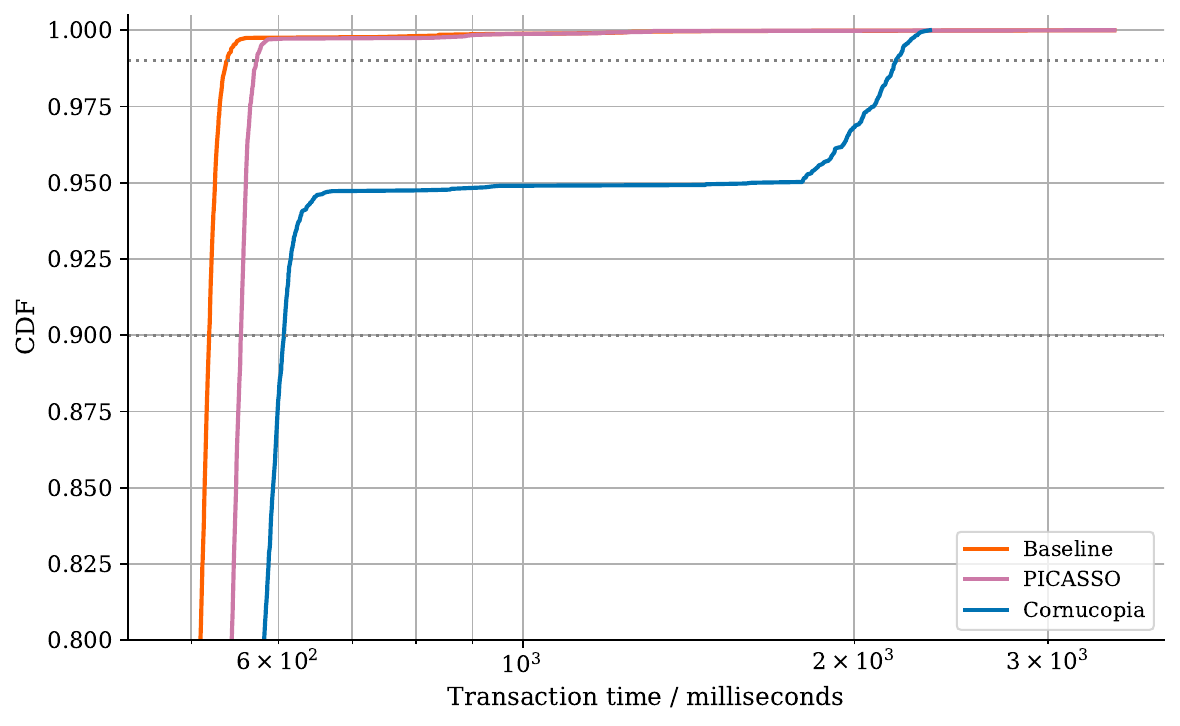}
    \vspace{12pt}
    \caption{PostgreSQL pgbench tail latency expressed similarly as in~\cite{Filardo24}.}\label{cc-pgbench-tail}
    \end{minipage}
    \hfill
    \begin{minipage}{.40\textwidth}
    \includegraphics[scale=0.6]{figures/spec_dram_traffic_overhead.pdf}
    \caption{Approximation of DRAM traffic overhead compare to baseline.}\label{fig:cc-specdram}
    \end{minipage}
\end{figure*}
\fi
\begin{table*}[t]
\centering
\caption{PostgreSQL pgbench: TPS and transaction latency percentiles (ms) with overhead vs.\ baseline}
\label{tab:pgbench-latency}
\begin{tabular}{lccccc}
\toprule
Scenario & TPS & P50 & P90 & P99 & P99.9 \\
\midrule
Baseline & 2.10 & 463.3 & 519.1 & 538.8 & 1129.9 \\
PICASSO & 1.97 (-6.2\%) & 494.6 (+6.7\%) & 554.7 (+6.9\%) & 573.5 (+6.4\%) & 1165.1 (+3.1\%) \\
Cornucopia & 1.61 (-23.1\%) & 538.9 (+16.3\%) & 606.8 (+16.9\%) & 2182.3 (+305.0\%) & 2291.3 (+102.8\%) \\
\bottomrule
\end{tabular}
\end{table*}
\ifarxiv
\begin{figure}[h!]
\centering

 \begin{minipage}{.45\textwidth}

\captionof{table}{PICASSO overhead on MiBench}\label{tab:mibench}
\begin{tabular}{lrrr}
\toprule
Benchmark & Baseline & PICASSO & Overhead (\%) \\
\midrule
adpcm\_decode & 3325012 & 3380972 & 1.68\% \\
adpcm\_encode & 3048620 & 3083658 & 1.15\% \\
aes & 74548 & 75133 & 0.78\% \\
bitcount & 2632864 & 2632886 & 0.00\% \\
blowfish & 1393763 & 1465512 & 5.15\% \\
crc & 14435 & 14509 & 0.51\% \\
dijkstra & 1427512 & 1432030 & 0.32\% \\
limits & 4434 & 4436 & 0.05\% \\
patricia & 1610777 & 1632902 & 1.37\% \\
picojpeg & 2038299 & 2074049 & 1.75\% \\
qsort & 514081 & 513281 & -0.16\% \\
randmath & 38221 & 38601 & 0.99\% \\
rc4 & 67938 & 68072 & 0.20\% \\
rsa & 49450 & 52004 & 5.16\% \\
sha & 1464582 & 1474223 & 0.66\% \\
\midrule
Average & & & 1.31\% \\
\bottomrule
\end{tabular}

\end{minipage}

 \begin{minipage}{.35\textwidth}
    \centering
    \includegraphics[width=\linewidth]{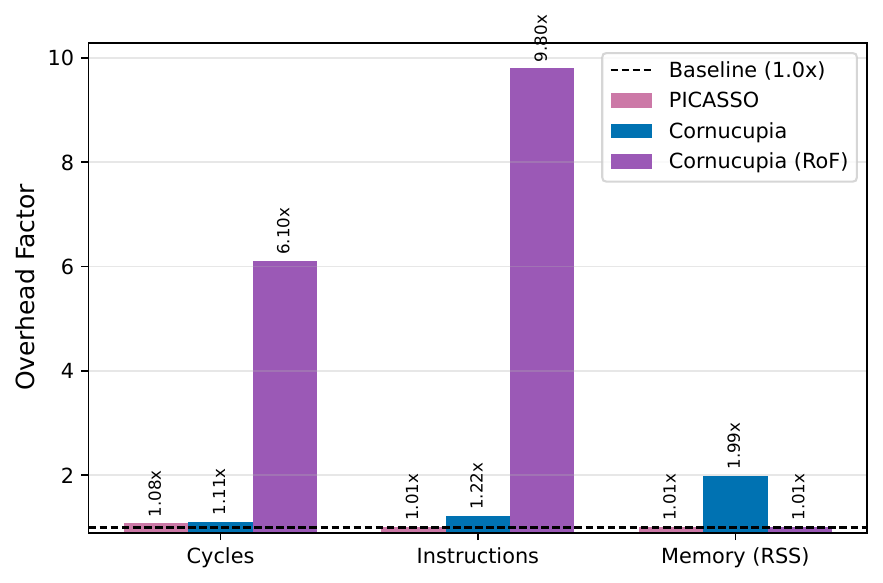}
    \caption{Normalized cycles, instructions, and memory overhead for SQLite speedtest1 for Cornucopia and Cornucopia-RoF and \PICASSO.}\label{fig:sqllitesystemmetric}
 \end{minipage}
\end{figure}
\fi

\begin{table*}[h!]
    \centering
    \caption{Performance cost on VCU118 @ 25MHz expressed as CoreMark test results for $5x10^3$ iterations. The CoreMark score for a processor is reported as CoreMark-iterations-per-second-per-core-MHz. The $\Delta$ is relative to CHERI-Toooba Core results.}\label{tab:coremark}
    \resizebox{\textwidth}{!}{
    \begin{tabular}{r r rrc rrr rrr  c}\toprule
              &           &         Total ticks & \multicolumn{2}{c}{$\Delta$} &      Total time (sec) &   \multicolumn{2}{c}{$\Delta$} & Iterations/sec &  \multicolumn{2}{c}{$\Delta$} &         Score \\ \midrule
        \multicolumn{2}{c}{\textbf{CHERI-Toooba}}&&&&&&& \\
        & baseline (nocap) & 938984056 & \multicolumn{2}{c}{--}
                                               &        37 & \multicolumn{2}{c}{--}
                                               &       135 & \multicolumn{2}{c}{--}
                                               &       5.4 \\

        \rowcolor{white}    & purecap          & 954633763 & +15649707 & $+1.67\%$
                                               &        38 &        +1 &    $+2.70\%$
                                               &       131 &        -4 &  $-2.96\%$
                                               &      5.24 \\
        \midrule 
         \multicolumn{2}{c}{\textbf{\PICASSO CHERI-Toooba}}&&&&&&& \\
         & baseline (nocap)           & {973452912} & {+34468856} & {+3.67\%}
                                               &        38   & +1 & $+2.70\%$
                                               & 131         & -4 & $-2.96\%$
                                               & 5.24 \\

        \rowcolor{white}   & purecap           & 995997744 & {+57013688} & +6.07\%
                                               & 39        & +2 & $+5.41\%$
                                               & 128 & -7 &  $-5.19\%$
                                               & 5.12 \\
        \bottomrule  
\end{tabular}
}
\end{table*}

\begin{table*}[h!]
\centering
\caption{gRPC benchmark: QPS and latency percentiles with overhead vs.\ baseline}
\label{tab:grpc-benchmark}
\resizebox{.8\textwidth}{!}{
\begin{tabular}{llccccc}
\toprule
 Scenario & QPS & P50 (ms) & P90 (ms) & P95 (ms) & P99 (ms) \\
\midrule
 Baseline & 15.24 & 515.1 & 569.0 & 580.4 & 598.0 \\
  PICASSO & 14.80 (-2.9\%) & 530.7 (+3.0\%) & 586.2 (+3.0\%) & 604.0 (+4.1\%) & 628.5 (+5.1\%) \\
  Cornucopia & 12.87 (-15.6\%) & 552.3 (+7.2\%) & 634.8 (+11.6\%) & 687.4 (+18.4\%) & 2202.0 (+268.2\%) \\
\bottomrule
\end{tabular}}
\end{table*}

\ifarxiv
\begin{table*}[t]
\centering
\caption{SQLite Speedtest1 Benchmark Results}\label{tab:sqlite-benchmark}\label{tab:sqlitephase}
\resizebox{.7\textwidth}{!}{
\begin{tabular}{@{}llrrrrr@{}}
\toprule
ID & Benchmark & \multicolumn{2}{c}{PICASSO} & \multicolumn{2}{c}{Cornucopia} \\
\cmidrule(lr){3-4} \cmidrule(lr){5-6}
 & & Time (s) & Overhead & Time (s) & Overhead \\
\midrule
100 & 50000 INSERTs into table with no index & 53.48 & 1.09$\times$ & 49.06 & 1.00$\times$ \\
110 & 50000 ordered INSERTS with one index/PK & 91.37 & 1.10$\times$ & 92.61 & 1.11$\times$ \\
120 & 50000 unordered INSERTS with one index/PK & 103.55 & 1.09$\times$ & 112.15 & 1.18$\times$ \\
130 & 25 SELECTS, numeric BETWEEN, unindexed & 104.66 & 1.10$\times$ & 96.00 & 1.01$\times$ \\
140 & 10 SELECTS, LIKE, unindexed & 125.58 & 1.04$\times$ & 129.20 & 1.06$\times$ \\
142 & 10 SELECTS w/ORDER BY, unindexed & 207.27 & 1.09$\times$ & 198.22 & 1.04$\times$ \\
145 & 10 SELECTS w/ORDER BY and LIMIT, unindexed & 165.48 & 1.09$\times$ & 151.94 & 1.00$\times$ \\
150 & CREATE INDEX five times & 143.72 & 1.12$\times$ & 163.38 & 1.27$\times$ \\
160 & 10000 SELECTS, numeric BETWEEN, indexed & 88.59 & 1.03$\times$ & 87.46 & 1.02$\times$ \\
161 & 10000 SELECTS, numeric BETWEEN, PK & 89.71 & 1.03$\times$ & 89.23 & 1.03$\times$ \\
170 & 10000 SELECTS, text BETWEEN, indexed & 175.43 & 1.04$\times$ & 167.54 & 0.99$\times$ \\
180 & 50000 INSERTS with three indexes & 156.29 & 1.08$\times$ & 200.89 & 1.39$\times$ \\
190 & DELETE and REFILL one table & 158.07 & 1.07$\times$ & 207.29 & 1.41$\times$ \\
200 & VACUUM & 120.56 & 1.07$\times$ & 182.75 & 1.62$\times$ \\
210 & ALTER TABLE ADD COLUMN, and query & 4.67 & 1.09$\times$ & 4.33 & 1.01$\times$ \\
230 & 10000 UPDATES, numeric BETWEEN, indexed & 86.99 & 1.10$\times$ & 80.19 & 1.02$\times$ \\
240 & 50000 UPDATES of individual rows & 120.51 & 1.07$\times$ & 111.52 & 0.99$\times$ \\
250 & One big UPDATE of the whole 50000-row table & 20.43 & 1.07$\times$ & 19.02 & 1.00$\times$ \\
260 & Query added column after filling & 5.02 & 1.09$\times$ & 4.57 & 0.99$\times$ \\
270 & 10000 DELETEs, numeric BETWEEN, indexed & 188.16 & 1.07$\times$ & 180.69 & 1.02$\times$ \\
280 & 50000 DELETEs of individual rows & 136.09 & 1.06$\times$ & 128.31 & 1.00$\times$ \\
290 & Refill two 50000-row tables using REPLACE & 338.96 & 1.08$\times$ & 399.53 & 1.27$\times$ \\
300 & Refill a 50000-row table using (b\&1)==(a\&1) & 159.50 & 1.07$\times$ & 219.51 & 1.48$\times$ \\
310 & 10000 four-ways joins & 347.18 & 1.07$\times$ & 319.78 & 0.99$\times$ \\
320 & subquery in result set & 707.26 & 1.08$\times$ & 649.39 & 1.00$\times$ \\
400 & 70000 REPLACE ops on an IPK & 110.89 & 1.08$\times$ & 126.69 & 1.23$\times$ \\
410 & 70000 SELECTS on an IPK & 78.81 & 1.11$\times$ & 71.51 & 1.00$\times$ \\
500 & 70000 REPLACE on TEXT PK & 119.23 & 1.09$\times$ & 130.57 & 1.19$\times$ \\
510 & 70000 SELECTS on a TEXT PK & 119.61 & 1.11$\times$ & 107.79 & 1.00$\times$ \\
520 & 70000 SELECT DISTINCT & 65.48 & 1.06$\times$ & 97.70 & 1.59$\times$ \\
980 & PRAGMA integrity\_check & 321.43 & 1.10$\times$ & 289.91 & 0.99$\times$ \\
990 & ANALYZE & 44.59 & 1.08$\times$ & 41.07 & 0.99$\times$ \\
\midrule
& \textbf{TOTAL} & 4758.56 & 1.08$\times$ & 4909.81 & 1.11$\times$ \\
\bottomrule
\end{tabular}}
\end{table*}
\fi

\end{document}